\documentclass[prodmode,acmjacm]{acmsmall}
\usepackage{amsmath,amssymb,amscd,epsfig,url}

\def\<{{\langle}} \def\>{{\rangle}}       
\def\Vmst{{V(G) \setminus \{s,t\}}}
\usepackage[ruled]{algorithm2e}

\SetAlFnt{\small}
\SetAlCapFnt{\small}
\SetAlCapNameFnt{\small}
\SetAlCapHSkip{0pt}
\IncMargin{-\parindent}

\acmVolume{9}
\acmNumber{4}
\acmArticle{39}
\acmYear{2014}
\acmMonth{3}

\begin{document}

\markboth{A. Potechin}{Bounds on monotone switching networks for directed connectivity}

\title{Bounds on Monotone Switching Networks for Directed Connectivity}
\author{AARON POTECHIN
\affil{Institute for Advanced Study}}

\begin{abstract}
We separate monotone analogues of $L$ and $NL$ by proving that any monotone switching network solving 
directed connectivity on $n$ vertices must have size at least $n^{\Omega{(\lg{n})}}$
\end{abstract}




\acmformat{Ptechin, A. 2014. 
Bounds on monotone switching networks for directed connectivity.}

\begin{bottomstuff}
This material is based on work supported by the National Science Foundation Graduate Research Fellowship 
under Grant No. 0645960.
Author's address: A. Potechin, Mathematics Department, MIT
\end{bottomstuff}

\maketitle

\section{Introduction}\label{intro}
$L$ versus $NL$, the problem of whether non-determinism helps in logarithmic space 
bounded computation, is a longstanding open question in computational complexity. At present, only 
a few results are known. It is known that the problem is equivalent to the question of whether there is 
a log-space algorithm for the \textit{directed connectivity} problem, namely given an $n$ vertex directed 
graph $G$ and pair of vertices $s,t$, find out if there is a directed path from $s$ to $t$ in $G$. 
Savitch \cite{savitch} gave an $O(\log^{2}n)$-space deterministic algorithm for 
directed connectivity, thus proving that $NSPACE(g(n)) \subseteq DSPACE((g(n)^2))$ for 
every space constructable function $g$. Immerman \cite{nlconlone} and 
Szelepcs\'{e}nyi \cite{nlconltwo} independently gave an $O(\log n)$-space non-deterministic algorithm 
for directed \textit{non-connectivity}, thus proving that $NL = co$-$NL$. For the problem of 
\textit{undirected connectivity} (i.e. where the input graph $G$ is undirected), a probabilistic 
algorithm was shown using random walks by Aleliunas, Karp, Lipton, Lov\'{a}sz, and Rackoff \cite{randomwalk}, 
and Reingold \cite{undirectedgraph} gave a deterministic 
$O(\log n)$-space algorithm for the same problem, showing that undirected connectivity is in $L$. 
Trifonov \cite{trifonov} independently gave an $O(\lg{n}\lg{\lg{n}})$ space algorithm for undirected connectivity.

So far, most of the work trying to show that $L \neq NL$ has been done 
using the JAG model or the branching program model. The JAG (Jumping Automata on Graphs) model was introduced by 
Cook and Rackoff \cite{cook} as a simple model for which we can prove good lower time and space bounds but which is still 
powerful enough to simulate most known algorithms for the st-connectivity problem. This implies that 
if there is an algorithm for st-connectivity breaking these bounds, it must use some new techniques which cannot be captured 
by the JAG model. Later work in this area has focused on extending this framework to additional algorithms and more 
powerful variants of the JAG model. Two relatively recent results in this area are the result of 
Edmonds, Poon, and Achlioptas \cite{JAGlowerbounds} which shows tight lower bounds for the more powerful 
NNJAG (Node-Named Jumping Automata on Graphs) model and the result of Lu, Zhang, Poon, and Cai \cite{JAGsimulations} showing 
that Reingold's algorithm for undirected connectivity and a few other algorithms for undirected connectivity can all be 
simulated by the RAM-NNJAG model, a uniform variant of the NNJAG model.

Ironically, the branching program model was originally developed as a way to make more efficient switching networks. In the 
early 20th century switching networks were used for practical applications, so the research focus was on finding good ways 
to construct switching networks for various problems. For example, see the pioneering papers of Shannon \cite{shannonone}, 
\cite{shannontwo}. Lee \cite{lee} developed the branching program model as a way to easily construct switching networks, 
as if we have a branching program for a problem, we can make a switching network for that problem simply by making all of 
the edges undirected. Masek \cite{masek} showed that the branching program model could be used to show space lower bounds 
and there has been a lot of research in this direction ever since. For a survey of some of the many results on 
branching programs, switching networks, and a related model, switching-and-rectifier networks, see 
Razborov \cite{razborov}.

In this paper, we explore trying to prove $L \neq NL$ using the switching network model. This may seem like a strange choice, 
as switching networks are less intuitive to think about than branching programs. However, switching networks have the very 
nice property of reversibility, which is crucial for our techniques and results.

While we eventually hope to prove strong lower bounds on general switching networks, such bounds are currently beyond our 
reach. Thus, for now we must place a restriction on the switching networks we analyze in order to obtain good lower 
bounds. We choose to restrict ourselves to monotone switching networks, which is a natural choice for two main reasons. First, 
to prove general lower bounds we must prove monotone lower bounds along the way, so we may as well start by trying to prove 
monotone lower bounds. Second, the restriction to monotone switching networks is simple and clean.

Indeed, monotone complexity theory is a very rich field and researchers have had great success in separating different monotone 
complexity classes. Razborov \cite{razborov} used an approximation method to show that any monotone circuit solving the k-clique 
problem on $n$ vertices (determining whether or not there is a set of $k$ pairwise adjacent vertices in a graph on $n$ vertices) 
when $k = \lceil \frac{\lg{n}}{4} \rceil$ must have size at least $n^{\Omega({\lg{n}})}$, thus proving that $mP \neq mNP$. 
This method was later improved by Alon and Boppana \cite{alonboppana} and by Haken \cite{haken}. Karchmer and 
Wigderson \cite{karchmerwigderson} showed that any monotone circuit solving undirected connectivity has depth at least 
$\Omega((\lg{n})^2)$, thus proving that undirected connectivity is not in monotone-$NC^1$ and separating monotone-$NC^1$ 
and monotone-$NC^2$. Raz and McKenzie \cite{razmckenzie} later separated the entire monotone $NC$ hierarchy, proving 
that monotone-$NC \neq$ monotone-$P$ and for any $i$, monotone-$NC^i \neq$ monotone-$NC^{i+1}$.

While our techniques are very different, our results build on this knowledge. We show that any monotone switching network 
solving the directed connectivity problem on a set of vertices $V(G)$ with $n = |V(G)|$ must have size at least 
$n^{\Omega{(\lg{n})}}$, which solves open problem 2 of 
Grigni and Sipser \cite{monotonecomplexity} (which is also open problem 4 of Razborov \cite{razborov}) and separates 
monotone analogues of $L$ and $NL$.
\begin{remark}
The question of whether we have separated monotone-L from monotone-NL depends on how monotone-L is defined. If we define (non-uniform) monotone-L to be the class of all functions computable by polynomial size monotone switching networks, then we indeed have this separation. However, as noted in Grigni and Sipser \cite{monotonecomplexity}, (non-unifrom) monotone-L can also be defined as the class of all functions computable by polynomial size, logarithmic width monotone circuits. The relationship of these two definitions of (non-uniform) monotone-L to each other is an open problem. 
\end{remark}
\begin{remark}
There have been several papers building on this work since it was first presented. In a follow-up work, Chan and Potechin \cite{chan} generalized the techniques used here to the iterated indexing and k-clique problems, showing tight monotone lower space bounds, giving an alternate proof of the separation of the monotone NC-hierarchy. Robere, Cook, Filmus, and Pitassi \cite{averagebounds} later showed an average case lower bound on monotone switching networks for directed connectivity over some distribution of inputs. Both of these papers provide an alternate presentation of the results here.
\end{remark}
\subsection{Notation and definitions}\label{definitions}
Throughout the paper, we will be dealing with two main graphs, the input graph $G$ and the switching network $G'$. To make it clear which one we are discussing at any given time, we use unprimed letters to denote objects related to the input graph $G$ and we use primed letters to denote objects related to the switching network $G'$. Also, we use lowercase letters for single objects like vertices, edges, and functions and we use capital letters for sets and more complicated objects like graphs, paths, and walks.

We now give several definitions which will be used throughout the paper and which will allow us easily state our results. Since we focus on switching networks for the directed connectivity problem, 
we start with a specialized definition of switching networks for directed connectivity. 
\begin{definition}\label{modifiedswitchingdefinition}
A switching network for directed connectivity on a set of vertices $V(G)$ with distinguished vertices $s,t$ is a tuple $< G', s', t', {\mu}'>$ where $G'$ is an undirected multi-graph with distinguished 
vertices $s'$,$t'$ and ${\mu}'$ is a labeling function giving each edge $e' \in E(G')$ a label of the form 
$v_1 \to v_2$ or $\neg(v_1 \to v_2)$ for some vertices $v_1, v_2 \in V(G)$ with $v_1 \neq v_2$.
\end{definition}
For the remainder of the paper, we will assume the following
\begin{enumerate}
\item We have a set of vertices $V(G)$ with distinguished vertices $s,t$.
\item All input graphs $G$ have vertex set $V(G)$.
\item All switching network are switching networks for directed connectivity on $V(G)$.
\end{enumerate}
\begin{remark}
Since we are always assuming our switching networks are switching networks for directed connectivity on a set of vertices $V(G)$ with distinguished vertices $s,t$, we will just write switching network for brevity.
\end{remark}
\begin{definition}
We take the size of the input graph to be $n = |\Vmst|$. We exclude $s,t$ when considering the size of $V(G)$ because it makes our calculations easier.
\end{definition}
\begin{definition} \ 
\begin{enumerate}
\item We say that a switching network $G'$ \textbf{accepts} an input graph $G$ if there is a path $P'$ in $G'$ from $s'$ to $t'$ such that for each edge $e' \in E(P')$, $\mu'(e')$ is consistent with the input graph $G$ 
(i.e. of the form $e$ for some edge $e \in E(G)$ or $\neg{e}$ for some $e \notin E(G)$).
\item We say that $G'$ is \textbf{sound} if it does not accept any input graphs $G$ which do not have a path from $s$ to $t$.
\item We say that $G'$ is \textbf{complete} if it accepts all input graphs $G$ which have a path from $s$ to $t$.
\item We say that $G'$ solves directed connectivity if $G'$ is both sound and complete. 
\item We take the \textbf{size} of $G'$ to be $|V(G')|$.
\item We say that $G'$ is \textbf{monotone} if it has no labels of the form $\neg(v_1 \to v_2)$.
\end{enumerate}
\end{definition}
\begin{figure}[ht]
\centerline{\includegraphics[height=4cm]{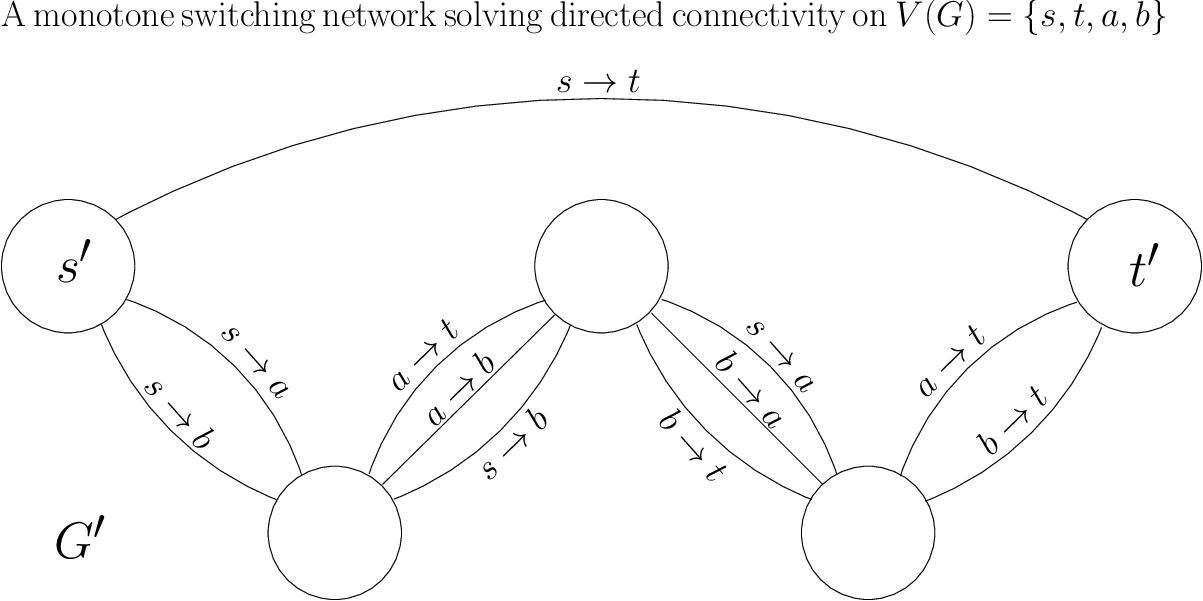}}
\caption{In this figure, we have a monotone switching network $G'$ that solves directed connectivity on 
$V(G) = \{s,t,a,b\}$, i.e. there is a path from $s'$ to $t'$ in $G'$ whose labels are consistent with 
the input graph $G$ if and only if there is a path from $s$ to $t$ in $G$. For example, if we have the edges 
$s \to a$, $a \to b$, and $b \to t$ in $G$, so there is a path from $s$ to $t$ in $G$, then in $G'$, 
starting from $s'$, we can take the edge labeled $s \to a$, then the edge labeled $a \to b$, 
then the edge labeled $s \to a$, and finally the 
edge labeled $b \to t$, and we will reach $t'$. If in $G$ we have the edges $s \to a$, $a \to b$, 
$b \to a$, and $s \to b$ and no other edges, so there is no path from $s$ to $t$, then in $G'$ there is 
no edge that we can take to $t'$, so there is no path from $s'$ to $t'$.}
\label{examplenetwork}
\end{figure}
\begin{figure}[ht]
\centerline{\includegraphics[height=5cm]{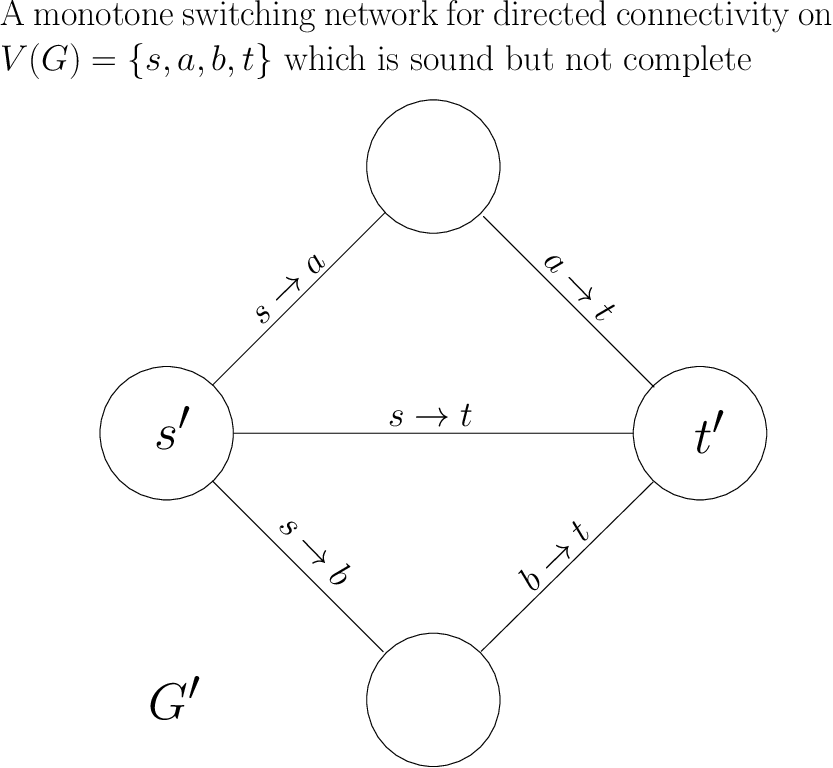}}
\caption{In this figure, we have another monotone switching network $G'$ for directed connectivity on 
$V(G) = \{s,t,a,b\}$. This $G'$ accepts the input graph $G$ if and only $G$ contains either the 
edge $s \to t$, the edges $s \to a$ and $a \to t$, or the edges $s \to b$ and $b \to t$. Thus, this $G'$ 
is sound but not complete.}
\label{soundbutnotcomplete}
\end{figure}
\newpage
\begin{figure}[ht]
\centerline{\includegraphics[height=6cm]{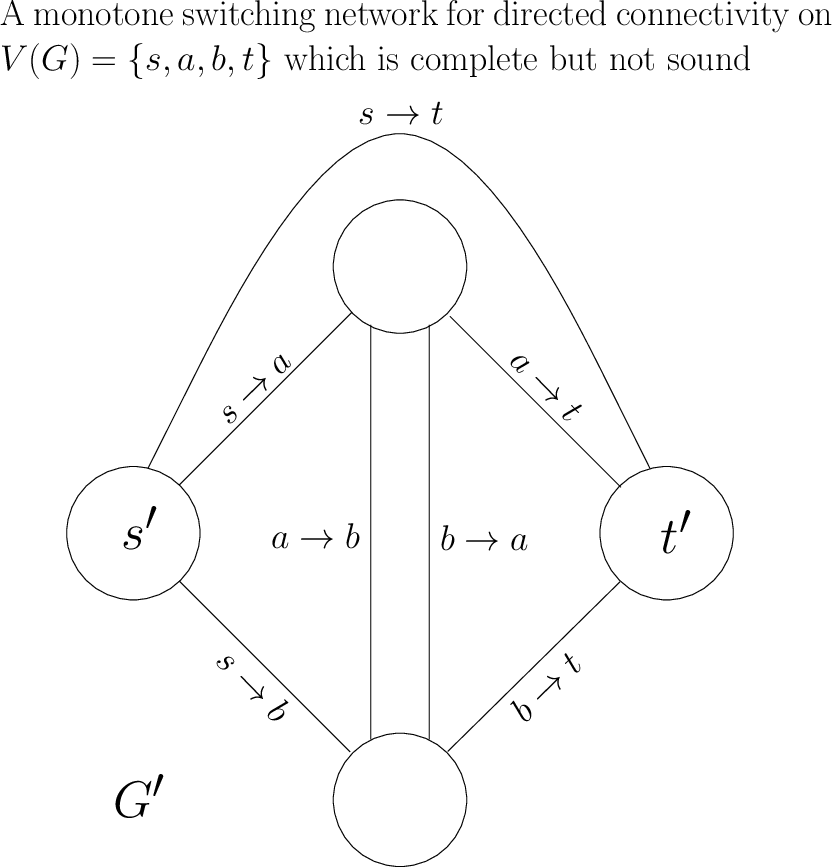}}
\caption{In this figure, we have another monotone switching network $G'$ for directed connectivity on 
$V(G) = \{s,t,a,b\}$. This $G'$ accepts the input graph $G$ whenever $G$ contains a path from $s$ to $t$, so it is 
complete. However, this $G'$ is not sound. To see this, consider the input graph $G$ with $E(G) = \{s \to a, b \to a, b \to t\}$. 
In $G'$, we can start at $s'$, take the edge labeled $s \to a$, then the edge labeled $b \to a$, then the edge labeled $b \to t$, 
and we will reach $t'$. Thus, this $G'$ accepts the given input graph $G$, but $G$ does not contain a path from $s$ to $t$ because 
the edge from $b$ to $a$ goes the wrong way.}
\label{completebutnotsound}
\end{figure}
\begin{remark}
Figures \ref{soundbutnotcomplete} and \ref{completebutnotsound} illustrate why we can't just have one vertex of the switching network $G'$ for each vertex of our original graph $G$. The 
reason is that we are trying to simulate a {\bf directed} graph with an {\bf undirected} graph.
\end{remark}
\begin{figure}[ht]
\centerline{\includegraphics[height=6cm]{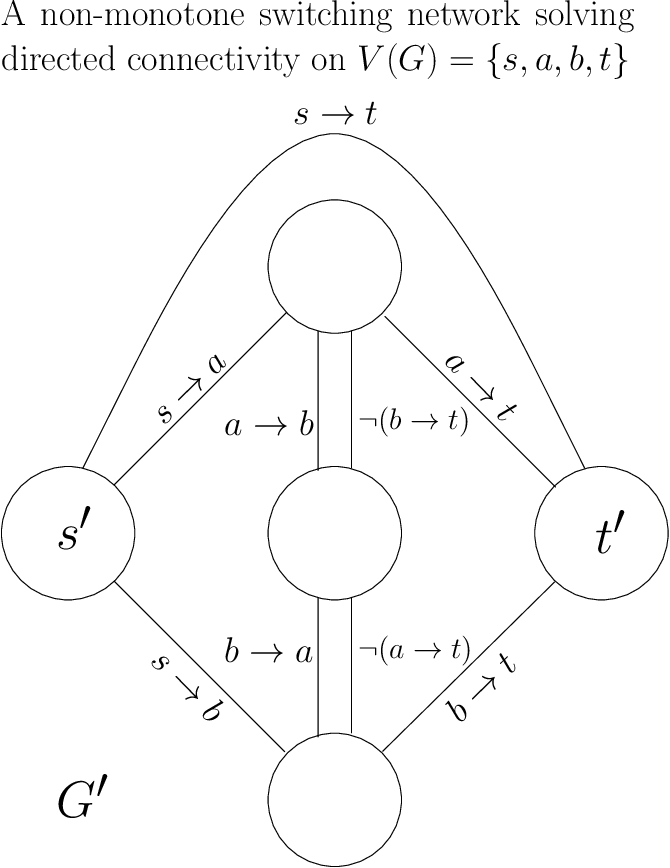}}
\caption{In this figure, we have a non-monotone switching network $G'$ solving directed connectivity on 
$V(G) = \{s,t,a,b\}$. Note that the edge with label $\neg{(b \to t)}$ and the edge with label $\neg{(a \to t)}$ are 
necessary for $G'$ to be complete. To see this, consider 
the input graph $G$ with $E(G) = \{s \to a, a \to b, b \to t\}$. To get from $s'$ to $t'$ in $G'$ we must first take the edge 
labeled $s \to a$, then take the edge labeled $a \to b$, then take the edge labeled $\neg{(a \to t)}$, and finally 
take the edge labeled $b \to t$.}
\label{nonmonotone}
\end{figure}
\begin{remark}
It is natural to ask where the examples of Figures \ref{examplenetwork} and \ref{nonmonotone} come from. 
As we will see, there are many different ways to construct switching networks solving directed connectivity 
on a set of vertices and we will give the particular constructions leading to the switching networks in 
Figures \ref{examplenetwork} and \ref{nonmonotone} later in the paper. For now, the reader should just 
make sure that he/she understands Definition \ref{modifiedswitchingdefinition}. 
That said, it is a good exercise to verify that these 
switching networks have the claimed properties and to try and figure out what they are doing.
\end{remark}
In this paper we analyze monotone switching networks. However, rather than looking 
at all possible input graphs, we focus on particular sets of input graphs. To do this, instead of assuming that the 
switching networks we analyze solve directed connectivity, we only assume that these switching networks solve the promise problem where the input graph $G$ is guaranteed to either be in some set $I$ of input graphs which 
contain a path from $s$ to $t$ or to not contain a path from $s$ to $t$.
\begin{definition}\label{inputdifficulty}
Given a set $I$ of input graphs which all contain a path from $s$ to $t$, 
let $m(I)$ be the size of the smallest sound monotone switching network which accepts all of the input 
graphs in $I$.
\end{definition}
In this paper, we focus on input graphs which contain a path from $s$ to $t$ and no other edges, as they are the minimal YES instances and are thus the hardest input graphs for a monotone switching network to accept. We have the following definitions.
\begin{definition} \ 
\begin{enumerate}
\item Define $\mathcal{P}_{n,l}$ (where $n = |\Vmst|$) to be the set of input graphs $G$ such that 
$E(G) = \{v_0 \to v_1, v_1 \to v_2, \cdots, v_{l-1} \to v_l\}$ where $v_0 = s$, $v_l = t$, and 
$v_0, \cdots, v_l$ are distinct vertices of $V(G)$. 
\item Define $\mathcal{P}_{n,\leq l} = \cup_{j = 1}^{l}{\mathcal{P}_{n,j}}$
\item Define $\mathcal{P}_{n} = \mathcal{P}_{n,\leq n+1} = \cup_{j = 1}^{n+1}{\mathcal{P}_{n,j}}$
\end{enumerate}
\end{definition}
\begin{proposition}
A monotone switching network $G'$ solves directed connectivity if and only if 
it is sound and accepts every input graph in $\mathcal{P}_{n}$.
\end{proposition}
\begin{corollary}
The size of the smallest monotone switching network solving directed connectivity on $n$ vertices (excluding $s,t$) is $m(\mathcal{P}_{n})$. 
\end{corollary}
\subsection{Paper outline and results}
Our main result is the following theorem
\begin{theorem}\label{bigresult}
If $n \geq 1$ and $l \geq 2$ then 
\begin{enumerate}
\item $\frac{1}{2}\left(\frac{n}{64(l-1)^2}\right)^{\frac{\lceil{\lg{l}}\rceil}{2}} \leq m(\mathcal{P}_{n,l}) 
\leq m(\mathcal{P}_{n,\leq l}) \leq n^{\lceil{\lg{l}}\rceil} + 2$
\item $\frac{1}{2}n^{\frac{\lg{n}}{16} - \frac{3}{4}} \leq m(\mathcal{P}_{n}) \leq n^{\lg{n} + 1}+2$
\end{enumerate}
\end{theorem}
We build up to this result step by step. In Section \ref{specialcase} we use a bottleneck argument to prove the result for an even more restricted class of switching networks, certain knowledge switching networks. This also provides the upper bounds for Theorem \ref{bigresult}. In Section \ref{introducefouriertransform}, we introduce a very different approach to the problem: Fourier analysis and invariants. While this approach is less intuitive, it allows us to obtain lower bounds on all sound monotone switching networks for directed connectivity, not just certain knowledge switching networks. Using this approach, we show a quadratic lower bound and give conditions sufficient for showing stronger lower bounds. In Section \ref{fourieranalogues}, we synthesize the two approaches. We show how our results about certain knowledge switching networks can be adapted to the Fourier analysis and invariants approach and deduce a superpolynomial lower bound. Finally, in Section \ref{nlgnlowerboundsection} we carry out the analysis more carefully to prove the lower bounds of Theorem \ref{bigresult}.
\section{Certain knowledge switching networks}\label{specialcase}
\noindent In this section, we introduce and analyze certain knowledge switching 
networks, a subclass of monotone switching networks for directed connectivity which are always sound and can 
be described by a simple reversible game for solving directed connectivity. The main results of this section are the following  
upper and lower bounds on the size of certain knowledge switching networks solving directed connectivity. These bounds show that 
certain knowledge switching networks can match the performance of Savitch's algorithm and this is tight.
\begin{definition}
Given a set $I$ of input graphs all of which contain a path from $s$ to $t$, let $c(I)$ be the size of the smallest certain-knowledge switching network which accepts all of the input graphs in $I$.
\end{definition}
\begin{theorem}\label{certainknowledgecleanbounds}
If $l \geq 2$ and $n \geq 2(l-1)^2$ then 
\begin{enumerate}
\item ${(\frac{n}{2(l-1)})}^{\lceil{\lg{l}}\rceil} \leq c(\mathcal{P}_{n,l}) \leq 
c(\mathcal{P}_{n,\leq l}) \leq n^{\lceil{\lg{l}}\rceil}+2$
\item $n^{\frac{1}{4}\lg{n} - \frac{1}{2}} \leq c(\mathcal{P}_{n}) \leq n^{\lg{n} + 1}+2$
\end{enumerate}
\end{theorem}
\subsection{The certain knowledge game for directed connectivity}\label{reversiblegame}
We will define certain knowledge switching networks using the following simple reversible game for determining whether there is a path from $s$ to $t$ in an input graph $G$.
\begin{definition}[Certain knowledge game]
We define a knowledge set $K$ to be a set of edges between vertices of $V(G)$. An edge $u \to v$ in $K$ represents the knowledge that there is a path from $u$ to $v$ in $G$. We do not allow knowledge sets to contain loops. 

In the certain knowledge game, we start with the empty knowledge set $K = \{\}$ and use the following types of moves:
\begin{enumerate}
\item If we directly see that $v_1 \to v_2 \in E(G)$, we may add or remove $v_1 \to v_2$ from $K$.
\item If edges $v_3 \to v_4$, $v_4 \to v_5$ are both in $K$ and $v_3 \neq v_5$, we may add or remove $v_3 \to v_5$ 
from $K$.
\end{enumerate}
We win the certain knowledge game if we obtain a knowledge set $K$ containing a path from $s$ to $t$.
\end{definition}
\begin{proposition}
The certain knowledge game is winnable for an input graph $G$ if and only if there is a path from $s$ to $t$ in $G$.
\end{proposition}
\subsection{Adapting the certain knowledge game for monotone switching networks}\label{adaptingreversiblegame}
Intuitively, certain knowledge switching networks are switching networks $G'$ where each vertex $v' \in V(G')$ 
corresponds to a knowledge set $K_{v'}$ and the edges between the vertices of $G'$ correspond to moves from 
one knowledge set to another. However, there are two issues that need to be addressed. First, if $G'$ is a 
switching network, $u',v',w' \in V(G')$, and there are edges with label $e$ between $u'$ and $v'$ and 
between $v'$ and $w'$, then we may as well add an edge with label $e$ between $u'$ and $w'$. This edge 
now represents not just one move in the game but rather several moves. Thus an edge in $G'$ with label $e$ 
should correspond to a sequence of moves from one knowledge set to another, each of which can be done with just 
the knowledge that $e \in E(G)$.

The second issue is that the basic certain knowledge game has many winning states but a switching network 
$G'$ has only one accepting vertex $t'$. To address this, we need to merge all of the winning states of 
the game into one state. To do this, we add a move to the game allowing us to go from any winning state to 
any other winning state.
\begin{definition}[Modified certain knowledge game]
In the modified certain knowledge game, we start with the empty knowledge set $K = \{\}$ and use 
the following types of moves:
\begin{enumerate}
\item If we directly see that $v_1 \to v_2 \in E(G)$, we may add or remove $v_1 \to v_2$ from $K$.
\item If edges $v_3 \to v_4$, $v_4 \to v_5$ are both in $K$ and $v_3 \neq v_5$, we may add or remove $v_3 \to v_5$ 
from $K$.
\item If $s \to t \in K$ then we can add or remove any other edge from $K$.
\end{enumerate}
We win the modified certain knowledge game if we obtain a knowledge set $K$ containing a path from 
$s$ to $t$.
\end{definition}
\begin{remark}
In the modified certain knowledge game, an edge $v_1 \to v_2$ in $K$ now represents knowing that either there 
is a path from $v_1$ to $v_2$ in $G$ or there is a path from $s$ to $t$ in $G$.
\end{remark}
\begin{proposition}
The modified certain knowledge game is winnable for an input graph $G$ if and only if there is a path from 
$s$ to $t$ in $G$.
\end{proposition}
With this modified certain knowledge game, we are now ready to formally define certain knowledge switching networks.
\begin{definition}\label{certainknowledgedef}
We say a monotone switching network $G'$ is a certain knowledge switching network if we can assign a knowledge 
set $K_{v'}$ to each vertex $v' \in V(G')$ so that the following conditions hold:
\begin{enumerate}
\item $K_{s'} = \{\}$
\item $K_{t'}$ contains a path from $s$ to $t$ (this may or may not be just the edge $s \to t$)
\item If there is an edge with label $e = v_1 \to v_2$ between vertices $v'$ and $w'$ in $G'$, then we can 
go from $K_{v'}$ to $K_{w'}$ with a sequence of moves in the modified certain knowledge game, all of which 
can be done using only the knowledge that $v_1 \to v_2 \in E(G)$.
\end{enumerate}
We call such an assignment of knowledge sets to vertices of $G'$ a certain knowledge description of $G'$.
\end{definition}
\begin{proposition}
Every certain knowledge switching network is sound.
\end{proposition}
\begin{proof}
If there is a path from $s'$ to $t'$ in $G'$ which is consistent with an input graph $G$ then it 
corresponds to a sequence of moves in the modified certain knowledge game from $K_{s'} = \{\}$ to a 
$K_{t'}$ containing a path from $s$ to $t$ where each move can be done with an edge in $G$. This implies that the 
modified certain knowledge game can be won for the input graph $G$, so there is a path from $s$ to $t$ in $G$.
\end{proof}
\begin{figure}[ht]
\centerline{\includegraphics[height=6cm]{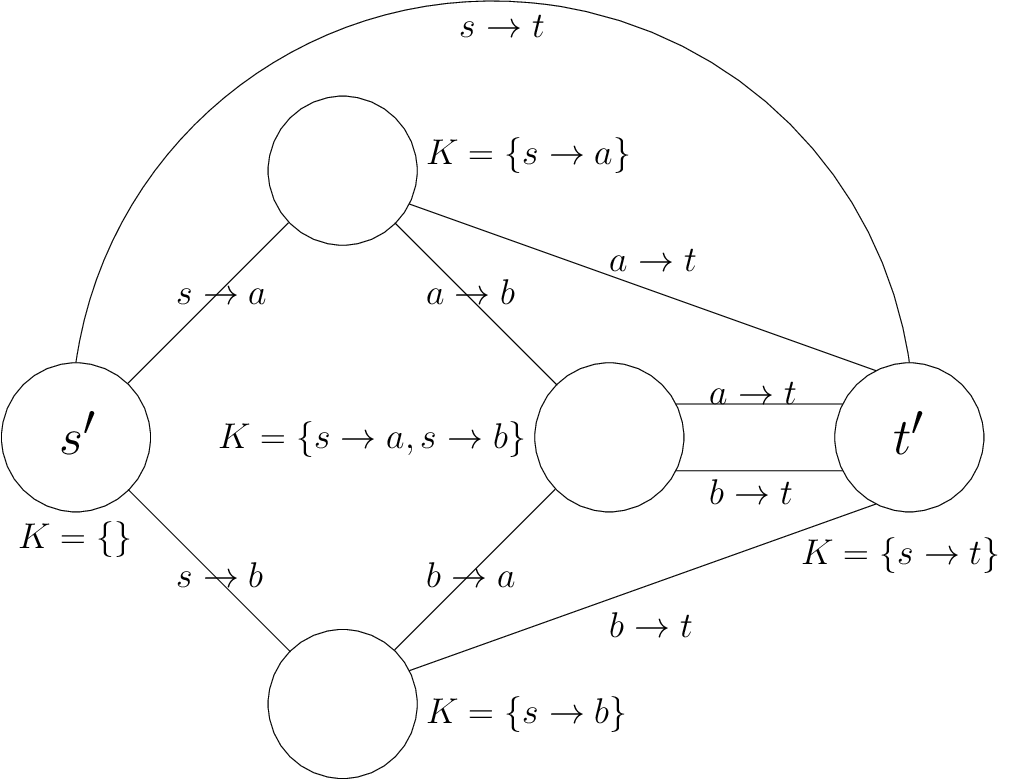}}
\caption[A certain knowledge switching network $G'$ solving directed connectivity on $V(G) = \{s,t,a,b\}$ together with a certain knowledge description of it]{In this figure, we have a certain knowledge switching network $G'$ solving directed connectivity (on 
$V(G) = \{s,t,a,b\}$) together with a certain knowledge description for it.}
\label{examplecertainknowledge}
\end{figure}
\subsection{Connection to the reversible pebbling game for directed connectivity}\label{reachabilityfroms}
While certain knowledge switching networks can consider all paths in $G$, most of our examples only consider reachability from $s$. In this case, the certain knowledge game for directed connectivity reduces to a slightly modified form of a reversible pebbling game for directed connectivity introduced by Bennet \cite{cbennet} to study time/space tradeoffs in computation.
\begin{definition}
For each subset $V \subseteq V(G) \setminus \{s\}$, define $K_V = \{s \to v: v \in V\}$
\end{definition}
\begin{lemma}\label{keysequenceofmoves}
For any $v_1,v_2 \in V(G) \setminus \{s\}$ and $V \subseteq V(G) \setminus \{s\}$ with $v_1 \in V \cup \{s\}$ there is a sequence of moves in 
the modified certain knowledge game from $K_{V}$ to $K_{V \cup \{v_2\}}$ which only requires the knowledge 
that $v_1 \to v_2 \in E(G)$
\end{lemma}
\begin{proof}
The result is trivial if $v_1 = s$ as we can then just add $s \to v_2$ to $K$ directly. Otherwise, starting from $K = K_{V}$, take the following sequence of moves:
\begin{enumerate}
\item Add $v_1 \to v_2$ to $K$.
\item Add $s \to v_2$ to $K$ (we already have $s \to v_1$ and $v_1 \to v_2$ in $K$)
\item Remove $v_1 \to v_2$ from $K$.
\end{enumerate}
We are now at $K = K_{V \cup \{v_2\}}$
\end{proof}
If all of our knowledge sets are of the form $K_V$ and we only use sequences of moves as described above then we can express the certain knowledge game as a pebbling game as follows. The knowledge set $K_V$ corresponds to having pebbles on $V \cup \{s\}$. Now the above sequence of moves corresponds to the following type of move:
\begin{enumerate}
\item If there is a pebble on $v_1$ and we have the edge $v_1 \to v_2$, add or remove a pebble from $v_2$.
\end{enumerate}
This is precisely Bennet's reversible pebbling game for directed connectivity. However, similar to before, it must be modified slightly to merge all accepting states into one state. The resulting reversible pebbling game has the following moves.
\begin{enumerate}
\item If there is a pebble on $v_1$ and we have the edge $v_1 \to v_2$, add or remove a pebble from $v_2$.
\item If there is a pebble on $t$, add or remove any other pebble except the one on $s$
\end{enumerate}
Before moving on, we make two important remarks to help the reader gain intuition for certain knowledge switching networks.
\begin{remark}
Note that in the reversible pebbling game, when we place a pebble on $v_2$ we are NOT allowed to remove the pebble from $v_1$. This is a key difference between this model and the JAG model. The reason is that there is no sequence of moves in the modified certain knowledge game from 
$K_{\{v_1\}}$ to $K_{\{v_2\}}$ which only requires the knowledge that $v_1 \to v_2 \in E(G)$. We are forgetting the fact that there is a 
path from $s$ to $v_1$ in $G$ or a path from $s$ to $t$ in $G$ and forgetting information is irreversible. Also, while 
we can deduce that there is a path from $s$ to $v_2$ in $G$ from the fact that there is a path from 
$s$ to $v_1$ in $G$ and $v_1 \to v_2 \in E(G)$, we cannot deduce that 
there is a path from $s$ to $v_1$ in $G$ from the fact that there is a path from $s$ to $v_2$ in $G$ and 
$v_1 \to v_2 \in E(G)$.
\end{remark}
\begin{figure}[ht]
\centerline{\includegraphics[height=9cm]{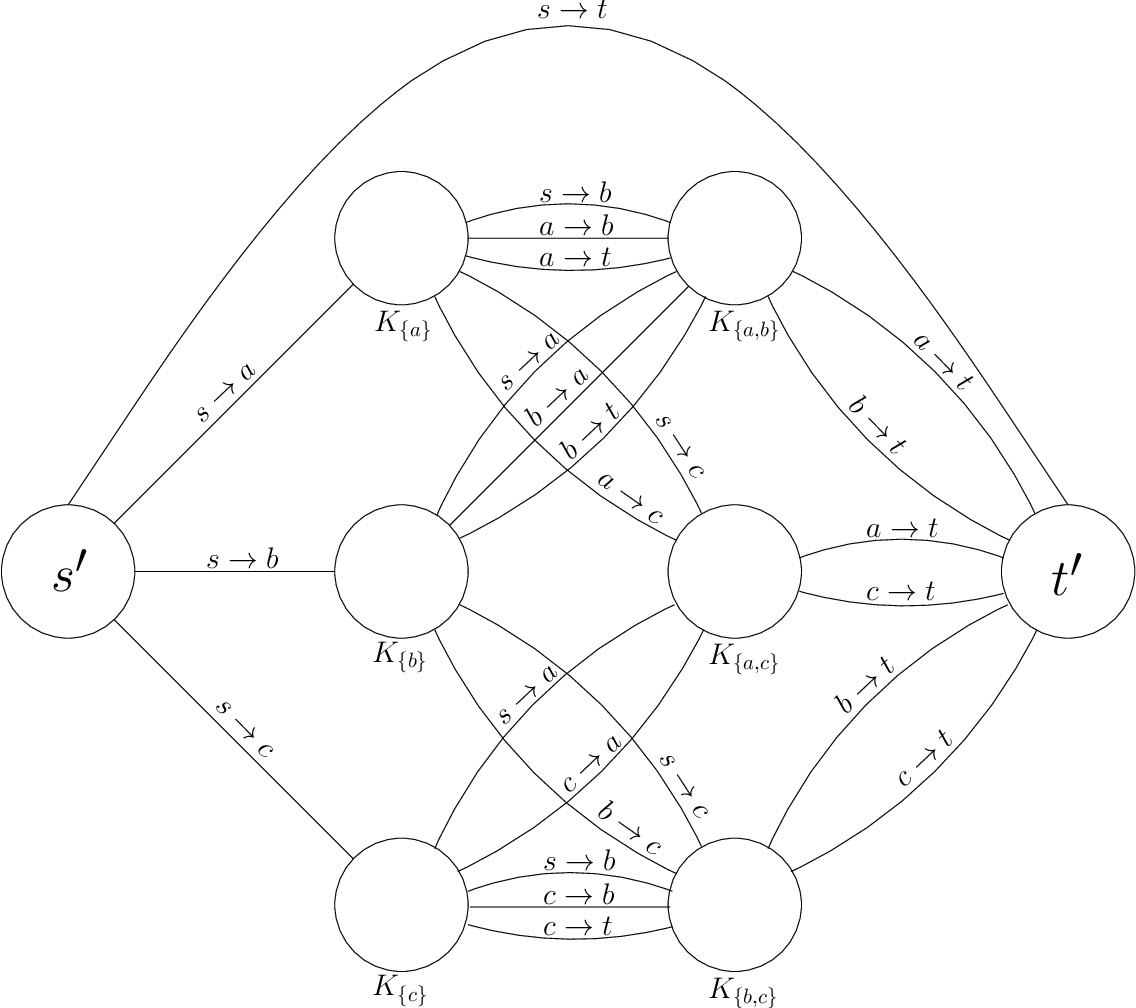}}
\caption[A certain knowledge switching network $G'$ solving directed connectivity on $V(G) = \{s,t,a,b,c\}$ together with a certain knowledge description of it]{In this figure, we have a certain knowledge switching network $G'$ solving directed connectivity (on 
$V(G) = \{s,t,a,b,c\}$) together with a certain-knowledge description for it. By default, we take 
$K_{s'} = \{\}$ and $K_{t'} = \{s \to t\}.$}
\label{biggercertainknowledge}
\end{figure}
\begin{remark}
Figure \ref{biggercertainknowledge} shows how removing old information can help in finding a path from $s$ to $t$ 
with the modified certain knowledge game. 
If $G$ is the input graph with $V(G) = \{s,t,a,b,c\}$ and $E(G) = \{s \to a, a \to b, b \to c, c \to t\}$, then to 
get from $s'$ to $t'$ in $G'$, we must first take the edge labeled $s \to a$ to reach $K = K_{\{a\}}$, then take 
the edge labeled $a \to b$ to reach $K = K_{\{a,b\}}$, then go ``backwards'' along the edge labeled $s \to a$ to 
reach $K = K_{\{b\}}$, then take the edge labeled $b \to c$ to reach $K = K_{\{b,c\}}$ and finally take 
the edge labeled $c \to t$ to reach $t'$.
\end{remark}
\subsection{An upper bound on certain-knowledge switching networks}
While the certain knowledge condition is restrictive, certain knowledge switching networks 
nevertheless have considerable power. In particular, the following certain knowledge 
switching networks match the power of Savitch's algorithm. 
\begin{definition}\label{universalcertainknowledge}
Given a set of vertices $V(G)$ with distinguished vertices $s,t$, let $G'_c(n,r)$ be the certain knowledge 
switching network with vertices $t' \cup \{v'_V: V \subseteq \Vmst, |V| \leq r\}$ and all 
labeled edges allowed by condition 3 of Definition \ref{certainknowledgedef}, 
where each $v'_V$ has knowledge set $K_V$, $s' = v'_{\{\}}$, and $K_{t'} = \{s \to t\}$. Define 
$G'_c(n) = G'_c(n,n)$.
\end{definition}
\begin{example}
The certain-knowledge switching network shown in Figure \ref{examplecertainknowledge} is $G'_c(2) = G'_c(2,2)$ with some edges missing. The certain-knowledge switching network shown in Figure \ref{biggercertainknowledge} is $G'_c(3,2)$ with some edges missing.
\end{example} 
\begin{theorem}\label{constantlcertainknowledgeupperbound}
For all $l \geq 1$, $c(\mathcal{P}_{n,\leq l}) \leq \sum_{j=1}^{\lceil{\lg{l}}\rceil}{{n \choose j}}+2$
\end{theorem}
\begin{proof}
Consider the switching network $G' = G'_c(n, \lceil{\lg{l}}\rceil)$. $G'$ is a certain-knowledge switching network 
with $|V(G')| = \sum_{j=1}^{\lceil{\lg{l}}\rceil}{{n \choose j}}+2$, so to prove the theorem 
it is sufficient to show that $G'$ accepts all of the input graphs in $\mathcal{P}_{n,\leq l}$.

Consider an input graph $G \in \mathcal{P}_{n,\leq l}$. $E(G) = \{v_i \to v_{i+1}: 0 \leq i \leq j-1\}$ where 
$v_0 = s'$, $v_j = t'$, and $j \leq l$. We will show that $G'$ accepts $G$ by showing that 
we can win at the reversible pebble game without placing more than $\lceil{\lg{l}\rceil}$ 
pebbles on the vertices $v_1, \cdots, v_{j-1}$. This result was first proved by Bennet \cite{cbennet}, we present a proof here for convenience.
\begin{definition} Let $G$ be the input graph with vertices $\{v_0,v_1,v_2, \cdots\}$ and edges 
$\{v_i \to v_{i+1}: i \geq 0\}$. Assume that we start with a pebble on $s = v_0$.
\begin{enumerate} 
\item Let $f(i)$ be the minimal number $m$ such that we can place a pebble on $v_i$ without ever 
having more than $m$ pebbles on the vertices $v_1, \cdots, v_{i-1}$.
\item Let $g(i)$ be the minimal number $m$ such that we can reach the game state where there is a pebble on 
$v_0$ and $v_i$ and no other pebbles without ever having more than $m$ pebbles on the vertices 
$v_1, \cdots, v_{i}$.
\end{enumerate}
\end{definition}
\begin{lemma}\label{fandgupperbounds}
For all integers $i \geq 1$, 
\begin{enumerate}
\item $f(i) \leq \lceil \log{(i)} \rceil$
\item $g(i) \leq \lceil \log{(i)} \rceil + 1$
\end{enumerate}
\end{lemma}
\begin{proof}
\begin{proposition}\label{cleanup}
For all $i \geq 1$, $g(i) \leq f(i) + 1$
\end{proposition}
\begin{proof}
We have a sequence of moves allowing us to place a pebble on $v_i$ while placing at most 
$f(i)$ pebbles on the vertices $v_1, \cdots, v_{i-1}$. After completing this sequence, run the sequence in 
reverse except that we do not remove the pebble on $v_i$. When we are done, we only have pebbles 
on $v_0$ and $v_i$ and at all times we have at most $f(i)+1$ pebbles on the vertices 
$v_1, \cdots, v_{i}$, as needed. 
\end{proof}
\begin{proposition}\label{moveforward}
For all $i,j \geq 1$, $f(i+j) \leq \max{\{g(i),f(j)+1\}}$
\end{proposition}
\begin{proof}
We first take a sequence of moves allowing us to reach the state where there are pebbles on $v_0$ and 
$v_i$ and no other pebbles without ever having more than $g(i)$ pebbles on $v_1, \cdots, v_i$. We then 
take the sequence of moves allowing us to put a pebble on $v_j$ without ever having more than 
$f(j)$ pebbles on $v_1,\cdots,v_{j-1}$ except that we shift the sequence of moves to the right by $i$. 
This sequence now allows us to start from the state where there are pebbles on $v_0$ and 
$v_i$ and no other pebbles and put a pebble on $v_{i+j}$ without ever having a pebble on 
$v_1,\cdots,v_{i-1}$ or having more than $f(j) + 1$ pebbles on $v_{i},\cdots,v_{i+j-1}$. Composing these 
two sequences of moves gives a sequence of moves putting a pebble on $v_{i+j}$ while never having 
more than $\max{\{g(i),f(j)+1\}}$ pebbles on $v_1,\cdots,v_{i+j-1}$
\end{proof}
Applying Proposition \ref{cleanup} to Proposition \ref{moveforward} we obtain that for all $i,j$, 
$f(i+j) \leq \max{\{f(i),f(j)\}} + 1$. $f(1) = 0$ so we can easily show by induction that for all 
$i \geq 1$, $f(i) \leq \lceil \log{(i)} \rceil$. Applying Proposition \ref{cleanup} again we obtain that for 
all $i \geq 1$, $g(i) \leq \lceil \log{(i)} \rceil + 1$, as needed.
\end{proof}
We now use Lemma \ref{fandgupperbounds} to prove Theorem \ref{constantlcertainknowledgeupperbound}. By Lemma \ref{fandgupperbounds} there is a sequence of moves for the reversible pebble game on $G$ allowing 
us to win without placing more than $\lceil{\lg{l}\rceil}$ pebbles on the vertices $v_1, \cdots, v_{j-1}$. 
We now translate this winning sequence of moves into a walk in $G'$ (which can then be shortened to a path). If we have pebbles 
on a set of vertices $V \subseteq \{v_0, \cdots, v_{j-1}\}$ with $v_i \in V$ and our move is to place a pebble on 
$v_{i+1}$, this corresponds to moving from $K_V$ to $K_{V \cup \{v_{i+1}\}}$ in $G'$ along an edge labeled 
$v_i \to v_{i+1}$ (we know such an edge exists because of Lemma \ref{keysequenceofmoves}). Similarly, 
if we have pebbles on a set of vertices $V \subseteq \{v_0, \cdots, v_{j-1}\}$ with $v_i,v_{i+1} \in V$ 
and our move is to remove a pebble from $v_{i+1}$, this corresponds to moving from 
$K_{V \cup \{v_{i+1}\}}$ to $K_V$ in $G'$ along an edge labeled $v_i \to v_{i+1}$. Finally, if we have 
pebbles on a set of vertices $V \subseteq \{v_0, \cdots, v_{j-1}\}$ with $v_{j-1} \in V$ 
and our move is to place a pebble on $v_{j} = t$, this corresponds to moving from 
$K_{V}$ to $K_{t'}$ in $G'$ along an edge labeled $v_{j-1} \to t$. The entire sequence of moves corresponds 
to a walk from $s'$ to $t'$ in $G'$ whose edge labels are all consistent with $G$ 
so $G'$ accepts the input graph $G$, as needed.
\end{proof}
\subsection{A lower size bound on certain-knowledge switching networks}
We now prove lower bounds on $c(\mathcal{P}_{n,\leq l})$.
\begin{definition}
For a knowledge set $K$ such that $K$ does not contain a path from $s$ to $t$, define 
$$V(K) = \{v: v \in \Vmst, \exists w \in V(G): w \neq v, v \to w \in E(G) \text{ or } w \to v \in E(G)\}$$
\end{definition}
Our lower bound argument is a bottleneck argument using the following lemma, which says that if we have an input graph $G$ containing a path $P$ from $s$ to $t$ and no other edges, then for any walk $W'$ from $s'$ to $t'$ in $G'$ whose edge labels are all in $E(P)$, there is one vertex $v'$ on $W'$ such that $V(K_{v'})$ contains many vertices of $P$ and no vertices not in $P$, which gives a lot of information about $P$.
\begin{lemma}\label{specialboundlemma}
Let $G'$ be a certain knowledge switching network. For any certain knowledge description of $G'$ and 
any path $P = s \to v_1 \to \cdots \to v_{l-1} \to t$, if $G$ is the input graph with vertex set $V(G)$ and 
$E(G) = E(P)$, if $W'$ is a walk in $G'$ whose edge labels are all in $G$ from a vertex $v'_{start}$ where 
$K_{v'_{start}} = \{\}$ to a vertex $v'_{end}$ where $K_{v'_{end}}$ contains a path from $s$ to $t$ then $W'$ passes through a vertex $v'$ such that $V(K_{v'}) \subseteq \{v_1, \cdots, v_{l-1}\}$ and $|V(K_{v'})| \geq \lceil{\lg(l)}\rceil$. 
\end{lemma}
However, the proof of this lemma is long, so we relegate it to Appendix \ref{specialboundlemmafullproof}. Instead, we show here that this lemma holds if all $K_{v'}$ are of the form $K_V$ where $V \subseteq V(G) \setminus \{s\}$. This is equivalent to proving the following result about the reversible pebbling game.
\begin{lemma}
$f(l) \geq \lceil{\lg{l}}\rceil$
\end{lemma}
This result was first proved by Li and Vitanyi \cite{reducabilitytwo}. We give a short alternative proof of this result here which emphasizes the role of reveresiblity.
\begin{proof}
We have that $f(1) = 0$, $f(2) = 1$, and $f(i)$ is an nondecreasing function of $i$, so this follows immediately from the following lemma. 
\begin{lemma}
For all $i \geq 2$, $f(2i-1) \geq f(i)+1$
\end{lemma}
\begin{proof}
Consider a sequence of moves in the reversible pebbling games which places a pebble on vertex $i$. Note that if we merge all of the vertices $v_1, \cdots, v_{i-1}$ with $s = v_0$, this squence becomes a sequence of moves in the reversible pebbling game with the vertices $s,v_{i},\cdots,v_{2i-1}$ that pebbles $v_{2i-1}$. By definition, this requires placing at least $f(i)$ pebbles on the vertices $v_i \cdots, v_{2i-2}$. Thus, at some point in our sequence of moves we must have at least $f(i)$ pebbles on the vertices $v_i \cdots, v_{2i-2}$.

If at this point, we have any pebble on the vertices $v_1, \cdots, v_{i-1}$, then we have used $f(i)+1$ pebbles. Thus, we may assume that we have no pebbles on the vertices $v_1, \cdots, v_{i-1}$. Now let $v_j$ be the leftmost vertex that is pebbled and run the sequence of moves we used to reach this state in reverse. At some point, we must remove a pebble on $v_j$ so that we can reach the initial state of no pebbles anywhere. However, to do this, we must have first had $f(j) \geq f(i)$ pebbles on the vertices $v_1, \cdots, v_{j-1}$. Moreover, at this point we still had a pebble on $v_j$ so we had a total of at least $f(i)+1$ pebbles placed, as needed.
\end{proof}
This completes the proof of Lemma \ref{specialboundlemma} when all $K_{v'}$ are of the form $K_V$ where $V \subseteq V(G) \setminus \{s\}$.
\end{proof}
The other part of our lower bound proof is finding a large collection of paths such that each 
pair of paths has very few vertices in common. We give a direct way to do this here using polynomials, this can be done using Nisan-Wigderson combinatorial designs \cite{nisanwigderson}.
\begin{lemma}\label{largecollectionslemma}
If $m,k_1,k_2$ are non-negative integers and there is a prime $p$ such that $k_2 < k_1 \leq p$ and 
$m \geq p{k_1}$, then there is a collection of $p^{k_2+1}$ subsets of $[0,m-1]$ of size $k_1$ such that each pair of subsets has at most $k_2$ elements in common.
\end{lemma}
\begin{proof}
To obtain our collection, first take the set of all polynomials in $\mathbb{F}_p[x]$ of degree at most $k_2$ where $\mathbb{F}_p$ is the integers modulo $p$. There are $p^{k_2+1}$ such polynomials. For each such polynomial $f(x)$, let $S_f = \{(x,f(x)): 0 \leq x < k_1\}$. For any two distinct polynomials $f_1$ and $f_2$ of degree at most $k_2$, $S_{f_1} \cap S_{f_2} = \{(x,f(x_1)): f_1(x) - f_2(x) = 0\}$. $f_1 - f_2$ is a non-zero polynomial in $\mathbb{F}_p[x]$ of degree at most $k_2$, so there are at most $k_2$ $x \in \mathbb{F}_p$ such that $f_1(x) - f_2(x) = 0$. Thus, $|S_{f_1} \cap S_{f_2}| \leq k_2$.

We now translate these sets into subsets of $[0,p{k_1}-1]$ by using the map \\
$\phi: [0,p{k_1} - 1] \to [0,k_1-1] \times \mathbb{F}_p$\\
$\phi(x) = (\lfloor \frac{x}{p} \rfloor, (x$ mod $p))$

Taking our subsets to be $\phi^{-1}(S_f)$ for all $f \in \mathbb{F}_p[x]$ of degree at most $k_2$, each such subset of $[0,p{k_1}-1] \subseteq [0,m-1]$ has $k_1$ elements. Since $\phi$ is injective and 
surjective, for any distinct $f_1,f_2 \in \mathbb{F}_p[x]$ of degree at most $k_2$, 
$|\phi^{-1}(S_{f_1}) \cap \phi^{-1}(S_{f_2})| = |S_{f_1} \cap S_{f_2}| \leq k_2$ and this completes the proof.
\end{proof}
\begin{corollary}\label{largecollections}
If $n \geq 2$ and $k_1,k_2$ are non-negative integers with $k_2 < k_1 \leq \sqrt{\frac{n}{2}}$ then 
there is a collection of at least $(\frac{n}{2k_1})^{k_2+1}$ paths from $s$ to $t$ of length $k_1 + 1$ on the 
vertices $V(G)$ such that each pair of paths has at most $k_2$ vertices in common (excluding $s$ and $t$).
\end{corollary}
\begin{proof}
We prove this result using Lemma \ref{largecollectionslemma} and a suitable prime number $p$ chosen using Bertrand's postulate.
\begin{theorem}[Bertrand's postulate]
For any integer $m > 3$ there is a prime $p$ such that $m < p < 2m-2$
\end{theorem}
\begin{corollary}\label{goodprimeexists}
For any real number $m \geq 1$ there is a prime $p$ such that $m \leq p \leq 2m$.
\end{corollary}
\begin{proof}
By Bertrand's postulate, for any real number $m > 3$ there is a prime 
$p$ such that $m \leq \lceil{m}\rceil < p < 2\lceil{m}\rceil - 2 < 2m$. If $m \in [1,2]$ then $m \leq 2 \leq 2m$. 
If $m \in [2,3]$ then $m \leq 3 \leq 2m$.
\end{proof}
By Corollary \ref{goodprimeexists} we can take a prime $p$ such that 
$\frac{n}{2k_1} \leq p \leq \frac{n}{k_1}$. We now have that $n \geq pk_1$ and since $k_1 \leq \sqrt{\frac{n}{2}}$ 
we have that $k^2_1 \leq \frac{n}{2}$ which implies that $k_1 \leq \frac{n}{2k_1} \leq p$. The result now follows 
from Lemma \ref{largecollectionslemma}.
\end{proof}
We now prove the following lower bound on certain knowledge switching networks:
\begin{theorem}\label{constantlcertainknowledgelowerbound}
If $l \geq 2$ and $n \geq 2(l-1)^2$, then $c(\mathcal{P}_{n,l}) \geq {(\frac{n}{2(l-1)})}^{\lceil{\lg{l}}\rceil}$
\end{theorem}
\begin{proof}
Taking $k_1 = l-1$ and $k_2 = \lceil{\lg{l}}\rceil - 1$ and using Corollary \ref{largecollections}, 
we have a collection of at least $(\frac{n}{2k_1})^{k_2+1} = {(\frac{n}{2(l-1)})}^{\lceil{\lg{l}}\rceil}$ 
paths of length $l$ from $s$ to $t$ on the set of vertices $V(G)$ such that each pair of paths $P_i,P_j$ 
has at most $k_2$ vertices in common (excluding $s$ and $t$). However, by Lemma \ref{specialboundlemma}, for any 
certain knowledge switching network $G'$ which accepts all of the input graphs in 
$\mathcal{P}_{n,l}$, we can associate a vertex $v'_i$ in $G'$ to each path $P_i$ in our collection such that 
$|V(K_{v'_i})| > k_2$ and $V(K_{v'_i})$ is a subset of the vertices of $P_i$. This implies that we cannot have 
$v'_i = v'_j$ for any $i \neq j$ as otherwise we would have that $|V(K_{v'_i})| = |V(K_{v'_j})| > k_2$ and 
$|V(K_{v'_i})|$ is a subset of the vertices of both $P_i$ and $P_j$, 
which is impossible as any two distinct paths in our collection have at most $k_2$ vertices in common. 
Thus, $|V(G')| \geq {(\frac{n}{2(l-1)})}^{\lceil{\lg{l}}\rceil}$, as needed.
\end{proof}
\subsection{Simplified bounds on certain-knowledge switching networks}
\noindent We now use Theorems \ref{constantlcertainknowledgeupperbound} and \ref{constantlcertainknowledgelowerbound} to prove 
Theorem \ref{certainknowledgecleanbounds}.
\vskip.1in
\noindent
{\bf Theorem \ref{certainknowledgecleanbounds}.}
{\it
Let $V(G)$ be a set of vertices with distinguished vertices $s,t$. Taking $n = |\Vmst|$, if 
$l \geq 2$ and $n \geq 2(l-1)^2$ then 
\begin{enumerate}
\item ${(\frac{n}{2(l-1)})}^{\lceil{\lg{l}}\rceil} \leq c(\mathcal{P}_{n,l}) \leq 
c(\mathcal{P}_{n,\leq l}) \leq n^{\lceil{\lg{l}}\rceil}+2$
\item $n^{\frac{1}{4}\lg{n} - \frac{1}{2}} \leq c(\mathcal{P}_{n}) \leq n^{\lg{n} + 1}+2$
\end{enumerate}
}
\begin{proof}
For the first statement, the lower bound is just Theorem \ref{constantlcertainknowledgelowerbound}. 
To prove the upper bound, note that by Theorem \ref{constantlcertainknowledgeupperbound} we have that 
$c(\mathcal{P}_{n,\leq l}) \leq \sum_{j=1}^{\lceil{\lg{l}}\rceil}{{n \choose j}}+2$. If $\lceil{\lg{l}}\rceil = 1$ then 
$l=2$ so $\sum_{j=1}^{\lceil{\lg{l}}\rceil}{{n \choose j}} = n^{\lceil{\lg{l}}\rceil} = n$. If $\lceil{\lg{l}}\rceil > 1$ then $l > 2$ so $n \geq 2(l-1)^2 > 2\lceil{\lg{l}}\rceil$. 
This implies that 
$\sum_{j=1}^{\lceil{\lg{l}}\rceil}{{n \choose j}} \leq \lceil{\lg{l}}\rceil{n \choose {\lceil{\lg{l}}\rceil}} 
\leq n^{\lceil{\lg{l}}\rceil}$, as needed.

For the second statment, the upper bound follows immediately from the upper bound of the first statement. For the 
lower bound, taking $l = \lceil{\sqrt{\frac{n}{2}}}\rceil$ by 
Theorem \ref{constantlcertainknowledgelowerbound} we have that 
\begin{align*}
c(\mathcal{P}_{n}) &\geq c(\mathcal{P}_{n,l}) \geq {\left(\frac{n}{2(l-1)}\right)}^{\lceil{\lg{l}}\rceil} \geq 
\left(\frac{n}{2\sqrt{\frac{n}{2}}}\right)^{\lg(\sqrt{\frac{n}{2}})} \\
&= \frac{n^{\frac{1}{4}\lg{n} - \frac{1}{4}}}{2^{\frac{1}{4}\lg{n} - \frac{1}{4}}} 
\geq \frac{n^{\frac{1}{4}\lg{n} - \frac{1}{4}}}{2^{\frac{1}{4}\lg{n}}} = n^{\frac{1}{4}\lg{n} - \frac{1}{2}}
\end{align*}
\end{proof}
\begin{remark}
A size bound of $\Theta(s(n))$ on switching networks solving a problem 
roughly corresponds to a space bound of $\Theta(\lg{(s(n))})$ on algorithms solving that problem. Thus, the size bounds of Theorem \ref{certainknowledgecleanbounds} correspond to a space bound 
of $\Theta({\lceil\lg{l}\rceil}\lg{n})$ for finding all paths of length at most $l$ and a space bound 
of $\Theta((\lg{n})^2)$ for finding all paths, which is exactly the performance of Savitch's algorithm.
\end{remark}
\section{Fourier Analysis and Invariants on Monotone Switiching Networks For Directed Connectivity}\label{introducefouriertransform}
To prove a strong lower size bound on general monotone switching networks solving 
directed connectivity, more sophisticated techniques are needed. In this section, we introduce a very different way of analyzing the problem: Fourier analysis and invariants. We first use 
Fourier analysis and invariants to prove a quadratic lower size bound and then show how more general lower size bounds can be obtained.
\subsection{Function descriptions of sound monotone switching networks}
The following tautology is trivial yet illuminating: For any yes/no question, the answer is yes if and only if it is not no.

Before, we analyzed each vertex of the switching network in terms of how much progress has been made towards showing 
directly that the answer to the question is yes. Here, we will analyze each vertex of the switching network in terms 
of which NO instances have been eliminated. For monotone switching networks, we only need to consider maximal 
NO instances, as once these have been eliminated all other NO instances must have been eliminated as well. For 
directed connectivity, the maximal NO instances correspond to cuts. Thus, we will analyze each vertex of the switching network 
in terms of which cuts have been crossed. We make this rigorous below.
\begin{definition}
We define an s-t cut (below we use cut for short) of $V(G)$ to be a partition of $V(G)$ into subsets $L(C),R(C)$ such that 
$s \in L(C)$ and $t \in R(C)$. We say an edge $v_1 \to v_2$ crosses $C$ if $v_1 \in L(C)$ and $v_2 \in R(C)$.
Let $\mathcal{C}$ denote the set of all cuts $C$ of $V(G)$.
\end{definition}
\begin{definition}\label{functiondescriptiondefinition}
We define a function description of a monotone switching network to be an 
assignment of a function $h_{v'}$ to each vertex $v' \in V(G')$ such that
\begin{enumerate}
\item Each $h'_{v'}$ is a function from $\mathcal{C}$ to $\{0,1\}$.
\item $\forall C \in \mathcal{C}, s'(C) = 1$ and $t'(C) = 0$.
\item If there is an edge $e' \in G'$ with label $e$ between vertices $v'$ and $w'$ in $G'$, 
for all $C \in \mathcal{C}$ such that $e$ does not cross $C$, $v'(C) = w'(C)$.
\end{enumerate}
For convenience we identify each vertex $v'$ with its associated function $h'_{v'}$ i.e. 
we take $v'(C) = h'_{v'}(C)$ for all $C \in \mathcal{C}$
\end{definition}
\begin{remark}
The fact that $v'(C)$ is invariant along any edge $e'$ in $G'$ whose label does not cross $C$ is the foundation for our lower bounds.
\end{remark}
\begin{proposition}\label{functiondescriptionsoundness}
Any monotone switching network which has a function description is sound.
\end{proposition}
\begin{proof}
Assume that $G'$ has a function description yet accepts some input graph $G$ which does not have a path from $s$ to $t$. Then there is some path $P'$ in $G'$ from $s'$ to $t'$ whose labels are all in $E(G)$. Now let $C$ be the cut such that $L(C) = \{v \in V(G):\text{ there is a path from s to v in G}\}$. Note that $E(G)$ cannot have any edge crossing $C$ as otherwise there would be a path in $G$ from $s$ to some vertex in $R(C)$. This implies that for any two adjacent vertices $v'$ and $w'$ in $P'$, $v'(C) = w'(C)$. But then we must have that $s'(C) = t'(C)$, contradicting the fact that $s'(C) = 1$ and $t'(C) = 0$.
\end{proof}
We now show the converse to this proposition, that every sound monotone switching network has a function description. 
\begin{definition}
For a cut $C$, define the input graph $G(C)$ to be the graph with vertex set $V(G)$ and edge set 
$E(G(C)) = \{e: e$ does not cross $C\}$
\end{definition}
\begin{definition}
Define the reachability function description for a sound monotone switching network $G'$ to be the 
assignment of the function $h_{v'}: \mathcal{C} \to \{0,1\}$ to each vertex $v' \in V(G')$ where $h_{v'}(C) = 1$ if there is a walk from $s'$ to $v'$ in $G'$ whose edge labels are all in $E(G(C))$ and $0$ otherwise.
\end{definition}
\begin{proposition}\label{reachabilitydescriptionvalidity}
For any sound monotone switching network $G'$, the reachability function description is a function description of $G'$.
\end{proposition}
\begin{proof}
Consider the reachability function description for $G'$. For all $C \in \mathcal{C}$, $s'(C) = 1$. Assume that $t'(C) = 1$ for some $C \in \mathcal{C}$. If so, there must be a path $P'$ in $G'$ from $s'$ to $t'$ such that no edge label in $P'$ crosses $C$. If so, then since $G(C)$ contains all edges which do not cross $C$, all edge labels in $P'$ are contained in $E(G(C))$ so $G'$ accepts $G(C)$ and is thus not sound. Contradiction. Thus, $t'(C) = 0$ for all $C$.

To see that the third condition for a knowledge description holds, assume that it does not hold. Then there 
is an edge $e'$ in $G'$ with endpoints $v'$ and $w'$ and a cut $C$ such that the label $e$ of $e'$ does not 
cross $C$ but $v'(C) \neq w'(C)$. Without loss of generality, $v'(C) = 1$ and $w'(C) = 0$. 
But then there is a walk $W'$ from $s'$ to $v'$ such that none of the labels of 
its edges cross $C$. If so, taking $W'_2$ to be the walk $W'$ with the edge $e'$ added at the end, $W'_2$ is a walk from $s'$ to $w'$ such that none of the labels of the edges of $W'_2$ cross $C$, so we should have $w'(C) = 1$. Contradiction.
\end{proof}
\begin{remark}
Reversibility is crucial here. If the edges of the switching network were instead directed and we had a similar reachability function description, we could have $v'(C) = 0$ but $w'(C) = 1$ if we have no directed walk from $s'$ to $v'$ whose edges are all in $G(C)$ but we do have a directed walk from $s'$ to $w'$ whose edges are all in $G(C)$.
\end{remark}
\begin{figure}[ht]
\centerline{\includegraphics[height=6cm]{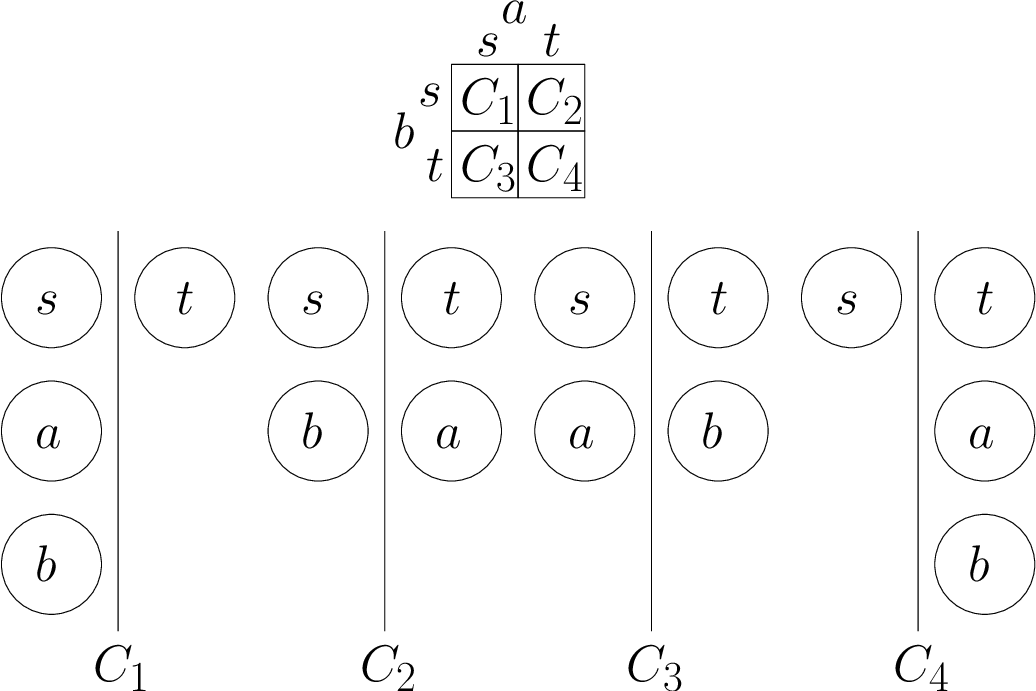}}
\caption[Representing cuts of $V(G)$]{Here we show how to represent all of the cuts of $V(G)$ simultaneously when $V(G) = \{s,a,b,t\}$. 
The column determines whether $a$ is with $s$ or $t$ and the row determines whether $b$ is with $s$ or $t$.}
\label{representingcuts}
\end{figure}
\begin{figure}[ht]
\centerline{\includegraphics[height=6cm]{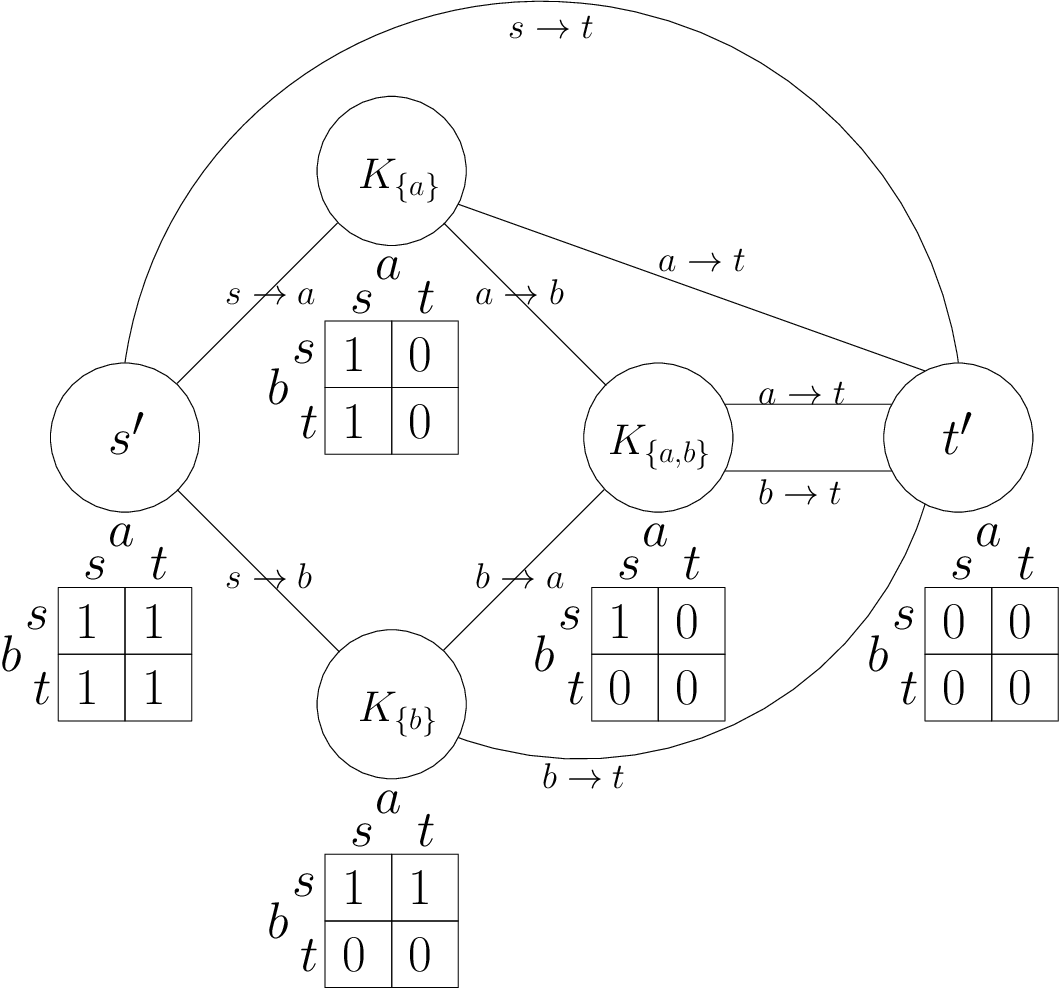}}
\caption[A certain knowledge switching network and the reachability function description for it.]{This is the certain-knowledge switching network $G'_c(2,2)$ together with its
certain-knowledge description and the reachability function description for it.}
\label{certainknowledgefunctiondescription}
\end{figure}
\subsection{Fourier analysis}
Now that we have assigned each vertex $v' \in V(G')$ a function $v':\mathcal{C} \to \{0,1\}$, we can use Fourier analysis to analyze our switching networks. We begin by defining a dot product, a Fourier basis, and Fourier coefficients.
\begin{definition}
Given two functions 
$f,g: \mathcal{C} \to \mathbb{R}$, $f \cdot g = 2^{-n}\sum_{C \in \mathcal{C}}{f(C)g(C)}$
\end{definition}
\begin{proposition}\label{vertexnorm}
If $G'$ is a sound monotone switching network for directed connectivity with a given function 
description, then for all $v' \in V(G')$, $||v'|| = \sqrt{v' \cdot v'} \leq 1$.
\end{proposition}
\begin{definition}
Given a set of vertices $V \subseteq \Vmst$, define $e_V: \mathcal{C} \to \mathbb{R}$ by 
$e_V(C) = {(-1)^{|V \cap L(C)|}}$.
\end{definition}
\begin{proposition}
The set $\{e_V, V \subseteq \Vmst\}$ is an orthonormal basis for the vector space of functions from 
$\mathcal{C}$ to $\mathbb{R}$. 
\end{proposition}
\begin{definition}
Given a function $f: \mathcal{C} \to \mathbb{R}$ and a set of vertices $V \subseteq \Vmst$, 
define $\hat{f_V} = f \cdot e_V$.
\end{definition}
\begin{proposition}[Fourier Decomposition and Parseval's Theorem]
For any function $f: \mathcal{C} \to \mathbb{R}$, $f = \sum_{V \subseteq \Vmst}{\hat{f_V}}{e_V}$ and 
$f \cdot f = \sum_{V \subseteq \Vmst}{\hat{f_V}^2}$
\end{proposition}
\begin{figure}[ht]
\centerline{\includegraphics[height=4cm]{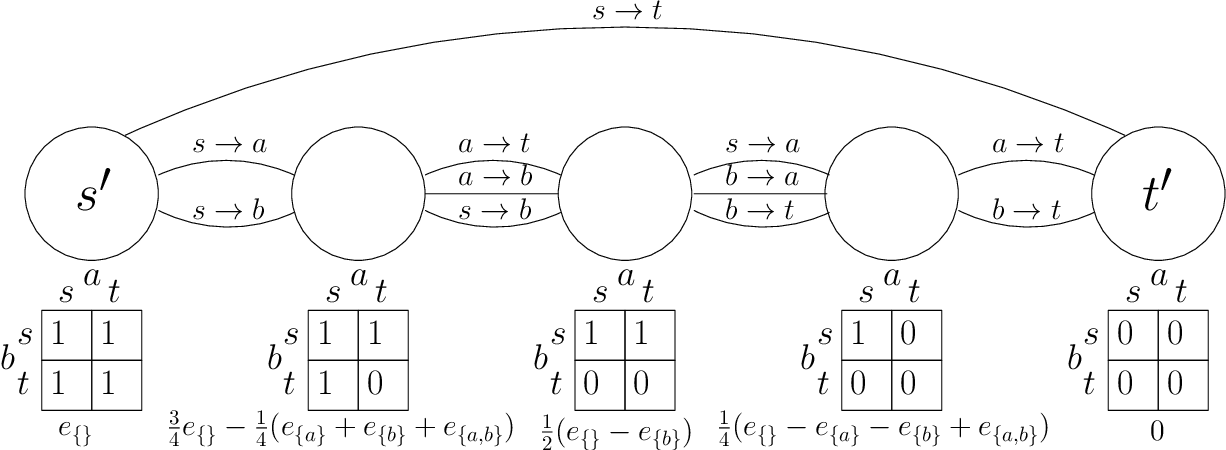}}
\caption[A monotone switching network together with the reachability function description for it and the Fourier decomposition of each function]{In this figure, we have the monotone switching network solving directed connectivity on $V(G) = \{s,a,b,t\}$ 
shown in Figure \ref{examplenetwork} together with the reachability function description for it and the 
Fourier decomposition of each function.}
\label{exampleonefunctiondescription}
\end{figure}
\begin{remark}
The switching network shown in Figures \ref{examplenetwork} and \ref{exampleonefunctiondescription} determines whether or not 
there is a path from $s$ to $t$ by checking all of the cuts one by one to determine if there is an edge in $G$ crossing that cut. This 
can be generalized to give a monotone switching network of size $2^{n}$ solving directed connectivity. While 
the size of such a switching network is enormous, this extreme construction is interesting because the switching network is 
analyzing all of the possible paths from $s$ to $t$ at the same time.
\end{remark}
\subsection{A quadratic lower bound}
\noindent We now show tight lower bounds on $m(\mathcal{P}_{n,\leq 2})$ and $m(\mathcal{P}_{n,\leq 3})$ by proving a corresponding lower bound on the dimension of the span of $V(G')$ whenever $G'$ is sound and accepts all paths from $s$ to $t$ of length 2 and 3 respectively.
\begin{theorem}\label{quadraticlowerbound}
$m(\mathcal{P}_{n,\leq 2}) = n + 2$ and $m(\mathcal{P}_{n,\leq 3}) = \binom{n}{2} + n + 2$.
\end{theorem}
The idea behind the proof is as follows. Given a path in $G'$ from $s'$ to $t'$ whose edge labels are all edges of some path $P$, we find a linear combination of the edges of $P'$ whose Fourier decomposition gives a lot of information about $P$. The coefficient of each edge $e' \in E(P')$ in this linear combination will be determined by what its label $\mu'(e')$ is.
\begin{definition}
Given a switching network $G'$, a directed path $P'$ from $s'$ to $t'$ in $G'$, and a set of edges $E$, 
recall that ${\mu}'$ is the labeling function for edges in $G'$ and define
$$\Delta_E(P') = \sum_{e'\in E(P'): \mu'(e') \in E}{e'}$$
where if $e'$ goes from $v'$ to $w'$ in $P'$ then we define $e' = w'-v'$. As a special case, define $$\Delta_{e}(P') = \Delta_{\{e\}}(P') = \sum_{e'\in E(P'): \mu'(e') = e}{e'}$$
\end{definition}
We now consider what must be true about the functions $\Delta_E(P')$.
\begin{proposition}
If $P'$ is a path from $s'$ to $t'$ in $G'$ and $E_1, \cdots, E_m$ is a partition of the edge labels of the edges in $P'$, then $\sum_{i=1}^{m}{\Delta_{E_i}(P')(C)} = -1$ for all $C \in \mathcal{C}$
\end{proposition}
\begin{proof}
\begin{align*}
\sum_{i=1}^{m}{\Delta_{E_i}(P')(C)} &= \sum_{i=1}^{m}{\sum_{e'\in E(P'): \mu'(e') \in E_i}{e'(C)}} \\
&= \sum_{e'\in E(P')}{e'(C)} = t'(C)-s'(C) = -1
\end{align*}
\end{proof}
\begin{definition}
Given a set of edges $E$, define $\mathcal{C}_E$ to be the set of cuts which are not crossed by any edge in $E$. As a special case, given an edge $e$, define $\mathcal{C}_{e} = \mathcal{C}_{\{e\}}$ to be the set of cuts which are not crossed by $e$
\end{definition}
\begin{proposition}
For any switching network $G'$, directed path $P'$ from $s'$ to $t'$ in $G'$, and set of edges $E$, $\Delta_E(P')(C) = 0$ for all $C \in \mathcal{C}_E$
\end{proposition}
\begin{proof}
If $C \in \mathcal{C}_E$ then $\Delta_E(P')(C) = \sum_{e' \in E(P'): \mu'(e') \in E}{e'(C)} = 0$ because whenever $\mu'(e') \in E$, $\mu'(e')$ does not cross $C$ so $e'(C) = 0$.
\end{proof}
\begin{corollary}\label{fixedfunction}
If $G$ is an input graph on the set of vertices $V(G)$ and $(E_1,E_2)$ is a partition of $E(G)$ then
\begin{enumerate}
\item If $C \in \mathcal{C}_{E_1}$, 
$(\Delta_{E_1}(P') - \Delta_{E_2}(P'))(C) = 1$.
\item If $C \in \mathcal{C}_{E_2}$,
$(\Delta_{E_1}(P') - \Delta_{E_2}(P'))(C) = -1$.
\end{enumerate}
\end{corollary}
\begin{proof}
If $C \in \mathcal{C}_{E_1}$, then $\Delta_{E_1}(P')(C) = 0$ so 
$$(\Delta_{E_1}(P') - \Delta_{E_2}(P'))(C) = -(\Delta_{E_1}(P') + \Delta_{E_2}(P'))(C) = 1$$
If $C \in \mathcal{C}_{E_2}$, then $\Delta_{E_2}(P')(C) = 0$ so 
$$(\Delta_{E_1}(P') - \Delta_{E_2}(P'))(C) = (\Delta_{E_1}(P') + \Delta_{E_2}(P'))(C) = -1$$
\end{proof}
We now ready to prove Theorem \ref{quadraticlowerbound}.
\begin{proof}[of Theorem \ref{quadraticlowerbound}] 
The upper bounds on $m(\mathcal{P}_{n,\leq 2})$ and $m(\mathcal{P}_{n,\leq 3})$ follow immediately 
from Theorem \ref{constantlcertainknowledgeupperbound}. We prove the lower bounds using the following 
proposition:
\begin{proposition}\label{dimensionargument}
$|V(G')| \geq dim(span\{V(G')\}) + 1$
\end{proposition}
\begin{proof}
$t' = 0$ so 
$$|V(G')| = |V(G') \setminus \{t'\}| + 1 \geq dim(span\{V(G') \setminus \{t'\}\}) + 1 = 
dim(span\{V(G')\}) + 1$$
\end{proof}
If $P$ is a path of length 2 in $G$ from $s$ to $t$, then $P$ has the form $P = s \to v \to t$. 
Take $E_1 = \{s \to v\}$ and $E_2 = \{v \to t\}$. For any cut $C$,
\begin{enumerate}
\item If $v \in L(C)$ then $C$ cannot be crossed by any edge in $E_1$ so by Corollary \ref{fixedfunction}, 
$$(\Delta_{E_1}(P') - \Delta_{E_2}(P'))(C) = 1$$
\item If $v \in R(C)$ then $C$ cannot be crossed by any edge in $E_2$ so by Corollary \ref{fixedfunction}, 
$$(\Delta_{E_1}(P') - \Delta_{E_2}(P'))(C) = -1$$
\end{enumerate}
This implies that $\Delta_{E_1}(P') - \Delta_{E_2}(P') = -e_{\{v\}}$. Note that 
$\Delta_{E_1}(P') - \Delta_{E_2}(P')$ is a linear combination of vertices of $G'$ so if 
$G'$ accepts all inputs in $\mathcal{P}_{n,2}$ then $e_{\{v\}} \in span\{V(G')\}$ for all $v \in \Vmst$. 
$e_{\{\}} = s' \in span\{V(G')\}$ as well so by Proposition \ref{dimensionargument}, 
$|V(G') \setminus \{s',t'\}| \geq n+1+1 = n+2$.

If $P$ is a path of length 3 in $G$ from $s$ to $t$, then $P$ has the form $P = s \to v_1 \to v_2 \to t$. 
Take $E_1 = \{s \to v_1, v_2 \to t\}$ and $E_2 = \{v_1 \to v_2\}$. For any cut $C$,
\begin{enumerate}
\item If $v_1,v_2 \in L(C)$ then $C$ cannot be crossed by any edge in $E_2$ so by Corollary \ref{fixedfunction}, 
$$(\Delta_{E_1}(P') - \Delta_{E_2}(P'))(C) = -1$$
\item If $v_1\in L(C)$, $v_2 \in R(C)$ then $C$ cannot be crossed by any edge in $E_1$ so by Corollary \ref{fixedfunction}, 
$$(\Delta_{E_1}(P') - \Delta_{E_2}(P'))(C) = 1$$
\item If $v_2 \in L(C)$, $v_1 \in R(C)$ then $C$ cannot be crossed by any edge in $E_2$ so by Corollary \ref{fixedfunction}, 
$$(\Delta_{E_1}(P') - \Delta_{E_2}(P'))(C) = -1$$
\item If $v_1,v_2 \in R(C)$ then $C$ cannot be crossed by any edge in $E_2$ so by Corollary \ref{fixedfunction}, 
$$(\Delta_{E_1}(P') - \Delta_{E_2}(P'))(C) = -1$$
\end{enumerate}
By direct computation, this implies that 
$$(\Delta_{E_1}(P') - \Delta_{E_2}(P')) = \frac{1}{2}(-e_{\{\}} - e_{\{v_1\}} + e_{\{v_2\}} - e_{\{v_1,v_2\}})$$
Thus, we have that if $G'$ accepts all inputs in $\mathcal{P}_{n,\leq 3}$, since we already have that $e_{\{\}} = t' \in span\{V(G')\}$ and $e_{\{v\}} \in span\{V(G')\}$ for all 
$v \in \Vmst$, we also have that for all $v_1,v_2 \in \Vmst$, $e_{\{v_1,v_2\}} \in span\{V(G')\}$. 
By Proposition \ref{dimensionargument}, 
$$|V(G')| \geq \binom{n}{2} + n + 1 + 1 = \binom{n}{2} + n + 2$$
as needed.
\end{proof}
\subsection{General lower bounds}
Unfortunately, the linear independence argument breaks down for longer paths. The problem is that 
for paths $P$ of length greater than $3$, we can no longer find a non-trivial partition $(E_1,E_2)$ of the edges of $P$ 
such that $\Delta_{E_1}(P')$ and $\Delta_{E_2}(P')$ are invariant over all sound monotone switching networks $G'$ and paths 
$P'$ from $s'$ to $t'$ whose edge labels are all in $P$. Thus, for longer paths we need a more sophisticated approach.

For this approach, we partition the edges of $E(G)$ into several sets $\{E_i\}$ and look at the dot product of vertices $v' \in V(G')$ with a carefully chosen set of functions $\{g_{G,E_i}\}$. These functions are chosen so that for all $i$, $g_{G,E_i} \cdot s' = 1$ and whenever there is an edge between vertices $v'$ and $w'$ in $G'$ with label $e \in E_i$, $v' \cdot g_{G,E_i} = w' \cdot g_{G,E_i}$.

We now imagine the following game. There are several players, one for each set of vertices $E_i$. At each vertex $v'$, the ith player has value $g_{G,E_i} \cdot v'$. The players are trying to go from all having value $1$ at $s'$ to all having value $0$ at $t'$ in a sound monotone switching network $G'$ while only taking edges in $G'$ whose labels are in $E(G)$. While doing this, they are trying to keep their values as close to each other as possible. 

However, since every edge the players take has label in $E_i$ for some $i$, for any given move there will be some player whose value remains fixed. This means that their values cannot all change at the same time so there will be some point where there is a significant discrepency between their values. This corresponds to a vertex $v'$ and $i,j$ such that $v' \cdot (g_{G,E_j} - g_{G,E_i})$ is non-negligible, which we can use to prove our lower bounds. We make this intuition rigorous below.
\begin{definition}
We say a function $g: \mathcal{C} \to \mathbb{R}$ is $E$-invariant for a set of edges $E$ if $g(C) = 0$ for all $C \notin \mathcal{C}_E$ (i.e. $g(C) = 0$ whenever $C$ can be crossed by an edge in $E$). As a special case, we say that a function $g: \mathcal{C} \to \mathbb{R}$ is $e$-invariant if $g(C) = 0$ for all $C \notin \mathcal{C}_{e}$
\end{definition}
\begin{proposition}\label{invariantproposition}
If $P'$ is a path from $s'$ to $t'$ in $G'$ and $E$ is a set of edges, then if $g$ is an $E$-invariant function, 
$\Delta_{E}(P') \cdot g = 0$
\end{proposition}
\begin{proof}
This follows immediately from the facts that $\Delta_{E}(P')(C) = 0$ whenever $C \in \mathcal{C}_E$ and $g(C) = 0$ whenever $C \notin \mathcal{C}_E$.
\end{proof}
\begin{lemma}\label{generalizedcruciallemma}
Let $G$ be an input graph containing a path from $s$ to $t$. If we have a partition 
$(E_1,\cdots,E_q)$ of the edges of $G$ and functions $g_{G,E_i}$ such that 
$g_{G,E_i}$ is $E_i$-invariant for all $i$ then for any 
sound monotone switching network $G'$, for any path $P'$ in $G'$ from $s'$ to $t'$ whose 
edge labels are all in $E(G)$, 
$$\sum_{i=2}^{q}{\Delta_{E(G) \setminus E_i}(P') \cdot (g_{G,E_i} - g_{G,E_1})} = 
(q-2)(g_{G,E_1} \cdot e_{\{\}}) - \sum_{i=2}^{q}{g_{G,E_i} \cdot e_{\{\}}}$$
\end{lemma}
\begin{proof}
Let $P'$ be a walk from $s'$ to $t'$ in $G'$ whose edge labels are all in $E(G)$. Since $g_{G,E_i}$ is $E_i$-invariant, 
$$\forall i, \Delta_{E(G) \setminus E_i}(P') \cdot g_{G,E_i} = 
\Delta_{E(G)}(P') \cdot g_{G,E_i} = g_{G,E_i} \cdot (t' - s') = -g_{G,E_i} \cdot e_{\{\}}$$
Since $g_{G,E_1}$ is $E_1$-invariant, 
\begin{align*}
&\sum_{i=2}^{q}{\Delta_{E(G) \setminus E_i}(P') \cdot g_{G,E_1}} = ((q-2)\sum_{i=2}^{q}{\Delta_{E_i}(P')} + 
(q-1)\Delta_{E_1}(P')) \cdot g_{G,E_1}\\
&= ((q-2)\sum_{i=1}^{q}{\Delta_{E_i}(P')}) \cdot g_{G,E_1} \\
&= (q-2)(g_{G,E_1} \cdot (t'-s')) = -(q-2)(g_{G,E_1} \cdot e_{\{\}})
\end{align*}
Putting all of these equations together gives the needed equality.
\end{proof}
\begin{corollary}\label{generalizedcrucialcorollary}
Let $G$ be an input graph containing a path from $s$ to $t$. If we have a partition 
$(E_1,\cdots,E_q)$ of the edges of $G$ and functions $g_{G,E_i}$ such that 
$g_{G,E_i}$ is $E_i$-invariant for all $i$ and $g_{G,E_i} \cdot e_{\{\}}$ is the same for all $i$, 
then for any sound monotone switching network $G'$ which accepts $G$, taking $z = g_{G,E_1} \cdot e_{\{\}}$, 
$$\sum_{i=2}^{q}{\sum_{v' \in V(P')}{|v' \cdot (g_{G,E_i} - g_{G,E_1})|}} \geq z$$
In particular, there must be some $i \in [2,q]$ such that 
$$\sum_{v' \in V(G')}{|v' \cdot (g_{G,E_i} - g_{G,E_1})|} \geq \frac{z}{q-1}$$
\end{corollary}
\begin{proof}
This follows immediately from Lemma \ref{generalizedcruciallemma} and the fact that for all $i$, 
$$\sum_{v' \in V(P')}{|v' \cdot (g_{G,E_i} - g_{G,E_1})|} \geq 
|\Delta_{E(G) \setminus E_i}(P') \cdot (g_{G,E_i} - g_{G,E_1})|$$
because $\Delta_{E(G) \setminus E_i}(P')$ is a linear combination of the vertices in $P'$ 
where each vertex has coefficient $-1$, $0$, or $1$.
\end{proof}
With this corollary in hand, we now show how a lower bound can be obtained by finding suitable collections of functions for a large number of input graphs.
\begin{theorem}\label{generalizedmanyinvariantslowerbound}
Let $I = \{G_j\}$ be a set of input graphs on $V(G)$ all of which contain a path from $s$ to $t$.
If for each $j$ we have a partition $(E_{1j}, \cdots, E_{{q_j}j})$ of the edges of $G_j$, functions 
$g_{G_j,E_{1j}}, \cdots, g_{G_j,E_{{q_j}j}}: \mathcal{C} \to \mathbb{R}$, and constants $\{z_j\}$ and $\{M_j\}$ such that 
\begin{enumerate}
\item For all $j$, $g_{G_j,E_{ij}}$ is $E_{ij}$-invariant for $i \in \{1,q_j\}$
\item For all $j$ and all $i \in [1,q_j]$ , $g_{G_j,E_{ij}} \cdot e_{\{\}} = z_j > 0$
\item For all $j_1,j_2$ where $j_1 \neq j_2$ and all $i_1,i_2$, 
$$(g_{G_{j_1},E_{{i_1}j_1}} - g_{G_{j_1},E_{{1}j_1}}) \cdot (g_{G_{j_2},E_{{i_2}j_2}} - g_{G_{j_2},E_{{1}j_2}}) = 0$$
\item For all $i,j$, $||g_{G_j,E_{ij}} - g_{G_j,E_{1j}}|| \leq M_j$
\end{enumerate}
then $m(I) \geq \sqrt{\sum_{j}{\frac{(\frac{z_j}{q_j-1})^2}{{M_j}^2}}}$
\end{theorem}
\begin{proof}
We prove Theorem \ref{generalizedmanyinvariantslowerbound} using Corollary \ref{generalizedcrucialcorollary}, 
an orthogonality argument, and the Cauchy-Schwarz inequality.
\begin{proposition}\label{crucialproposition}
If $\{g_j\}$ is a collection of nonzero orthogonal functions from $\mathcal{C}$ to $\mathbb{R}$, then for any function 
$h: \mathcal{C} \to \mathbb{R}$ where $||h|| = \sqrt{h \cdot h} \leq 1$, $\sum_{j}{\frac{(g_j \cdot h)^2}{{||g_j||}^2}} \leq 1$
\end{proposition}
\begin{proof}
If $\{g_j\}$ is a collection of nonzero orthogonal functions, we can extend it to an orthogonal basis $\{g_j\} \cup \{f_i\}$ 
for the vector space of functions from $\mathcal{C}$ to $\mathbb{R}$. Now 
$h = \sum_{j}{\frac{(g_j \cdot h)}{(g_j \cdot g_j)}g_j} + \sum_{i}{\frac{(f_i \cdot h)}{(f_i \cdot f_i)}f_i}$, so 
$1 \geq h \cdot h = \sum_{j}{\frac{(g_j \cdot h)^2}{(g_j \cdot g_j)}} + \sum_{i}{\frac{(f_i \cdot h)^2}{(f_i \cdot f_i)}} \geq 
\sum_{j}{\frac{(g_j \cdot h)^2}{{||g_j||}^2}}$, as needed.
\end{proof}
Now let $G'$ be a sound monotone switching network which accepts all of the inputs in $I = \{G_j\}$. 
By Corollary \ref{generalizedcrucialcorollary}, 
$\forall j, \exists i_j:\sum_{v' \in V(G')}{|(g_{G_j,E_{{i_j}j}} - g_{G_j,E_{1j}}) \cdot v'|} \geq \frac{z_j}{q_j-1}$. Using the Cauchy Schwarz inequality 
$(\sum_{v'}{f(v')g(v')})^2 \leq \sum_{v'}{f(v')^2}\sum_{v'}{g(v')^2}$ with $f(v') = 1$ and 
$g(v') = |(g_{G_j,E_{{i_j}j}} - g_{G_j,E_{1j}}) \cdot v'|$, we have that 
$$\forall j, \left(\frac{z_j}{q_j-1}\right)^2 \leq |V(G')|\sum_{v' \in V(G')}{((g_{G_j,E_{{i_j}j}} - g_{G_j,E_{1j}}) \cdot v')^2}$$
This implies that  
$$ \sum_{j}\frac{\left(\frac{z_j}{q_j-1}\right)^2}{{M_j}^2} \leq |V(G')|\sum_{j}{\sum_{v' \in V(G')}{\frac{((g_{G_j,E_{{i_j}j}} - g_{G_j,E_{1j}}) \cdot v')^2}{{||g_{G_j,E_{{i_j}j}} - g_{G_j,E_{1j}}||}^2}}}$$ 
However, by Proposition \ref{crucialproposition} applied to $v'$, 
$$\sum_{j}{\sum_{v' \in V(G')}{\frac{((g_{G_j,E_{ij}} - g_{G_j,E_{1j}}) \cdot v')^2}{{||g_{G_j,E_{ij}} - g_{G_j,E_{1j}}||}^2}}} \leq \sum_{v' \in V(G')}{1} = |V(G')|$$
Putting these inequalities together, 
${|V(G')|}^2 \geq \sum_{j}\frac{(\frac{z_j}{q_j-1})^2}{M^2}$, so 
$|V(G')| \geq \sqrt{\sum_{j}{\frac{(\frac{z_j}{q_j-1})^2}{{M_j}^2}}}$, as needed.
\end{proof}
\subsection{Conditions for a good set of functions}
\noindent The simplest way to use Theorem \ref{generalizedmanyinvariantslowerbound} is to take one input graph $G$, find a set of functions $\{g_{G,E_{i}}\}$ and then obtain the other input graphs and sets of functions by symmetry. We now give conditions which are sufficient to ensure that we can do this and deduce that if such sets of functions exist for paths $P$ of arbitrary length then any monotone switching network solving 
directed connectivity must have superpolynomial size.
\begin{theorem}\label{goodinvarianttolowerbound}
Let $V(G) = \{s,t,v_1, \cdots, v_m\}$. If there is a partition $E_1,\cdots,E_{q}$ of the edges of $G$, functions $\{g_{G,E_i}\}$, a value $z > 0$, a value $M$, and a value $r \leq m$ such that:
\begin{enumerate}
\item $g_{G,E_i}$ is $E_i$-invariant for $i \in [1,q]$
\item For all $i \in [1,q]$ and all $V \subseteq \Vmst$ with $|V| < r$, $(g_{G,E_i} - g_{G,E_1}) \cdot e_V = 0$
\item $g_{G,E_1} \cdot e_{\{\}} = z$
\item For all $i \in [1,q]$, $||g_{G,E_i} - g_{G,E_1}|| \leq M$
\end{enumerate}
then for all $n \geq 2m^2$, if $W$ is a set of vertices such that $V(G) \subseteq W$ and $|W \setminus \{s,t\}| = n$ then letting $H$ be the input graph with $V(H) = W$ and $E(H) = E(G)$ and letting $I$ be the set of all input graphs which are isomorphic to $H$ (keeping $s$ and $t$ fixed), $m(I) \geq \frac{z}{(q-1)M}(\frac{n}{2m})^{\frac{r}{2}}$
\end{theorem}
\begin{proof}
\noindent We first show that we can add additional isolated vertices to the input graph $G$ while still 
keeping the same functions (expressed in terms of their Fourier coefficients).
\begin{proposition}
For any $U,V \subseteq \Vmst$, ${e_U}{e_V} = e_{V \Delta U}$ where $\Delta$ is the set-symmetric difference function, i.e. $V \Delta U = (U \cup V) \setminus (U \cap V)$
\end{proposition}
\begin{proposition} For all $v,w \in \Vmst$, for all $C \in \mathcal{C}$,
\begin{enumerate}
\item $(e_{\{\}} + e_{\{w\}})(C) = 2$ if $w \in R(C)$ and $0$ if $w \in L(C)$.
\item $(e_{\{\}} - e_{\{v\}})(C) = 2$ if $v \in L(C)$ and $0$ if $v \in R(C)$.
\item $((e_{\{\}} - e_{\{v\}})(e_{\{\}} + e_{\{w\}}))(C) = 4$ if $v \in L(C)$ and $w \in R(C)$ and $0$ otherwise.
\end{enumerate}
\end{proposition}
\begin{corollary}\label{einvariancecondition} \ 
\begin{enumerate}
\item If $e = s \to w$ for some $w \in \Vmst$ then $g$ is $e$-invariant if and only if $(e_{\{\}} + e_{\{w\}})g = 0$.
Equivalently, $g$ is $e$-invariant if and only if $\hat{g}_{V \cup \{w\}} = -\hat{g}_V$ whenever 
$w \notin V$.
\item If $e = v \to t$ for some $v \in \Vmst$ then $g$ is $e$-invariant if and only if $(e_{\{\}} - e_{\{v\}})g = 0$.
Equivalently, $g$ is $e$-invariant if and only if $\hat{g}_{V \cup \{v\}} = \hat{g}_V$ whenever 
$v \notin V$.
\item  If $e = v \to w$ for some $v,w \in \Vmst$ then $g$ is $e$-invariant if and only if $(e_{\{\}} - e_{\{v\}})(e_{\{\}} + e_{\{w\}})g = 0$.
Equivalently, $g$ is $e$-invariant if and only if 
$\hat{g}_{V \cup \{v,w\}} = -\hat{g}_{V \cup \{v\}} + \hat{g}_{V \cup \{w\}} + \hat{g}_{V}$ whenever $v,w \notin V$.
\end{enumerate}
\end{corollary}
We now write $g_{G,E_i} = \sum_{V \subseteq \Vmst}{c_{iV}e_V}$. By Corollary 
\ref{einvariancecondition} if we have the input graph $H$ and take $g_{H,E_i} = \sum_{V \subseteq \Vmst}{c_{iV}e_V}$ then all conditions of Theorem \ref{goodinvarianttolowerbound} are still satisfied by $\{g_{H,E_i}\}$. Moreover, for all $i$ and all $V \nsubseteq \Vmst$, $g_{H,E_i} \cdot e_V = 0$.

We now take a set input graphs $I = \{H_j\}$ such that 
\begin{enumerate}
\item Each $H_j$ is obtained from $H$ by applying some permutation $\sigma_j$ to the vertices $W \setminus \{s,t\}$.
\item For all distinct $j_1$ and $j_2$, $\sigma_{j_1}(\Vmst) \cap \sigma_{j_1}(\Vmst) < r$
\end{enumerate} 
By Corollary \ref{largecollections}, we can take at least ${(\frac{n}{2m})}^{r}$ such graphs.
\begin{proposition}
If we take $E_{ij} = \sigma_j(E_i)$ and $g_{H_j,E_{ij}} = \sum_{V \subseteq \Vmst}{c_{iV}e_{\sigma_j(V)}}$ then
\begin{enumerate}
\item For all $j$, $g_{H_j,E_{ij}}$ is $E_{ij}$-invariant for $i \in [1,q]$
\item For all $j$ and all $i \in [1,q]$ , $g_{G_j,E_{ij}} \cdot e_{\{\}} = z$
\item For all $i,j$, $(g_{G_{j},E_{{i}j}} - g_{G_{j},E_{{1}j}}) \cdot e_V = 0$ whenever $|V| < r$ or $V \nsubseteq \sigma_j(\Vmst)$
\item For all $i,j$, $||g_{G_j,E_{{i}j}} - g_{G_j,E_{{1}j}}|| \leq M$
\end{enumerate}
\end{proposition}
\begin{proof}
This follows immediately from the properties of the functions $\{g_{H,E_i}\}$ and the fact that for all $i,j$ and all $V$,
$g_{H_j,E_{ij}} \cdot e_{\sigma_j(V)} = g_{H,E_i} \cdot e_V$
\end{proof}
Now note that since $(g_{H_j,E_{ij}} - g_{H_j,E_{1j}}) \cdot e_{V} = 0$ whenever $|V| < r$ or $V \nsubseteq \sigma_j(\Vmst)$ and for all distinct $j_1$ and $j_2$, $\sigma_{j_1}(\Vmst) \cap \sigma_{j_1}(\Vmst) < r$, we have that for all $i_1,i_2,j_1,j_2$ where $j_1 \neq j_2$,  
$$(g_{H_{j_1},E_{{i_1}{j_1}}} - g_{H_{j_1},E_{1{j_1}}}) \cdot (g_{H_{j_2},E_{{i_2}{j_2}}} - g_{H_{j_2},E_{1{j_2}}})= 0$$
Applying Corollary \ref{generalizedmanyinvariantslowerbound}, 
$$m(I) \geq \sqrt{\sum_{j}{\frac{\left(\frac{z}{q-1}\right)^2}{M^2}}} \geq \frac{z}{(q-1)M}\left(\frac{n}{2m}\right)^{\frac{r}{2}}$$
Adding the remaining input graphs which are isomorphic to $H$ to $I$ can only increase $m(I)$ and this completes the proof.
\end{proof}
\begin{corollary}\label{goodinvarianttolowerboundcorollary}
Take $V(P) = \{s,v_1, \cdots, v_{l-1},t\}$ and let $P$ be the path $s \to v_1 \to \cdots \to v_{l-1} \to t$. If $n \geq 2(l-1)^2$ and we can find a partition $\{E_1,\cdots,E_q\}$ of the edges of $P$, functions $\{g_{P,E_i}\}$, values $z,M$, and a value $r < l$ such that:
\begin{enumerate}
\item $g_{P,E_i}$ is $E_i$-invariant for $i \in [1,q]$
\item $(g_{P,E_i} - g_{P,E_1}) \cdot e_V = 0$ for all $i \in [1,q]$ and all $V \subseteq \Vmst$ with $|V| < r$ 
\item $g_{P,E_1} \cdot e_{\{\}} = z > 0$
\item For all $i$, $||g_{P,E_i} - g_{P,E_1}|| \leq M$
\end{enumerate}
then $m(\mathcal{P}_{n,l}) \geq \frac{z}{(q-1)M}(\frac{n}{2(l-1)})^{\frac{r}{2}}$.
\end{corollary}
\begin{proof}
This follows immediately from Theorem \ref{goodinvarianttolowerbound}.
\end{proof}
\begin{example}\label{invariantexampleone}
For $l = 2$ and $r = 1$ we can take $P = s \to v_1 \to t$, $E_1 = \{s \to v_1\}$, $E_2 = \{v_1 \to t\}$, 
$g_{P,E_1} = \frac{1}{2}(e_{\{\}} - e_{\{v_1\}})$, and $g_{P,E_2} = \frac{1}{2}(e_{\{\}} + e_{\{v_1\}})$ This gives $g_{P,E_2} - g_{P,E_1} = -e_{v_1}$. Using Proposition \ref{einvariancecondition} it can be verified directly that $g_{P,E_i}$ is $E_i$-invariant for 
$i \in \{1,2\}$. $||g_P|| = 1$ and $z = g_{P,E_1} \cdot e_{\{\}} = \frac{1}{2}$ 
so by Theorem \ref{goodinvarianttolowerbound}, for all $n \geq 2$, $m(\mathcal{P}_{n,1}) \geq \frac{\sqrt{n}}{2\sqrt{2}}$
\end{example}
\begin{example}\label{invariantexampletwo}
For $l = 3$ and $r = 2$ we can take $P = s \to v_1 \to v_2 \to t$, $E_1 = \{s \to v_1, v_2 \to t\}$, $E_2 = \{v_1 \to v_2\}$, 
$g_{P,E_2} = \frac{1}{4}(e_{\{\}} - e_{\{v_1\}} + e_{\{v_2\}} + 3e_{\{v_1,v_2\}})$, and 
$g_{P,E_1} = \frac{1}{4}(e_{\{\}} - e_{\{v_1\}} + e_{\{v_2\}} - e_{\{v_1,v_2\}})$. This gives 
$g_{P,E_2} - g_{P,E_1} = -e_{\{v_1,v_2\}}$. Using Proposition \ref{einvariancecondition} it can be verified directly that $g_{P,E_i}$ is $E_i$-invariant for $i \in \{1,2\}$. $||g_P|| = 1$ and $z = g_{P,E_1} \cdot e_{\{\}} = \frac{1}{4}$ so by Theorem \ref{goodinvarianttolowerbound}, 
for all $n \geq 8$, $m(\mathcal{P}_{n,2}) \geq \frac{n}{8}$.
\end{example}
\begin{remark}
These bounds are around the square root of the bounds obtained from the linear independence argument. This square root 
comes from the Cauchy-Schwarz inequality and so far we have not found a way to avoid having this square root. 
Nevertheless, getting a lower bound for $m(\mathcal{P}_{n})$ which is around $m(\mathcal{P}_{n})^c$ 
for some $c > 0$ is sufficient for our purposes and unlike the linear independence argument, we can use these techniques for longer paths.
\end{remark}
\section{A Superpolynomial Lower Bound}\label{fourieranalogues}
\noindent While the Fourier analysis and invariant approach of Section \ref{introducefouriertransform} is powerful, we need to actually find suitable functions $\{g_{P,E_i}\}$. There are several possibilities for how we 
could do this. One possibility is to look directly at the values $g_{P,E_i}(C)$ for all $C \in \mathcal{C}$. 
However, this gives us very little control over the Fourier coefficients of each $g_{P,E_i}$. 
A second possibility is to work directly with the Fourier coefficients of each $g_{P,E_i}$. This approach is 
viable, but it would involve analyzing how to satisfy many equations for $E_i$-invariance simultaneously. 
Here we take a third approach. We look at the dot products of each $g_{P,E_i}$ with the vertices of 
the certain knowledge switching network $G'_c(n)$ (see Definition \ref{universalcertainknowledge}). 
It turns out that all of the conditions of Corollary \ref{goodinvarianttolowerboundcorollary} correspond 
to simple conditions on the values of these dot products. Furthermore, we have complete freedom in choosing the values of these dot products, which enables us to construct suitable $\{g_{P,E_i}\}$ and 
thus prove the following theorem.
\begin{theorem}\label{keytheorem}
For all $l \geq 2$, if we have $V(G) = \{s,t,v_1, \cdots, v_{l-1}\}$ and let 
$P$ be the path $s \to v_1 \to \cdots \to v_{l-1} \to t$ then taking $r = \lceil{\lg{l}}\rceil$ and taking the partition $E_{i} = \{v_{i-1} \to v_{i}\}$ of the edges of $E$ (where $v_0 = s$ and $v_l = t$), we can find functions $\{g_{P,E_i}\}$ such that:
\begin{enumerate}
\item $g_{P,E_i}$ is $E_i$-invariant for all $i$
\item $(g_{P,E_i} - g_{P,E_1}) \cdot e_V = 0$ for all $i$ and all $V \subseteq \Vmst$ with $|V| < r$ 
\item $g_{P,E_1} \cdot e_{\{\}} > 0$
\end{enumerate}
\end{theorem}
\noindent Combined with Corollary \ref{goodinvarianttolowerboundcorollary}, this immediately proves the following theorem, 
which implies a superpolynomial lower bound on monotone switching networks solving directed connectivity on $n$ vertices.
\begin{theorem}\label{weaksuperpolynomial}
For any integer $l \geq 2$, there is a constant $c_l$ such that for all $n \geq 0$,
$m(\mathcal{P}_{n,l}) \geq {c_l}n^{\frac{\lceil{\lg{l}}\rceil}{2}}$
\end{theorem}
\subsection{From certain knowledge decriptions to function descriptions}
For the proof of Theorem \ref{keytheorem}, we show how results from Section \ref{specialcase}, in particular Lemma \ref{specialboundlemma}, can be adapted to the Fourier analysis and invariants approach. We begin by taking certain knowledge descriptions and giving a corresponding function description.
\begin{definition}
For a given knowledge set $K$, define the function $K: \mathcal{C} \to \{0,1\}$ to be the function where
$K(C) = 1$ if there is no edge in $K$ which crosses $C$ and $0$ otherwise.
\end{definition}
\begin{proposition}\label{knowledgesetfunctionprop}
If we can get from $K_1$ to $K_2$ in the certain knowledge game using only the knowledge that some 
edge $e$ is in $G$ and $e$ does not cross some cut $C$ then $K_2(C) = K_1(C)$.
\end{proposition}
\begin{proof}
This follows immediately from the fact that if $e$ does not cross $C$, then for any knowledge set $K$, 
no individual move on $K$ in the certain knowledge game which can be done with only the knowledge that 
$e$ is in $G$ changes the value of $K(C)$.
\end{proof}
\begin{corollary}
If a monotone switching network $G'$ has a certain-knowledge description 
$\{K_{v'}\}$ where each vertex $v'$ is assigned the knowledge set $K_{v'}$ then if we assign each 
vertex $v'$ the function $K_{v'}$, this is a function description of $G'$.
\end{corollary}
\subsection{A criterion for $E$-invariance}
Now that we have translated certain knowledge descriptions into function descriptions, 
we prove the following criterion for $E$-invariance. This criterion shows that to check $E$-invariance, it is sufficient to check $E$-invariance on certain knowledge switching networks.
\begin{theorem}\label{keyinvariancecondition}
If $g$ is a function from $\mathcal{C}$ to $\mathbb{R}$ and $E$ is a set of edges between vertices in $V(G)$ then 
$g$ is $E$-invariant if and only if $g \cdot v'_1 = g \cdot v'_2$ whenever $v'_1,v'_2$ are vertices of $G'_c(n)$ and there is an edge between $v'_1$ and $v'_2$ in $G'_c(n)$ whose edge label is in $E$.
\end{theorem}
\begin{proof}
We give a short direct proof here using inclusion/exclusion. For a deeper but longer and more technical proof, see Appendix \ref{elementarymonotone} and Appendix \ref{invarianceconditiondeepproof}.

For the only if direction, note that if $g$ is $E$-invariant then for all monotone switching networks $G'$ and $v',w' \in V(G')$, whenever there is an edge $e'$ between $v'$ and $w'$ whose label is in $E$, $(w' - v') \cdot g = 0$. To see this, note that $w'(C) - v'(C) = 0$ whenever $C \in \mathcal{C}_E$ (recall that this is the set of cuts which cannot be crossed by an edge in $E$) and $g(C) = 0$ for all $C \notin \mathcal{C}_E$.

For the if direction, consider a cut $C \notin \mathcal{C}_E$. There must be an edge $e = u \to v \in E$ such that $u \in L(C)$ and $v \in R(C)$. Consider the expression 
$$1_{C} = -\sum_{W: L(C) \subseteq W \subseteq V(G) \setminus \{v,t\}}{(-1)^{|W| - |L(C)|}(K_{W \cup \{v\}} - K_W)}$$

Given a cut $C_2$, if $L(C) \nsubseteq L(C_2)$ then there is a vetex $w \in L(C) \setminus L(C_2)$. If so then whenever $L(C) \subseteq W$, $s \to w \in K_W$ and $s \to w$ crosses $C_2$ so $K_{W \cup \{v\}}(C_2) = K_W(C_2) = 0$. This implies that $1_C(C_2) = 0$.

If $R(C) \nsubseteq R(C_2)$ then there is a vertex $w \in R(C) \setminus R(C_2)$. If $w = v$ then for all $W$ such that $L(C) \subseteq W \subseteq V(G) \setminus \{v,t\}$, $K_{W \cup \{v\}}(C_2) = K_{W}(C_2)$ so $1_C(C_2) = 0$. If $w \neq v$ then for all $W$ such that $L(C) \subseteq W \subseteq V(G) \setminus \{v,t,w\}$, $K_{W \cup \{w\}}(C_2) = K_{W}(C_2)$ and $K_{W \cup \{w,v\}}(C_2) = K_{W \cup \{v\}}(C_2)$. Since $K_{W \cup \{w\}}(C_2)$ and $K_{W}(C_2)$ always have opposite signs in the expression for $1_C(C_2)$ and $K_{W \cup \{v,w\}}(C_2)$ and $K_{W \cup \{v\}}(C_2)$ always have opposite signs in the expression for $1_C(C_2)$, this implies that $1_C(C_2) = 0$.

Finally, note that $K_{L(C)}(C) = 1$, $K_{L(C) \cup \{v\}}(C) = 0$ and for all $W$ such that $L(C) \subseteq W \subseteq V(G) \setminus \{v,t\}$ and $W \neq L(C)$, $K_{W \cup \{v\}}(C) = K_{W}(C) = 0$. Putting everything together, we have that $1_C(C_2) = 1$ if $C_2 = C$ and $0$ otherwise.

Using this, the result follows easily. For all $C \notin \mathcal{C}_E$.
$$g(C) = 2^{n}(g \cdot 1_C) = -2^{n}\sum_{W: L(C) \subseteq W \subseteq V(G) \setminus \{v,t\}}{(-1)^{|W| - |L(C)|}((K_{W \cup \{v\}} - K_W) \cdot g)} = 0$$
so $g$ is $E$-invariant, as needed.
\end{proof}
\subsection{Choosing Fourier coefficients via dot products}
Now that we have shown how to check $E$-invariance of a function $g$ by looking at values of $g \cdot K_V$ for $V \subseteq \Vmst$, we show in picking a function $g$, we can choose these values arbitrarily.
\begin{theorem}\label{dotproductfouriercoconnection}
For any set of values $\{a_V: V \subseteq \Vmst\}$, there is a unique function $g: \mathcal{C} \to \mathbb{R}$ 
such that for all $V \subseteq \Vmst$, $g \cdot K_V = a_V$. Furthermore, for any $r$, 
$\hat{g}_V = 0$ for all $V$ such that $|V| < r$ if and only if $a_V = g \cdot K_V = 0$ for all $V$ such that 
$|V| < r$.
\end{theorem}
\begin{proof}
\begin{proposition}\label{certainknowledgefouriercoprop}
For any $V \subseteq \Vmst$, $K_V = \frac{1}{2^{|V|}}\sum_{U \subseteq V}{(-1)^{|U|}e_{U}}$
\end{proposition}
\begin{proof}
Note that $K_V(C) = 1$ if $V \subseteq L(C)$ and $0$ otherwise. Now for all cuts $C \in \mathcal{C}$, 
$$\frac{1}{2^{|V|}}\sum_{U \subseteq V}{(-1)^{|U|}e_{U}}(C) 
= \frac{1}{2^{|V|}}\sum_{U \subseteq V}{(-1)^{|U|}(-1)^{|U \cap L(C)|}}$$
If $V \nsubseteq L(C)$ then all terms will cancel so $\frac{1}{2^{|V|}}\sum_{U \subseteq V}{(-1)^{|U|}e_{U}}(C) = 0$. 
If $V \subseteq L(C)$ then 
$$\frac{1}{2^{|V|}}\sum_{U \subseteq V}{(-1)^{|U|}(-1)^{|U \cap L(C)|}} = \frac{1}{2^{|V|}}\sum_{U \subseteq V}{1} = 1$$
This implies that $K_V = \frac{1}{2^{|V|}}\sum_{U \subseteq V}{(-1)^{|U|}e_{U}}$, as needed.
\end{proof}
\begin{corollary}\label{certainknowledgefouriercocor}
\noindent \newline
1. For all $V \subseteq \Vmst$, $e_{V} \cdot K_V \neq 0$\\
2. For all subsets $U,V$ of $\Vmst$, if $U \not\subset V$ then $e_{U} \cdot K_V = 0$.
\end{corollary}
To see the first part of Theorem \ref{dotproductfouriercoconnection}, pick an ordering $\{V_i\}$ of the subsets $V \subseteq \Vmst$ such that if $i < j$ then $V_j \not\subset V_i$. We now pick the Fourier coefficients $\hat{g}_{V_i}$ in increasing order of $i$. By statement 2 of Corollary \ref{certainknowledgefouriercocor}, for all subsets $U,V$ of $\Vmst$, if $U \not\subset V$ then $e_{U} \cdot K_V = 0$. This means that for each $i$, once we pick $\hat{g}_{V_i}$, this determines the value of $g \cdot K_{V_i} = a_{V_i}$ as for any $j > i$, $V_j \not\subset V_i$ so $e_{V_j} \cdot K_{V_i} = 0$. By statement 1 of Corollary \ref{certainknowledgefouriercocor}, for all $i$, $e_{V_i} \cdot K_{V_i} \neq 0$. This means that we always have a unique choice for each coefficient $\hat{g}_{V_i}$ which gives $g \cdot K_{V_i} = a_{V_i}$. Putting everything together, there is a unique function $g: \mathcal{C} \to \mathbb{R}$ such that for all $V \subseteq \Vmst$, $g \cdot K_V = a_V$, as needed.

Now we just need to show that if $a_V = g \cdot K_V = 0$ for all $V$ such that 
$|V| < r$ then $\hat{g}_V = 0$ for all $V$ such that $|V| < r$. To see this, assume it is false. Take a minimal subset 
$V$ of $\Vmst$ such that $\hat{g}_{V} \neq 0$ and $a_U = g \cdot K_U = 0$ for all $U \subseteq V$. Then 
$\hat{g}_{V} \neq 0$, $\hat{g}_{U} = 0$ for all $U \subsetneq V$, and $g \cdot K_U = 0$ for all $U \subseteq V$. 
However, by Corollary \ref{certainknowledgefouriercocor}, if $\hat{g}_{V} \neq 0$ and $\hat{g}_U = 0$ 
for all $U \subsetneq V$ then $g \cdot K_V \neq 0$. Contradiction.
\end{proof}
\subsection{Proof of Theorem \ref{keytheorem}}
\noindent We are now ready to construct the functions $\{g_{P,E_i}\}$ and prove Theorem \ref{keytheorem}, which we recall 
below for convenience.
\vskip.1in
\noindent
{\bf Theorem \ref{keytheorem}.}
{\it
For all $l \geq 2$, if we have $V(G) = \{s,t,v_1, \cdots, v_{l-1}\}$ and let 
$P$ be the path $s \to v_1 \to \cdots \to v_{l-1} \to t$ then taking $r = \lceil{\lg{l}}\rceil$ and taking the partition $E_{i} = \{v_{i-1} \to v_{i}\}$ of the edges of $E$ (where $v_0 = s$ and $v_l = t$), we can find functions $\{g_{P,E_i}\}$ such that:
\begin{enumerate}
\item $g_{P,E_i}$ is $E_i$-invariant for all $i$
\item $(g_{P,E_i} - g_{P,E_1}) \cdot e_V = 0$ for all $i$ and all $V \subseteq \Vmst$ with $|V| < r$ 
\item $g_{P,E_1} \cdot e_{\{\}} > 0$
\end{enumerate}
}
\begin{proof}
Note that by Theorem \ref{keyinvariancecondition} and Theorem \ref{dotproductfouriercoconnection}, the three 
conditions of Theorem \ref{keytheorem} are equivalent to the following three conditions:
\begin{enumerate}
\item For all $i$, $g_{P,E_i} \cdot u' = g_{P,E_i} \cdot v'$ for any vertices $u',v'$ of $G'_c(n)$ which have an edge between them whose label is $e_i$.
\item $g_{P,E_i} \cdot K_V = g_{P,E_1} \cdot K_V$ for all $i$ and $V \subseteq \Vmst$ with $|V| < r = \lceil{\lg{l}}\rceil$
\item $g_{P,E_1} \cdot K_{\{\}} > 0$
\end{enumerate}
By Theorem \ref{dotproductfouriercoconnection}, we can choose the values 
$\{g_{P,E_i} \cdot K_V: V \subseteq \Vmst\}$ 
freely, so it is sufficient to give a function $b': V(G'_c(n)) \times \{[1,l]\} \to \mathbb{R}$ such that 
\begin{enumerate}
\item If there is an edge between vertices $u'$ and $v'$ whose label is in $E_i$ then $b'(u',i) = b'(v',i)$
\item $b'(v',i) = b'(v',1)$ for all $i$ whenever $K_{v'} \in \{K_V: V \subseteq \Vmst, |V| < r\} \cup \{K_{t'}\}$
\item $b'(s',1) = 1$ and $b'(t',1) = 0$
\end{enumerate}
We choose the values $\{b'(v',i)\}$ by looking at connected components of certain graphs which 
are closely related to $V(G'_c(n))$. Let $H'$ be the graph with 
\begin{enumerate}
\item $V(H') = \{v' \in V(G'_c(n)): K_{v'} \in 
\{K_V: V \subseteq \Vmst, |V| < r\} \cup \{K_{t'}\}\}$
\item $E(H') = \{(u',v'): u',v' \in V(H'),$ there is an edge between $u'$ and $v'$ whose label is in $E(P)\}$
\end{enumerate}
For each $i$ let $H'_i$ be the graph with 
\begin{enumerate}
\item $V(H'_{i}) = V(G'_c(n))$
\item $E(H'_i) = E(H) \cup \{e' \in E(G'_c(n)): \mu'(e') \in E_i\}$
\end{enumerate}
\begin{proposition}\label{certainknowledgecomponents}
If $u', v' \in V(H')$ and $u'$ and $v'$ are in the same connected component of $H'_i$ for some $i$ then 
$u'$ and $v'$ are in the same connected component of $H'$.
\end{proposition}
\begin{proof}
Assume that we have $u'$ and $v'$ which are in different components of $H'$ but are in the same component of $H'_i$ for some $i$. If so, choose $u'$ and $v'$ to minimize the length of the shortest path in $H'_i$ from $u'$ to $v'$. Note that there cannot be any $w' \in V(H')$ on this path, as otherwise $w'$ is either in a different component of $H'$ than $u'$ in which case we could have taken the shorter path from $u'$ to $w'$ instead or $w'$ is in a different component of $H'$ than $v'$ in which case we could have taken the shorter path from $w'$ to $v'$ instead. Thus all edges of the path between $u'$ and $v'$ in $H'_i$ are not edges of $H'$ and thus must have label $e_i$. But then since $G'_c(n)$ has all allowable edges, there must be an edge between $u'$ and $v'$ with label $e_i$ so $u'$ and $v'$ are actually in the same connected component of $H'$. Contradiction.
\end{proof}
This proposition implies that we may first choose any set of values $\{b'(v')\}$ such that  
$b'(u') = b'(v')$ whenever $u'$ and $v'$ are in the same connected component of $H'$ and 
then choose any set of values $\{b'(v',i)\}$ such that if $v' \in V(H')$ then $b'(v',i) = b'(v')$ 
for all $i$ and $b'(u',i) = b'(v',i)$ whenever $u'$ and $v'$ are in the same connected component 
of $H'_i$.

One way to do this is to first take $b'(u') = 1$ if $u'$ is in the same connected component of $H'$ as $s'$ and $b'(u') = 0$ otherwise, then take $b'(v',i) = b'(u')$ whenever $v'$ is in the same connected component of $H'_i$ as $u'$ for some $u' \in V(H')$ and take $b'(v',i) = 0$ whenever $v'$ is not in the same connected component as any $u' \in V(H')$. This is guaranteed to satisfy the first two conditions of Theorem \ref{keytheorem}. 

For the third condition, we need to check that $s'$ and $t'$ are in different connected components of $H'$ as we then have that $b'(s') = 1$ and $b'(t') = 0$. To check this, 
assume that $s'$ and $t'$ are in the same connected component of $H'$. Then there is a path from 
$s'$ to $t'$ in $H'$. However, this is impossible by Lemma \ref{specialboundlemma}. Contradiction.
\end{proof} 
\begin{remark}
In choosing the values $b'(v')$ for $v' \in H'$, we are essentially picking the Fourier coefficients $\hat{g}_V: V < r$ for a ''base function'' $g$ which we then extend to an $E_i$-invariant function $g_i$ for every $i$. The crucial idea is that if we only look at the Fourier coefficients $\hat{(g_i)}_{V}$ for $|V| < r$, all of the $g_i$ look identical to $g$ and thus look identical to each other. 
\end{remark}
\section{An $n^{\Omega(\lg{n})}$ lower size bound}\label{nlgnlowerboundsection}
In this section, we prove an $n^{\Omega(\lg{n})}$ lower size bound on monotone switching networks 
solving directed connectivity by explicitly finding the functions $\{g_{P,E_i}\}$ given by Theorem \ref{keytheorem} and then modifying them by ''cutting off'' high Fourier coefficients.
\begin{theorem}\label{nlgnlowerbound}
$$m(\mathcal{P}_{n,l}) \geq \frac{1}{2}{(\frac{n}{64(l-1)^2})}^{\frac{\lceil{\lg{l}}\rceil}{2}}$$
$$m(\mathcal{P}_{n}) \geq  \frac{1}{2}n^{\frac{\lg{n}}{16} - \frac{3}{4}}$$
\end{theorem}
\begin{proof}
The first step in proving this lower bound is to gain a better understanding of the functions given by 
Theorem \ref{keytheorem}.
\begin{definition}\label{vertexdeltafunctions} For all $V \subseteq \Vmst$, define the function $g_V: \mathcal{C} \to \mathbb{R}$ so that 
\begin{enumerate}
\item $g_V(C) = 0$ if $(L(C) \setminus \{s\}) \nsubseteq V$
\item $g_V(C) = 2^{n}(-1)^{|V \setminus L(C)|}$ if $(L(C) \setminus \{s\}) \subseteq V$
\end{enumerate}
\end{definition}
The most important property of these functions is as follows.
\begin{lemma}\label{gvproperty}
If $V_1,V_2 \subseteq \Vmst$, $K_{V_1} \cdot g_{V_2} = 1$ if $V_1 = V_2$ and $0$ otherwise.
\end{lemma}
\begin{proof}
We have that $K_V(C) = 1$ if $V \subseteq L(C)$ and $K_V(C) = 0$ otherwise. Now note that 
$$K_{V_1} \cdot g_{V_2} = \sum_{C \in \mathcal{C}: V_1 \subseteq (L(C) \setminus \{s\}) \subseteq V_2}{(-1)^{|V_2 \setminus L(C)|}}$$
Ths implies that $K_{V_1} \cdot g_{V_2} = 1$ if $V_1 = V_2$ and $0$ otherwise, as needed.
\end{proof}
We can now construct the functions $\{g_{P,E_i}\}$ in terms of the functions $\{g_V\}$ and analyze their Fourier coefficients.
\begin{lemma}\label{gvlemma}
If $\{g_{P,E_i}\}$ are the functions given by Theorem \ref{keytheorem} then we have that
\begin{enumerate}
\item $g_{P,E_i} = \sum_{V \subseteq \Vmst}{b(V,i)g_V}$
\item $g_{P,E_i} \cdot e_V = \sum_{U \subseteq V}{b(U,i)(g_U \cdot e_V)} = 
\sum_{U \subseteq V}{b(U,i)(-2)^{|U|}}$
\end{enumerate}
\end{lemma}
\begin{proof}
The first statement follows from Lemma \ref{gvproperty} and the definition of $b(V,i)$ as the value of $g_{P,E_i} \cdot e_V$. For the second statement, we use the following proposition.
\begin{proposition}
For all $U,V \subseteq \Vmst$, $g_U \cdot e_V = (-2)^{|U|}$ if $U \subseteq V$ and $0$ otherwise
\end{proposition}
\begin{proof}
$$g_U \cdot e_V = \sum_{C \in \mathcal{C}: (L(C) \setminus \{s\}) \subseteq U}{(-1)^{|U \setminus L(C)|}(-1)^{|V \cap L(C)|}}$$
If there is some $i \in U \setminus V$, then shifting $i$ from $L(C)$ to $R(C)$ or vice versa changes $(-1)^{|U \setminus L(C)|}(-1)^{|V \cap L(C)|}$ by a factor of $-1$. Thus, everything cancels and we have $g_U \cdot e_V = 0$. If $U \subseteq V$ then 
$$g_U \cdot e_V = \sum_{C \in \mathcal{C}: (L(C) \setminus \{s\}) \subseteq U}{(-1)^{U \setminus L(C)}(-1)^{|V \cap L(C)|}} = \sum_{C \in \mathcal{C}: (L(C) \setminus \{s\}) \subseteq U}{(-1)^{|U|}} = (-2)^{|U|}$$
\end{proof}
The completes the proof of Lemma \ref{gvlemma}
\end{proof} 
\begin{corollary}
If we take each $b(V,i)$ to be $0$ or $1$ when choosing the functions $\{g_{P,E_i}\}$ then for all $V \subseteq \Vmst$,  $|g_{P,E_i} \cdot e_V| \leq 2^{2|V|}$.
\end{corollary}
If we take the functions $\{g_{P,E_i}\}$ given by Theorem \ref{keytheorem} directly, then 
$||g_{P,E_i} - g_{P,E_1}||$ may be very large. The key observation is that as shown below using 
Corollary \ref{einvariancecondition}, we can cut off all of the Fourier coefficients $g_{P,E_i} \cdot e_V$ where $|V| > r = \lceil{\lg{l}}\rceil$. 
\begin{lemma}
Taking $r = \lceil{\lg{l}}\rceil$, there exist functions $g_i$ such that 
\begin{enumerate}
\item For all $i$, $g_i$ is $E_i$-invariant.
\item For all $i$ and all $V$ such that $|V| < r$, $(g_i - g_1) \cdot e_V = 0$
\item $g_1 \cdot e_{\{\}} = 1$
\item For all $i$, $||g_i - g_1|| \leq (l-1)^{\frac{r-1}{2}}2^{2r+1}$
\end{enumerate}
\end{lemma}
\begin{proof}
We repeat Corollary \ref{einvariancecondition} here for convenience.
\vskip.1in
\noindent
{\bf Corollary \ref{einvariancecondition}.}
{\it \noindent
\begin{enumerate}
\item If $e = s \to w$ for some $w \in \Vmst$ then $g$ is $e$-invariant if and only if $(e_{\{\}} + e_{\{w\}})g = 0$.
Equivalently, $g$ is $e$-invariant if and only if $\hat{g}_{V \cup \{w\}} = -\hat{g}_V$ whenever 
$w \notin V$.
\item If $e = v \to t$ for some $v \in \Vmst$ then $g$ is $e$-invariant if and only if $(e_{\{\}} - e_{\{v\}})g = 0$.
Equivalently, $g$ is $e$-invariant if and only if $\hat{g}_{V \cup \{v\}} = \hat{g}_V$ whenever 
$v \notin V$.
\item  If $e = v \to w$ for some $v,w \in \Vmst$ then $g$ is $e$-invariant if and only if \\
$(e_{\{\}} - e_{\{v\}})(e_{\{\}} + e_{\{w\}})g = 0$.
Equivalently, $g$ is $e$-invariant if and only if \\
$\hat{g}_{V \cup \{v,w\}} = 
-\hat{g}_{V \cup \{v\}} + \hat{g}_{V \cup \{w\}} + \hat{g}_{V}$ whenever $v,w \notin V$.
\end{enumerate}
}
\begin{definition}
Define the functions $\{g_i\}$ so that 
\begin{enumerate}
\item $g_i \cdot e_V = g_{P,E_i} \cdot e_V$ if $|V| < r$
\item $g_i \cdot e_V = 0$ if $|V| > r$
\item If $i = 1$ (so that $E_i = \{s \to v_1\}$), and $|V| = r$ then 
\begin{enumerate}
\item If $v_1 \in V$ then $g_i \cdot e_V = -g_i \cdot e_{V \setminus \{v_1\}}$ if $v_1 \in V$
\item If $v_1 \notin V$ then $g_i \cdot e_V = 0$ 
\end{enumerate}
\item If $i = l$ (so that $E_i = \{v_{l-1} \to t\}$) and $|V| = r$ then 
\begin{enumerate}
\item If $v_{l-1} \in V$ then $g_i \cdot e_V = g_i \cdot e_{V \setminus \{v_{l-1}\}}$ 
\item If $v_{l-1} \notin V$ then $g_i \cdot e_V = 0$
\end{enumerate}
\item If $i \notin \{1,l\}$ (so $E_i = \{v_{i-1} \to v_{i}\}$) and $|V| =r$ then 
\begin{enumerate}
\item If $v_{i-1},v_i \in V$ then $g_i \cdot e_V = g_i \cdot e_{V \setminus \{v_{i-1},v_i\}} - 
g_i \cdot e_{V \setminus \{v_i\}} + g_i \cdot e_{V \setminus \{v_{i-1}\}}$
\item If $v_{i} \in V$ and $v_{i-1} \notin V$ then 
$g_i \cdot e_V = -g_i \cdot e_{V \setminus \{v_i\}}$
\item If $v_{i} \notin V$ then $g_i \cdot e_V = 0$
\end{enumerate}
\end{enumerate} 
\end{definition}
\begin{proposition}\label{checkinginvarianceprop}
$g_i$ is $E_i$ invariant for all $i$.
\end{proposition}
\begin{proof}
We can show that $g_i$ is $E_i$ invariant using Corollary \ref{einvariancecondition}. When
$v_{i-1},v_i \notin V$ and $|V \cup \{v_{i-1},v_i\} \setminus \{s,t\}| \geq r$ we can 
check directly that the associated equation in Corollary \ref{einvariancecondition} holds. When 
$v_{i-1},v_i \notin V$ and $|V \cup \{v_{i-1},v_i\} \setminus \{s,t\}| < r$ we 
use the fact that the associated equation in Corollary \ref{einvariancecondition} must hold for 
$g_{P,E_i}$ and thus holds for $g_i$ as well.
\end{proof}
\begin{corollary}
The functions $\{g_i\}$ have the following properties:
\begin{enumerate}
\item For all $i$, $g_i$ is $E_i$-invariant.
\item For all $i$ and all $V \subseteq \Vmst$ where $|V| < r$, 
$g_{i} \cdot e_V = g_{1} \cdot e_V$
\item For all $i$, $g_i \cdot e_V = 0$ whenever $V \subseteq \Vmst$ and $|V| > r$
\item For all $i$, $g_i \cdot e_{\{\}} = 1$
\item For all $i$, $g_i \cdot e_V \neq 0$ for at most ${l-1} \choose {\lceil{\lg{l}}\rceil - 1}$ $V$ with $|V| = r$
\item $|g_i \cdot e_V| \leq 3 \cdot 2^{2r - 2} \leq 2^{2r}$ for all $V$ with $|V| = r$
\end{enumerate}
\end{corollary}
\begin{proof}
The first statment is just Proposition \ref{checkinginvarianceprop}. The second, third, and fourth statements all follow 
from the definition of the functions $\{g_i\}$ and the properties of the functions $\{g_{P,E_i}\}$. For the fifth statement, 
note that every time we fix a nonzero Fourier coefficient $g_i \cdot e_V$ where $|V| = \lceil{\lg{l}}\rceil$ we use Fourier 
coefficients of the form $g_i \cdot e_W$ where $|W| < r$ to determine its value. Moreover, we 
never use the same Fourier coefficient twice, so the number of nonzero Fourier coefficients $g_i \cdot e_V$ where 
$|V| = r$ is at most $\binom{l-1}{r - 1}$. Finally, the sixth statement 
follows from the definition of $g_i$ and our bounds on the Fourier coefficients of the functions $\{g_{P,E_i}\}$.
\end{proof}
Note that when we look at $g_{i} - g_{1}$, all Fourier coefficients with $V < r$ cancel. From this, it follows that for any $i$, 
$||g_{i} - g_{1}||^2 \leq 2||g_{i}||^2 + 2||g_{1}||^2 \leq (l-1)^{r-1}2^{4r+2}$.
\end{proof}
We now prove Theorem \ref{nlgnlowerbound} using Corollary \ref{goodinvarianttolowerboundcorollary} which we repeat here for convenience.
\vskip.1in
\noindent
{\bf Corollary \ref{goodinvarianttolowerboundcorollary}.}
{\it
Take $V(P) = \{s,v_1, \cdots, v_{l-1},t\}$ and let $P$ be the path $s \to v_1 \to \cdots \to v_{l-1} \to t$. If $n \geq 2(l-1)^2$ and we can find a partition $\{E_1,\cdots,E_q\}$ of the edges of $P$, functions $\{g_{P,E_i}\}$, values $z,M$, and a value $r < l$ such that:
\begin{enumerate}
\item $g_{P,E_i}$ is $E_i$-invariant for $i \in [1,q]$
\item $(g_{P,E_i} - g_{P,E_1}) \cdot e_V = 0$ for all $i \in [1,q]$ and all $V \subseteq \Vmst$ with $|V| < r$ 
\item $g_{P,E_1} \cdot e_{\{\}} = z > 0$
\item For all $i$, $||g_{P,E_i} - g_{P,E_1}|| \leq M$
\end{enumerate}
then $m(\mathcal{P}_{n,l}) \geq \frac{z}{(q-1)M}(\frac{n}{2(l-1)})^{\frac{r}{2}}$.
}
By Corollary \ref{goodinvarianttolowerboundcorollary}, taking $r = \lceil{\lg{l}}\rceil$, 
$M = (l-1)^{\frac{r-1}{2}}2^{2r+1}$, and $q = l$, 
for all $n \geq 2(l-1)^2$, 
$$m(\mathcal{P}_{n,l}) \geq 
\frac{1}{M(l-1)}\left(\frac{n}{2(l-1)}\right)^{\frac{r}{2}} \geq \frac{n^{\frac{r}{2}}}{2^{\frac{5r}{2}+1}(l-1)^{r + \frac{1}{2}}}$$
Using the fact that $2^{r} \geq l$, we may reexpress this bound as 
 $$m(\mathcal{P}_{n,l}) \geq 
\frac{n^{\frac{\lceil{\lg{l}}\rceil}{2}}}{2^{6\frac{\lceil{\lg{l}}\rceil}{2}+1}(l-1)^{\lceil{\lg{l}}\rceil}} \geq \frac{1}{2}\left(\frac{n}{64(l-1)^2}\right)^{\frac{\lceil{\lg{l}}\rceil}{2}}$$
Note that we may ignore the condition that $n \geq 2(l-1)^2$ because the bound is trivial if $n < 2(l-1)^2$. 
Taking $l = \lceil{\frac{1}{8}n^{\frac{1}{4}}}\rceil$, we have that 
$$m(\mathcal{P}_{n}) \geq \frac{1}{2}(\sqrt{n})^{\frac{\frac{1}{8}\lg{n} - 3}{2}} = \frac{1}{2}n^{\frac{\lg{n}}{16} - \frac{3}{4}}$$
as needed.
\end{proof}
\section{Further work and open problems}\label{futureworkone}
In this paper, we have shown almost tight upper and lower bounds on the size of sound monotone switching networks solving directed connectivity. However, there are several limitations to this result.

The most important limitation is that the lower bounds only apply to monotone switching networks. Removing this limitation would almost certainly be extremely difficult, as it would show that L is not equal to NL, solving a major open problem in theoretical computer science.

Even in the monotone case, there are several limitations. Most importantly, this is a worst-case result showing that accepting all minimal YES instances and rejecting all maximum NO instances is hard. This limitation has been addressed in follow up work. Robere, Cook, Filmus, and Pitassi \cite{averagebounds} showed an average case lower bound when we take a distribution over minimal YES instances and maximal NO instances. In \cite{improvedbounds}, we consider the monotone space complexity of solving directed connectivity on other input graphs. More precisely, we define $m(G)$ to be the minimal size of a sound monotone switching network which accepts all input graphs isomorphic to $G$. Letting $l$ be the length of the shortest path from $s$ to $t$, we show that $m(G)$ is $n^{\Omega(\lg{l})}$ whenever no vertex of $G$ is connected by shorter paths to too many other vertices of $G$. We also show an upper bound, showing that $m(G)$ is small whenever almost all vertices $v$ in $G$ are directly reachable from $s$ or can directly reach $t$, i.e. $s \to v \in E(G)$ or $v \to t \in E(G)$. Building on this work, Brakensiek and Potechin \cite{brakensiek} proved almost tight bounds on $m(G)$ whenever $m(G)$ is an acyclic directed tree. A natural open problem is to obtain almost tight bounds on $m(G)$ whenever $G$ is an acyclic directed graph.

Another direction is to extend this result to other problems besides directed connectivity. Chan and Potechin \cite{chan} extended the techniques of this paper to show tight monotone space lower bounds for the GEN problem, giving an alternate proof of the separation of the monotone NC-hierarchy, as well as the k-clique problem. However, showing corresponding monotone space lower bounds for other problems, including k-matching (where monotone circuit depth lower bounds are known) remains open.

A third direction is to look at monotone circuits with logarithmic width and polynomial size, which is an alternative way to define (non-uniform) monotone-L. As noted in the introduction, it is an open problem how this definition of (non-uniform) monotone-L and the definition of (non-unifrom) montone-L in terms of polynomial-size monotone switching networks are related. 

Finally, we can aim to tighten our results further. We have determined $c(\mathcal{P}_{n})$ and $m(\mathcal{P}_{n})$ up to a constant in the exponent, what is the exact constant? Answering this question for certain knowledge switching networks would require sharper combinatorial analysis while answering this question for monotone switching networks would almost certainly require finding an alternative to the Cauchy-Schwarz argument. We can also ask whether monotone switching networks are better at solving directed connectivity than certain knowledge switching networks, i.e. is $c(\mathcal{P}_{n}) = m(\mathcal{P}_{n})$? From our follow-up work we know that certain knowledge switching networks are less effective than monotone switching networks for some input graphs but we have no reason to believe this is the case for minimal YES instances.
\section{Conclusion}\label{conclusionone}
In this paper, we developed powerful tools for analyzing monotone switching networks for directed connectivity and used them to prove that the minimum size of a monotone switching network solving directed connectivity is $n^{\Theta{(\lg{n})}}$, separating monotone analogues of L and NL. Since this work was first presented there have been several follow-up papers, which shows that using switching networks to analyze space complexity is a fruitful approach. That said, there are many open questions remaining and only time will tell how far this approach will take us.

\begin{acks}
The author would like to thank Boaz Barak, Eli-Ben Sasson, Yuan Li, Siuman Chan, and Jonathan Kelner for their help in editing the article. 
The author would also like to thank Boaz Barak for his advice on this research.\end{acks}

\bibliographystyle{acmsmall}
\bibliography{acmsmall-sam}
\begin{appendix}
\section{Elementary results on monotone switching networks}\label{elementarymonotone}
In this appendix, we analyze general monotone switching networks, showing how our ideas and results about certain knowledge switching networks generalize to this setting. The most important thing to note is that for general monotone switching networks, at any given vertex $v'$ we may not be certain of which paths are in $G$. Instead, we will have several possibilities for which paths are in $G$ and will only know that at least one of them holds.

To take this into account, we first define a knowledge game for directed conectivity which generalizes the certain knowledge game. We show that any sound monotone switching network can be described using this game. We then show two further results about monotone switching networks. First, by increasing the size by at most a linear factor, it is sufficient to only consider reachability from $s$. Secondly, we show a partial reduction of sound monotone switching networks to certain knowledge switching networks. While this reduction is not strong enough to prove good lower size bounds, as shown in Appendix \ref{invarianceconditiondeepproof} it is the fundamental reason behind Theorem \ref{keyinvariancecondition}
\subsection{The knowledge game for directed connectivity}\label{knowledge game}
Just as we thought of certain knowledge switching networks in terms of a game, we can think of general monotone switching networks in terms of a game, which is as follows.
\begin{definition}
A state of knowledge $J$ is a multi-set $\{K_1, \cdots, K_m\}$ of knowledge sets 
(we can have duplicates in $J$). In the knowledge game for directed connectivity, 
$J$ represents knowing that for at least one $i \in [1,m]$ the knowledge about $G$ represented by 
$K_i$ is true.
\end{definition}
\begin{example}
If $J = \{K_{\{a\}}, K_{\{b\}}, K_{\{c\}}\}$ then $J$ represents knowing that either 
there is a path from $s$ to $a$ in $G$, a path from $s$ to $b$ in $G$, a path from $s$ to $c$ in $G$, or a path 
from $s$ to $t$ in $G$.
\end{example}
\begin{definition}
In the knowledge game for directed connectivity, we start at $J_{s'} = \{\{\}\}$ and we win 
if we can get to $J_{t'} = \{\{s \to t\}\}$. We are allowed to use the following types of moves. If 
$J = \{K_1, \cdots, K_m\}$ then 
\begin{enumerate}
\item If we directly see that an edge $v_1 \to v_2$ is in $G$ we may add or remove $v_1 \to v_2$ from any $K_i$.
\item If $v_3 \to v_4, v_4 \to v_5 \in K_i$ and $v_3 \neq v_5$ we may add or remove $v_3 \to v_5$ from $K_i$.
\item If $s \to t \in K_i$ we may add or remove any other edge from $K_i$.
\item If $i,j \in [1,m]$, $i \neq j$, and $K_i \subseteq K_j$ then we may remove $K_j$ from $J$.
\item If $K$ is a knowledge set such that $K_i \subseteq K$ for some $i \in [1,m]$ then we may add $K$ to $J$.
\end{enumerate}
\end{definition}
\begin{remark}
The knowledge game for directed connectivity is a generalization of the modified certain knowledge game for 
directed connectivity. The moves which are new are the moves of types 4 and 5. Moves of type 4 make sense 
because if we know that $K_j$ implies $K_i$ and have the statement that $K_i$ OR $K_j$ is true then this 
statement is equivalent to the statement that $K_i$ is true. Moves of type 5 are the inverse of moves of 
type 4 so we still have reversibility.
\end{remark}
\begin{proposition}
It is possible to win the knowledge game for directed connectivity for an input graph $G$ if and only 
if there is a path from $s$ to $t$ in $G$.
\end{proposition}
\subsection{A partial order on knowledge sets and states of knowledge}\label{statesofknowledgepartialorder}
In the remainder of this section, it will be useful to have a partial order on states of knowledge. The intuitive idea behind this partial order is that $J_1 \leq J_2$ if the information represented by $J_1$ is contained in the information represented by $J_2$. We first define this partial order for knowledge sets and then generalize it to states of knowledge.
\begin{definition}
Define the transitive closure $\bar{K}$ of a knowledge set $K$ to be 
\begin{enumerate}
\item $\bar{K} = \{v_1 \to v_2: v_1,v_2 \in V(G), v_1 \neq v_2,$ there is a path from $v_1$ to $v_2$ whose edges 
are all in $K\}$ if $s \to t \notin K$
\item $\bar{K} = \{v_1 \to v_2: v_1,v_2 \in V(G), v_1 \neq v_2\}$ if $s \to t \in K$
\end{enumerate}
\end{definition}
\begin{definition}\label{knowledgesetpartialorderdef} \noindent
\begin{enumerate}
\item We say that $K_1 \leq K_2$ if $\bar{K}_1 \subseteq \bar{K}_2$.
\item We say that $K_1 \equiv K_2$ if $\bar{K}_1 = \bar{K}_2$.
\end{enumerate}
\end{definition}
\begin{proposition}
If $K_1, K_2, K_3$ are knowledge sets for $V(G)$, then 
\begin{enumerate}
\item $K_1 \leq K_1$ (reflexivity)
\item If $K_1 \leq K_2$ and $K_2 \leq K_1$ then $K_1 \equiv K_2$ (antisymmetry)
\item If $K_1 \leq K_2$ and $K_2 \leq K_3$ then $K_1 \leq K_3$ (transitivity)
\end{enumerate}
\end{proposition}
With this partial order, we can reexpress the definition of certain knowledge swtiching networks more cleanly.
\begin{proposition}\label{sequencetoequivalence} \noindent
\begin{enumerate}
\item For knowledge sets $K_1,K_2$, there is a sequence of moves from $K_1$ to $K_2$ in the modified certain knowledge game which does not require any information about the input graph $G$ if and only if $K_1 \equiv K_2$.
\item For knowledge sets $K_1,K_2$ and a possible edge $e$ of $G$, there is a sequence of moves from $K_1$ to $K_2$ in the modified certain knowledge game which only requires the information that $e \in E(G)$ if and only if $K_1 \cup \{e\} \equiv K_2 \cup \{e\}$.
\end{enumerate}
\end{proposition}
\begin{corollary}
We can restate the definition of certain knowledge switching networks as follows. A monotone switching network $G'$ is a certain knowledge switching network if we can assign a knowledge 
set $K_{v'}$ to each vertex $v' \in V(G')$ so that the following conditions hold:
\begin{enumerate}
\item $K_{s'} = \{\}$
\item $K_{t'} \equiv \{s \to t\}$
\item If there is an edge with label $e = v_1 \to v_2$ between vertices $v'$ and $w'$ in $G'$, then 
$K_{v'} \cup \{e\} \equiv K_{w'} \cup \{e\}$
\end{enumerate}
\end{corollary}
We generalize this partial order for states of knowledge as follows.
\begin{definition} \noindent
\begin{enumerate}
\item We say that $J_1 = \{K_{11}, \cdots, K_{1m_1}\} \leq J_2 = \{K_{21}, \cdots, K_{2m_2}\}$ 
if for all $j \in [1,m_2]$ there is an $i \in [1,m_1]$ such that $K_{1i} \leq K_{2j}$
\item We say that $J_1 \equiv J_2$ if $J_1 \leq J_2$ and $J_2 \leq J_1$.
\end{enumerate}
\end{definition}
\begin{proposition}
If $J_1, J_2, J_3$ are states of knowledge then 
\begin{enumerate}
\item $J_1 \leq J_1$ (reflexivity)
\item If $J_1 \leq J_2$ and $J_2 \leq J_1$ then $J_1 \equiv J_2$ (antisymmetry)
\item If $J_1 \leq J_2$ and $J_2 \leq J_3$ then $J_1 \leq J_3$ (transitivity)
\end{enumerate}
\end{proposition}
We have the same connection between this partial order and the knowledge game for directed connectivity, though the proof is non-trivial.
\begin{proposition}\label{statesofknowledgeequivalenceprop}
For states of knowledge $J_1,J_2$, if we can go from $J_1$ to $J_2$ 
in the knowledge game for directed connectivity with no information about the input graph $G$ 
then $J_1 \equiv J_2$.
\end{proposition}
\begin{proof}
By transitivity, to show that if we can get from $J_1$ to $J_2$ in the 
knowledge game for directed connectivity with no information about the input graph $G$ then $J_1 \equiv J_2$ 
it is sufficient to show that if we can get from $J_1$ to $J_2$ in the knowledge game for directed 
connectivity with a single move then $J_1 \equiv J_2$. $J_1$ can be written as 
$J_1 = \{K_{11}, \cdots, K_{1m}\}$ and we have the following cases:
\begin{enumerate}
\item If we use a move of type 2 or 3 altering some knowledge set $K_j$ 
to reach $J_2$ then $J_2 = \{K_{21}, \cdots, K_{2m}\}$ where $K_{2i} = K_{1i}$ for all $i \neq j$ and 
$K_{1j} \equiv K_{2j}$. For all $i$, $K_{1i} \equiv K_{2i}$ so $J_1 \equiv J_2$
\item If we use a move of type 4 to delete some knowledge set $K_j$ from $J_1$ then \\
$J_2 = \{K_{21}, \cdots,K_{2(j-1)},K_{2(j+1)}, \cdots, K_{2m}\}$ where for all $i \neq j$, $K_{2i} = K_{1i}$ 
and there exists a $j_2 \neq j$ such that $K_{2j_2} = K_{1j_2} \leq K_{1j}$. For all $i \neq j$, 
$K_{1i} \leq K_{2i}$ so $J_1 \leq J_2$. For all $i \neq j$, $K_{2i} \leq K_{1i}$ and $K_{2j_2} \leq K_{1j}$ so 
$J_2 \leq J_1$. Thus, $J_1 \equiv J_2$ as needed. Moves of type 5 are the reverse of moves of type 4, so by symmetry the result holds for these types of moves as well.
\end{enumerate}
\end{proof}
To show the converse to Proposition \ref{statesofknowledgeequivalenceprop}, we use the following lemma.
\begin{lemma}\label{statesofknowledgeequivalencelemma}
If $J_1 = \{K_{11}, \cdots, K_{1m_1}\} \equiv J_2 = \{K_{21}, \cdots, K_{2m_2}\}$ then there is 
a set $I_1 \subseteq [1,m_1]$, a set $I_2 \subseteq [1,m_2]$ of equal size to $I_1$, a function
$f_1: [1,m_1] \setminus I_1 \to I_1$, a function $f_2: [1,m_2] \setminus I_2 \to I_2$, and a perfect 
matching $\phi: I_1 \to I_2$ such that 
\begin{enumerate}
\item For all $i \in [1,m_1] \setminus I_1$, $K_{1f_1(i)} \leq K_{1i}$
\item For all $j \in [1,m_2] \setminus I_2$, $K_{2f_2(j)} \leq K_{2j}$
\item For all $i \in I_1$, $K_{1i} \equiv K_{2\phi(i)}$.
\end{enumerate}
\end{lemma}
\begin{proof}
Consider the graph formed as follows. The vertices of this graph will be the knowledge sets 
$\{K_{1i}, i \in [1,m_1]\} \cup \{K_{2j}, j \in [1,m_2]\}$. Since $J_1 \leq J_2$, for each $j \in [1,m_2]$ there 
is an $i \in [1,m_1]$ such that $K_{1i} \leq K_{2j}$. Draw a directed edge from each 
$K_{2j}$ to the corresponding $K_{1i}$ (if there are more then one possible $i$, just choose one of them). 
Since $J_2 \leq J_1$, for each $i \in [1,m_1]$ there is a $j \in [1,m_2]$ such that $K_{2j} \leq K_{1i}$. 
Draw a directed edge from each $K_{1i}$ to the corresponding $K_{2j}$ 
(if there are more then one possible $j$, just choose one of them). 

After adding all of these edge, we have a bipartite graph where each vertex has outdegree 1. This graph 
must have the structure of a set of cycles along with paths leading into the cycles. Choose $I_1$ and $I_2$ such that 
for each cycle $C$ there is exactly one $i_C \in I_1$ and exactly one $j_C \in I_2$ such that $K_{1i_C}$ is in $C$ 
and $K_{2j_C}$ is in $C$. Then for all cycles $C$ set $\phi(i_C) = j_C$. We know we can do this because 
there cannot be a cycle consisting entirely of vertices of the form $K_{1i}$ or a cycle consisting entirely 
of vertices of the form $K_{2j}$. We then choose the functions $f_1$ and $f_2$ as follows.
\begin{enumerate}
\item For all $i \in [1,m_1] \setminus I_1$ there is a cycle $C$ such that there is a path from $K_{1i}$ to $C$ and 
thus a path from $K_{1i}$ to $K_{1i_C}$. We take $f_1(i) = i_C$.
\item For all $j \in [1,m_2] \setminus I_2$ there is a cycle $C$ such that there is a path from $K_{2j}$ to $C$ and 
thus a path from $K_{2j}$ to $K_{2j_C}$. We take $f_2(j) = j_C$.
\end{enumerate}
Now note that an edge from a knowledge set $K_1$ to a knowledge set $K_2$ implies that $K_2 \leq K_1$. 
By transitivity, a path from a knowledge $K_1$ to a knowledge set $K_2$ also implies that $K_2 \leq K_1$. 
This implies that for any cycle all knowledge sets in the cycle are equivalent. The result now follows 
immediately because 
\begin{enumerate}
\item For all cycles $C$, $K_{1i_C}$ and $K_{2\phi(i_C)} = K_{2j_C}$ are in the same cycle so 
$K_{1i_C} \equiv K_{2j_C}$
\item For all $i \in [1,m_1] \setminus I_1$ there is a path from $K_{1i}$ to $K_{1f_1(i)}$ so $K_{1f_1(i)} \leq K_{1i}$
\item For all $j \in [1,m_2] \setminus I_2$ there is a path from $K_{2j}$ to $K_{2f_2(j)}$ so $K_{2f_2(j)} \leq K_{2j}$
\end{enumerate}
\end{proof}
\begin{corollary}\label{statesofknowledgeequivalencecorollary}
For states of knowledge $J_1,J_2$, we can go from $J_1$ to $J_2$ in the knowledge game 
for directed connectivity with no information about the input graph $G$ if and only if $J_1 \equiv J_2$.
\end{corollary}
\begin{proof}
The only if part is just Proposition \ref{statesofknowledgeequivalenceprop}. For the if part, 
assume that $J_1 = \{K_{11}, \cdots, K_{1m_1}\} \equiv J_2 = \{K_{21}, \cdots, K_{2m_2}\}$. 
By Lemma \ref{statesofknowledgeequivalencelemma} there is 
a set $I_1 \subseteq [1,m_1]$, a set $I_2 \subseteq [1,m_2]$ of equal size to $I_1$, a function
$f_1: [1,m_1] \setminus I_1 \to I_1$, a function $f_2: [1,m_2] \setminus I_2 \to I_2$, and a perfect 
matching $\phi: I_1 \to I_2$ such that 
\begin{enumerate}
\item For all $i \in [1,m_1] \setminus I_1$, $K_{1f_1(i)} \leq K_{1i}$
\item For all $j \in [1,m_2] \setminus I_2$, $K_{2f_2(j)} \leq K_{2j}$
\item For all $i \in I_1$, $K_{1i} \equiv K_{2\phi(i)}$.
\end{enumerate}
We will go from $J = J_1$ to $J_2$ in the knowledge game for directed connectivity using the following steps.
\begin{enumerate}
\item Use moves of type 2 and 3 to replace each $K_{1i}$ with $\bar{K}_{1i}$. 
\item For all $i \in [1,m_1] \setminus I_1$, $K_{1f_1(i)} \leq K_{1i}$ which implies that 
$\bar{K}_{1f(i)} \subseteq \bar{K}_{1i}$. We can thus use moves of type 4 to delete $\bar{K}_{1i}$ for all 
$i \in [1,m_1] \setminus I_1$.
\item We are now at $J = \{\bar{K}_{1i}: i \in I_1\}$. For all $i \in I_1$, $K_{1i} \equiv K_{2\phi(i)}$ so 
$\bar{K}_{1i} = \bar{K}_{2\phi(i)}$. Thus, $J = \{\bar{K}_{2j}: j \in I_2\}$. For all $j \in [1,m_2] \setminus I_2$, 
$K_{2f_2(j)} \leq K_{2j}$ which implies that 
$\bar{K}_{2f_2(j)} \subseteq \bar{K}_{2j}$. We can thus use moves of type 5 to add $\bar{K}_{2j}$ for all 
$j \in [1,m_2] \setminus I_2$.
\item We finish by using type 2 and 3 to replace each $\bar{K}_{2j}$ with $K_{2j}$ and obtain $J = J_2$
\end{enumerate}
\end{proof}
We have similar results when we directly see that an edge $e$ is in the input graph $G$.
\begin{definition}
For a state of knowledge $J = \{K_1, \cdots, K_m\}$ and an edge $e$, define \\
$J \cup \{e\} = \{K_1 \cup \{e\}, \cdots, K_m \cup \{e\}\}$
\end{definition}
\begin{lemma}\label{statesofknowledgeequivalencecorollarytwo} 
For states of knowledge $J_1,J_2$, we can go from $J_1$ to $J_2$ in the 
knowledge game for directed connectivity with the information that $v_1 \to v_2 \in E(G)$ 
if and only if $J_1 \cup \{v_1 \to v_2\} \equiv J_2 \cup \{v_1 \to v_2\}$.
\end{lemma}
\begin{proof}
If $J_1 \cup \{v_1 \to v_2\} \equiv J_1 \cup \{v_1 \to v_2\}$ then by Lemma 
\ref{statesofknowledgeequivalencelemma} we can go from $J_1 \cup \{v_1 \to v_2\}$ to 
$J_2 \cup \{v_1 \to v_2\}$ in the knowledge game for directed connectivity with no information about 
the input graph $G$. With the information that $v_1 \to v_2 \in E(G)$ we can go from 
$J_1$ to $J_1 \cup \{v_1 \to v_2\}$ and from $J_2 \cup \{v_1 \to v_2\}$ to $J_2$ in the knowledge game 
for directed connectivity using moves of type 1. Thus, we can go from $J_1$ to $J_2$ 
in the knowledge game for directed connectivity, as needed.

For the converse, note that for any states of knowledge $J_1,J_2$, for any sequence of moves 
in the knowledge game for directed connectivity to go from $J = J_1$ to $J = J_2$, if we replace 
$J$ with $J \cup \{v_1 \to v_2\}$ at each step we will still have a correct sequence of moves. 
Moreover, all moves of type 1 now correspond to doing nothing. This implies that we can get from 
$J_1 \cup \{v_1 \to v_2\}$ to $J_2 \cup \{v_1 \to v_2\}$ in the knowledge game for directed connectivity 
without knowing anything about the input graph $G$ so by Proposition 
\ref{statesofknowledgeequivalenceprop} we have that 
$J_1 \cup \{v_1 \to v_2\} \equiv J_2 \cup \{v_1 \to v_2\}$, as needed.
\end{proof}
\subsection{Knowledge description of monotone switching networks}
In this subsection, we show that all sound monotone switching networks can be described in terms of the knowledge game.
\begin{definition}\label{monotonedef}
If $G'$ is a monotone switching network, we call an assignment of states of 
knowledge $J_{v'}$ to vertices $v'$ of $G'$ a knowledge description if the following conditions hold:
\begin{enumerate}
\item $J_{s'} \equiv \{\{\}\}$
\item $J_{t'} \equiv \{\{s \to t\}\}$ or $J_{t'} = \{\}$
\item If there is an edge with label $e = v_1 \to v_2$ between vertices $v'$ and $w'$ in $G'$ then 
$J_{v'} \cup \{e\} \equiv J_{w'} \cup \{e\}$.
\end{enumerate}
\end{definition}
\begin{remark}
It is impossible to reach the state of knowledge $J = \{\}$ from $J_{s'} = \{\{\}\}$ in the knowledge 
game for directed connectivity. If $J_{v'} = \{\}$ this says that the vertex $v'$ is impossible to reach 
from $s'$ regardless of the input graph $G$.
\end{remark}
\begin{proposition}\label{knowledgedescriptionprop}
A monotone switching network $G'$ has a knowledge description if and only if it is sound.
\end{proposition}
\begin{proof}
If $G'$ has a knowledge description then it is sound because we can only win the knowledge 
game for directed connectivity if the input graph $G$ has a path from $s$ to $t$. 
Conversely, given a sound monotone switching network $G'$, set 
$J_{v'} = \{E:$ there is a walk from $s'$ to $v'$ in $G'$ whose edge labels are all in $E\}$.

If there is an edge with label $e$ between vertices $v'$ and $w'$ in $G'$ then for every 
$K \in J_{v'}$, $K \cup \{e\} \in J_{w'}$. $K \cup \{e\} \leq K \cup \{e\}$ 
so this implies that $J_{w'} \cup \{e\} \leq J_{v'} \cup \{e\}$. By a symmetrical argument, 
we also have that $J_{v'} \cup \{e\} \leq J_{w'} \cup \{e\}$. Thus, $J_{v'} \cup \{e\} \equiv J_{w'} \cup \{e\}$, as needed.

Now we just need to check that $J_{s'} \equiv \{\{\}\}$ and $J_{t'} \equiv \{\{s \to t\}\}$ or $J_{t'} \equiv \{\}$. 
$\{\} \in J_{s'}$ and can be used to delete everything else in $J_{s'}$. Thus $J_{s'} \equiv \{\{\}\}$. Since $G'$ is sound, for every $K \in J_{t'}$, 
$K$ contains a path from $s$ to $t$ so $K \equiv \{s \to t\}$. Using moves of type 2 and 3 we can transform every $K$ in $J_{t'}$ into 
$K = \{s \to t\}$ and then we can use moves of type 4 to delete all but one copy of $\{s \to t\}$, so either we originally had $J_{t'} = \{\}$ or we are left with $\{\{s \to t\}\}$. Thus $J_{t'} \equiv \{\{s \to t\}\}$ or $J_{t'} = \{\}$, as needed.
\end{proof}
\begin{figure}[ht]
\centerline{\includegraphics[height=7cm]{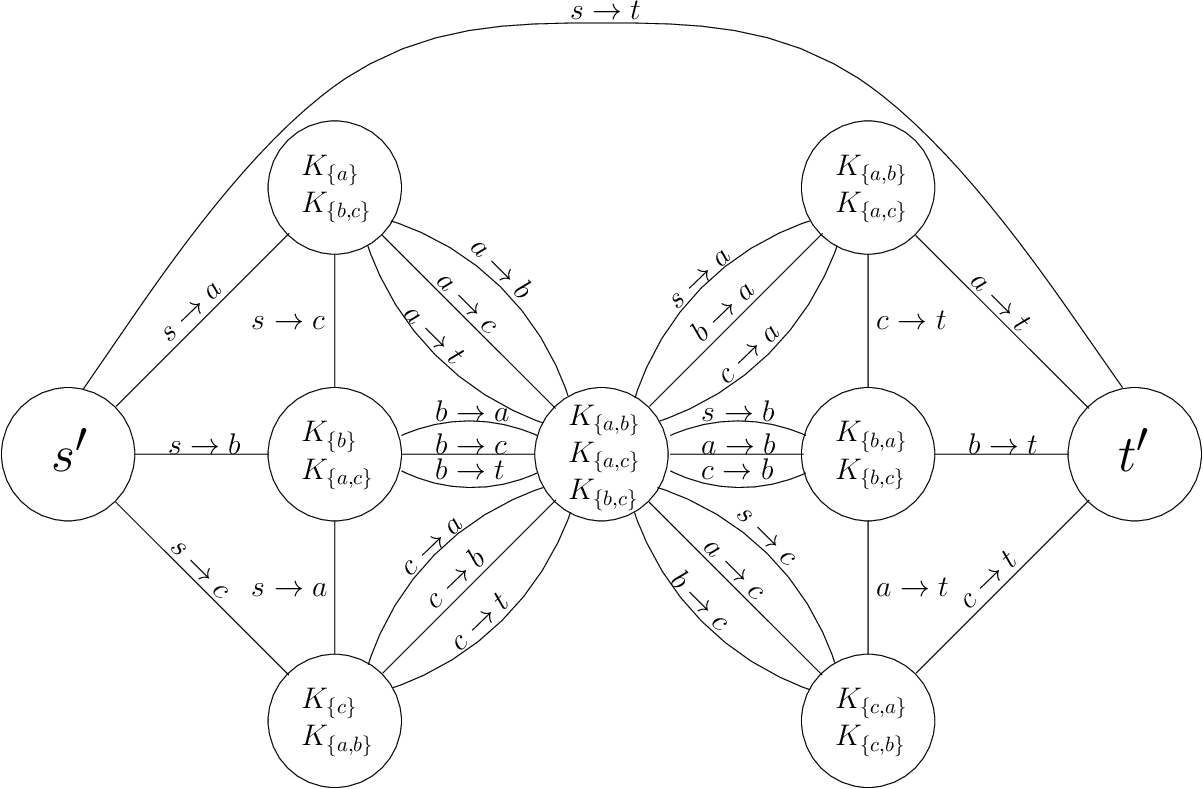}}
\caption[A monotone switching network $G'$ solving directed connectivity on $V(G) = \{s,t,a,b,c\}$ together with a certain knowledge description of it]{A monotone switching network that solves directed connectivity on $V(G) = \{s,a,b,c,t\}$ together with a 
knowledge description of it. The label inside each vertex gives the $J$ for that vertex, 
with each line corresponding to one of its $K$. By default we take 
$J_{s'} = \{\{\}\}$ and $J_{t'} = \{\{s \to t\}\}$.}
\label{monotonenetwork}
\end{figure}
\subsection{Reduction to reachability from $s$}
In this subsection, we prove the following theorem which shows that there is little loss in only considering monotone switching networks $G'$ which only make deductions based on reachability from $s$.
\begin{theorem}\label{simplification}
If $(G', s', t', \mu')$ is a sound monotone switching network, then there is a sound monotone switching 
network $(G'_2, s', t', \mu'_2)$ such that $G'_2$ accepts exactly the same inputs as $G'$, 
$|V(G'_2)| \leq (n+1)|V(G')|$, and $G'_2$ has a knowledge description where for any vertex $v'$ of $G'_2$, for any $K$ in $J_{v'}$, 
$K \in \{K_V: V \subseteq \Vmst\} \cup \{K_{t'}\} $
\end{theorem}
\begin{proof}
We construct $G'_2$ by taking $n+1$ copies of $G'$ and making the $s'$ of each copy equal 
to the $t'$ of the previous copy. We take $s'$ for $G'_2$ to be the $s'$ of the first copy of $G'$ and 
$t'$ for $G'_2$ to be the $t'$ of the last copy of $G'$. Clearly, $G'_2$ accepts exactly the same inputs 
as $G'$ and we have that $|V(G'_2)| \leq (n+1)|V(G')|$.

Now for a vertex $v'$ we construct $J_{v'}$ as follows. For each walk $W'$ from 
$s'$ to $v'$ in $G'_2$, create a $K$ for that walk as follows:
\begin{enumerate}
\item Start with the set $X_0 = \{s\}$ of vertices in $G$.
\item Let $e_i = v_i \to w_i$ be the edge in $G$ which is the label of the ith edge in $W'$. 
Take $X_i = X_{i-1}$ if $v_i \notin X_{i-1}$ and take $X_i = X_{i-1} \cup \{w_i\}$ if $v_i \in X_{i-1}$. Let $X$ be 
the set obtained after taking the final edge in $W'$.
\item Set $K = \cup_{v \in X \backslash \{s\}}{\{s \to v\}}$.
\end{enumerate}
Now take $J_{v'}$ to be the set of all such $K$.\\
Following similar logic as was used to prove Proposition \ref{knowledgedescriptionprop}, it can be
verified that this assignment of states of knowledge to vertices of $G'_2$ satisfies condition 3 of Definition \ref{monotonedef} and that $J_{s'} \equiv \{\{\}\}$. We just need to show $J_{t'} \equiv \{\{s \to t\}\}$ or $J_{t'} = \{\}$.

To show that $J_{t'} \equiv \{\{s \to t\}\}$ or $J_{t'} = \{\}$, consider a given walk $W'$ from $s'$ to $t'$ in $G'_2$ 
and look at how the set $\{X_i\}$ changes as we go along $W'$. Let $Y_j$ be the set of vertices we have when we first reach 
the vertex $t'_j$ which was the $t'$ of the ith copy of $G'$. $Y_0 = \{s\}$. Since $G'$ is sound, 
if $t \notin Y_j$ then the portion of $W'$ from $t'_j$ to $t'_{j+1}$ must have at least one edge which 
crosses the cut between $Y_j$ and $V(G) \setminus Y_j$. If $e_k$ is the first edge on this portion of $W'$ 
crossing this cut, then $Y_j \subseteq X_{k-1} \subsetneq X_k \subseteq Y_{j+1}$. Thus either 
$t \in Y_j \subseteq Y_{j+1}$ or $Y_j \subsetneq Y_{j+1}$. There are only $n$ vertices except for $s$ and $t$ so this 
implies that $t \in Y_{n+1}$. Thus for all $K \in J_{t'}$, $s \to t \in K$. Using the same logic as before, $J_{t'} \equiv \{\{s \to t\}\}$ or $J_{t'} = \{\}$, as needed.
\end{proof}
\subsection{Reduction to certain knowledge switching networks}
Finally, we prove a theorem that shows that in some sense, monotone switching networks can be 
reduced to certain-knowledge switching networks. Although this theorem is not strong enough 
to prove any lower size bounds, as shown in Appendix \ref{invarianceconditiondeepproof} it gives a deep reason why it is sufficient to consider certain knowledge switching networks when checking $E$-invariance.
\begin{definition}\label{monotonetocertainknowledge}
Given a sound monotone switching network $G'$ for directed connectivity together with a knowledge 
description and a path $P' = \{s' \to v'_1,v'_1 \to v'_2, \cdots, v'_{l'-2} \to v'_{l'-1}, v'_{l'-1} \to t'\}$ 
from $s'$ to $t'$ in $G'$, define the certain-knowledge switching network $H'(G',P')$ as follows:

First, if we do not already have that $J_{s'} = \{\{\}\}$ and $J_{t'} = \{\{s \to t\}\}$, then take 
$J_{s'} = \{\{\}\}$ and $J_{t'} = \{\{s \to t\}\}$. Now let $v'_{0} = s'$ and let $v'_{l'} = t'$. 
For each $k \in [0,l']$, $J_{v'_k} = \{K_{{v'_k}1}, \cdots, K_{{v'_k}{m_k}}\}$ 
for some positive integer $m_k$ and some knowledge sets $K_{{v'_k}1}, \cdots, K_{{v'_k}{m_k}}$. For each 
non-empty subset $S$ of $[1,m_k]$ let $K_{{v'_k}S} = \cup_{j \in S}{K_{{v'_k}j}}$.\\
We take $V(H'(G',P')) = \{w'_{{v'_k}S}: k \in [0,l'], S \subseteq [1,m_k], S \neq \emptyset\}$ where 
each $w'_{{v'_k}S}$ has knowledge 
set $K_{{v'_k}S}$. $J_{s'} = \{\{\}\}$ and $J_{t'} = \{\{s \to t\}\}$ so we take $s' = w'_{{v'_0}\{1\}}$ and 
$t' = w'_{{v'_{l'}}\{1\}}$ in $H'(G',P')$. We take all possible edges which are allowed by condition 3 of 
Definition \ref{certainknowledgedef}.
\end{definition}
\begin{theorem}\label{reductiontosimple}
If $G'$ is a sound monotone switching network for directed connectivity with a given knowledge 
description and $P' = \{s' \to v'_1,v'_1 \to v'_2, \cdots, v'_{l'-2} \to v'_{l'-1}, v'_{l'-1} \to t'\}$ is a path
from $s'$ to $t'$ in $G'$, then it is possible to take a subset of the edges of $H'(G',P')$ and 
assign a direction to each edge to obtain a directed graph $H'_{red}(G',P')$ for which the following is true:
\begin{enumerate}
\item $H'_{red}(G',P')$ consists of a directed path from $s'$ to $t'$ and directed cycles. 
\item Every vertex in $H'_{red}(G',P')$ is on a path or cycle.
\item For all vertices $w'_{v'_{k}S}$ where $|S|$ is odd,
\begin{enumerate}
\item If $w'_{v'_{k}S} \neq s'$ then the incoming edge for $w'_{v'_{k}S}$ has the same label as the edge from 
$v'_{k-1}$ to $v'_{k}$ in $P'$ and its other endpoint is either of the form $w'_{v'_{k-1}T}$ where 
$|T| = |S|$ or the form $w'_{v'_{k}S_2}$ where $S_2$ is obtained by adding or deleting one element from $S$.
\item If $w'_{v'_{k}S} \neq t'$ then the outgoing edge for $w'_{v'_{k}S}$ has the same label as the edge from 
$v'_{k}$ to $v'_{k+1}$ in $P'$ and its other endpoint is either of the form $w'_{v'_{k+1}T}$ where 
$|T| = |S|$ or the form $w'_{v'_{k}S_2}$ where $S_2$ is obtained by adding or deleting one element from $S$.
\end{enumerate}
\item For all vertices $w'_{v'_{k}S}$ where $|S|$ is even,
\begin{enumerate}
\item If $w'_{v'_{k}S} \neq t'$ then the incoming edge for $w'_{v'_{k}S}$ has the same label as the edge from 
$v'_{k}$ to $v'_{k+1}$ in $P'$ and its other endpoint is either of the form $w'_{v'_{k+1}T}$ where 
$|T| = |S|$ or the form $w'_{v'_{k}S_2}$ where $S_2$ is obtained by adding or deleting one element from $S$.
\item If $w'_{v'_{k}S} \neq s'$ then the outgoing edge for $w'_{v'_{k}S}$ has the same label as the edge from 
$v'_{k-1}$ to $v'_{k}$ in $P'$ and its other endpoint is either of the form $w'_{v'_{k-1}T}$ where 
$|T| = |S|$ or the form $w'_{v'_{k}S_2}$ where $S_2$ is obtained by adding or deleting one element from $S$.
\end{enumerate}
\end{enumerate}
\end{theorem}
\begin{proof}
For all $k$, letting $e_k$ be the label of the edge from $v'_{k}$ to $v'_{k+1}$ 
we apply Lemma \ref{statesofknowledgeequivalencelemma} to the states of knowledge $J_{v'_k} \cup {\{e_k\}}$ and 
$J_{v'_{k+1}} \cup {\{e_k\}}$. This gives us a set $I_{k1} \subseteq [1,m_k]$, a set $I_{k2} \subseteq [1,m_{k+1}]$ of 
equal size to $I_{k1}$, a function
$f_{k1}: [1,m_k] \setminus I_{k1} \to I_{k1}$, a function $f_{k2}: [1,m_{k+1}] \setminus I_{k2} \to I_{k2}$, and a perfect 
matching $\phi_k: I_{k1} \to I_{k2}$ such that 
\begin{enumerate}
\item For all $i \in [1,m_k] \setminus I_{k1}$, $K_{{v'_k}f_{k1}(i)} \cup \{e_k\} \leq K_{{v'_k}i}  \cup \{e_k\}$
\item For all $j \in [1,m_{k+1}] \setminus I_2$, $K_{{v'_{k+1}}f_{k2}(j)}  \cup \{e_k\}  \leq K_{{v'_{k+1}}j} \cup \{e_k\}$
\item For all $i \in I_{k1}$, $K_{{v'_k}i} \cup \{e_k\}  \equiv K_{{v'_{k+1}}\phi_k(i)} \cup \{e_k\}$.
\end{enumerate}
\begin{proposition}\label{subsuming} \noindent 
\begin{enumerate}
\item For all $S \subseteq I_{k1}$, $K_{{v'_{k}}S}  \cup \{e_k\}  \equiv K_{v'_{k+1}\phi(S)}  \cup \{e_k\}$
\item For all $S \subseteq [1,m_k]$ and $i \in S \setminus I_{k1}$, if 
$f_{k1}(i) \notin S$ then $K_{{v'_{k}}S}  \cup \{e_k\} \equiv K_{v'_{k}{(S \cup \{f_{k1}(i)\})}} \cup \{e_k\}$ and 
if $f_{k1}(i) \in S$ then $K_{{v'_{k}}S}  \cup \{e_k\}  \equiv K_{v'_{k}{(S \setminus \{f_{k1}(i)\})}} \cup \{e_k\}$
\item For all $T \subseteq [1,m_{k+1}]$ and $j \in T \setminus I_{k2}$, if $f_{k2}(j) \notin T$ then 
$K_{{v'_{k+1}}T} \cup \{e_k\}  \equiv K_{v'_{k+1}{(T \cup \{f_{k2}(j)\})}} \cup \{e_k\}$ and if $f_{k2}(j) \in T$ then 
$K_{{v'_{k+1}}T} \cup \{e_k\}  \equiv K_{v'_{k+1}{(T \setminus \{f_{k2}(j)\})}} \cup \{e_k\}$
\end{enumerate}
\end{proposition}
We now choose the edges of $H'_{red}(G',P')$ and assign directions to them as follows. For each vertex $w'_{v'_{k}S}$,
\begin{enumerate}
\item If $S \subseteq I_{k1}$ then take the edge with label $e_k$ between $w'_{v'_{k}S}$ and $w'_{v'_{k+1}\phi(S)}$. If $|S|$ is odd 
then have this edge go from $w'_{v'_{k}S}$ to $w'_{v'_{k+1}\phi(S)}$. If $|S|$ is even 
then have this edge go from $w'_{v'_{k+1}\phi(S)}$ to $w'_{v'_{k}S}$. 
\item If $S \nsubseteq I_{k1}$ then take the first $i \in S \setminus I_{k1}$ 
and take the edge with label $e_k$ between $w'_{v'_{k}S}$ and $w'_{v'_{k}(S \Delta \{f_{k1}(i)\})}$
where $S \Delta \{f_{k1}(i)\} = S \cup \{f_{k1}(i)\}$ if $f_{k1}(i) \notin S$ and 
$S \Delta \{f_{k1}(i)\} = S \setminus \{f_{k1}(i)\}$ if $f_{k1}(i) \in S$. Have this edge go 
from $w'_{v'_{k}S}$ to $w'_{v'_{k}(S \Delta \{f_{k1}(i)\})}$ if $|S|$ is odd and have this edge go 
from $w'_{v'_{k}(S \Delta \{f_{k1}(i)\})}$ to $w'_{v'_{k}S}$ if $|S|$ is even.
\end{enumerate}
For each vertex $w'_{v'_{k+1}T}$,
\begin{enumerate}
\item If $T \subseteq I_{k2}$ then take the edge with label $e_k$ between $w'_{v'_{k+1}T}$ and $w'_{v'_{k}\phi^{-1}(T)}$. If $|T|$ is odd 
then have this edge go from $w'_{v'_{k}\phi^{-1}(T)}$ to $w'_{v'_{k+1}T}$. If $|T|$ is even 
then have this edge go from $w'_{v'_{k+1}T}$ to $w'_{v'_{k}\phi^{-1}(T)}$. 
\item If $T \nsubseteq I_{k2}$ then take the first $j_2 \in T \setminus I_{k2}$ and take the edge with label $e_k$ between 
$w'_{v'_{k+1}T}$ and $w'_{v'_{k+1}(T \Delta \{f_{k2}(j)\})}$
where $T \Delta \{f_{k2}(j)\} = T \cup \{f_{k2}(j)\}$ if $f_{k2}(j) \notin T$ and 
$T \Delta \{f_{k2}(j)\} = T \setminus \{f_{k2}(j)\}$ if $f_{k2}(j) \in T$. Have this edge go 
from $w'_{v'_{k+1}(T \Delta \{f_{k2}(j)\})}$ to $w'_{v'_{k+1}T}$ if $|T|$ is odd and have this edge go 
from $w'_{v'_{k+1}(T \Delta \{f_{k2}(j)\})}$ to $w'_{v'_{k+1}T}$ if $|T|$ is even.
\end{enumerate}
Conditions 3 and 4 of Theorem \ref{reductiontosimple} are now satisfied by the edges we have chosen. All vertices have indegree one except 
for $s'$ and all vertices have outdegree one except for $t'$. This implies that $H'_{red}(G',P')$ consists of a path from $s'$ to $t'$ and 
directed cycles and that every vertex is on a path or cycle, as needed.
\end{proof}
\begin{figure}[ht]
\centerline{\includegraphics[height=6cm]{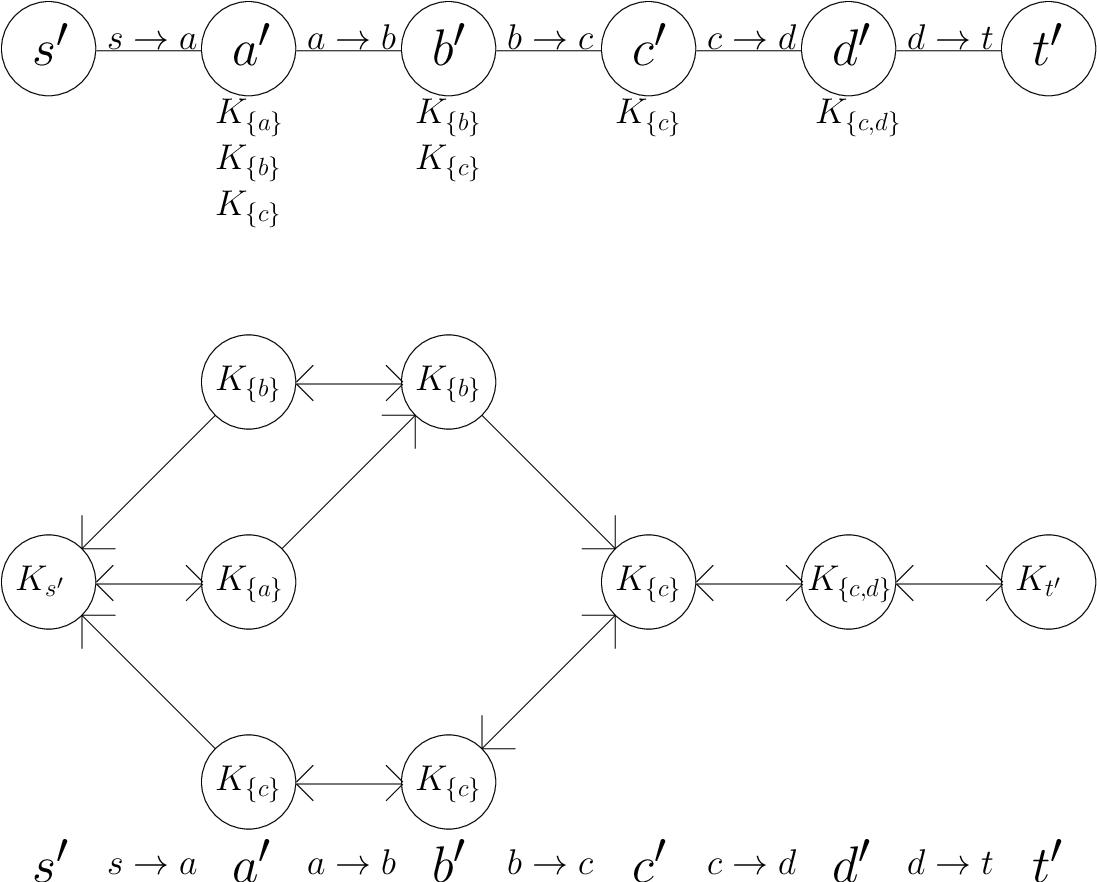}}
\caption[An illustation of the partial reduction of monotone switching networks to certain knowledge switching networks]{This is an illustration of the ideas used in the proof of 
Theorem \ref{reductiontosimple}. Above, we have the path $P'$ from 
$s'$ to $t'$ in $G'$, where the $J$ for each vertex is given below that vertex with each line corresponding to one of its $K$. 
Below, we have the arrows between all of the knowledge sets from the argument used to prove 
Lemma \ref{statesofknowledgeequivalencelemma}. Here the functions $\{\phi_k\}$ correspond to going along a bidirectional edge. 
The functions $\{f_{k1}\}$ and $\{f_{k2}\}$ correspond to going along a unidirectional edge and then going the opposite 
direction along a bidirectional edge. To get from 
$s'$ to $t'$ in $H'_{red}(G',P')$ we have the following sequence (not shown):
$K_{s'\{1\}} = \{\}$, $K_{a'\{1\}} = \{s \to a\}$, $K_{a'\{1,2\}} = \{s \to a, s \to b\}$, 
$K_{a'\{2\}} = \{s \to b\}$, $K_{b'\{1\}} = \{s \to b\}$, $K_{b'\{1,2\}} = \{s \to b, s \to c\}$, 
$K_{a'\{2,3\}} = \{s \to b, s \to c\}$, $K_{a'\{1,2,3\}} = \{s \to a, s \to b, s \to c\}$, 
$K_{a'\{1,3\}} = \{s \to a, s \to c\}$, $K_{a'\{3\}} = \{s \to c\}$, 
$K_{b'\{2\}} = \{s \to c\}$, $K_{c'\{1\}} = \{s \to c\}$, 
$K_{d'\{1\}} = \{s \to c, s \to d\}$, $K_{t'\{1\}} = \{s \to t\}$.}
\label{monotonetocertainknowledgepicture}
\end{figure}
\begin{corollary}
If $G'$ is a sound monotone switching network for directed connectivity with a given knowledge description and 
$P = \{s \to v_1,v_1 \to v_2,\cdots,v_{l-1} \to t\}$ is a path from $s$ to $t$ in $G$, then any path $P'$ in $G'$ from $s'$ to $t'$ 
using only the edges of $P$ must pass through at least one vertex $a'$ such that $J_{a'} \neq J_{t'}$ and if 
$J_{a'} = \{K_{a'1}, \cdots, K_{a'm}\}$ then $V = \cup_{i=1}^{m}{V(K_{a'i})}$ contains at least $\lceil{\lg{l}}\rceil$ of $v_1,\cdots,v_{l-1}$.
\end{corollary}
\begin{proof}
This follows immediately from Theorem \ref{reductiontosimple} and Lemma \ref{specialboundlemma}.
\end{proof}
\section{Alternate Proof of Theorem \ref{keyinvariancecondition}}\label{invarianceconditiondeepproof}
Before proving Theorem \ref{keyinvariancecondition}, we first show how a knowledge description of a monotone switching network can be translated into a function description of a monotone switching network.
\begin{definition}
For a given state of knowledge $J$, define the function $J: \mathcal{C} \to \{0,1\}$ so that
$J(C) = 0$ if there is no $K$ in $J$ such that $K(C) = 1$ and $1$ otherwise.
\end{definition}
\begin{proposition}\label{stateofknowledgefunctionprop}
If we can get from $J_1$ to $J_2$ in the knowledge game for directed connectivity using only the knowledge 
that some edge $e$ is in $G$ and $e$ does not cross some cut $C$ then $J_2(C) = J_1(C)$.
\end{proposition}
\begin{proof}
This follows immediately from the fact that if $e$ does not cross $C$, then for any state of knowledge $J$, 
no individual move on $J$ in the knowledge game for directed connectivity which can be done with only the knowledge that 
$e$ is in $G$ changes the value of $J(C)$.
\end{proof}
\begin{corollary}
If a monotone switching network $G'$ has a knowledge description where each vertex 
$v'$ is assigned the state of knowledge $J_{v'}$ then if we assign each $v'$ the 
function $J_{v'}$, we have a function description of $G'$.
\end{corollary}
\begin{remark}
If we take the knowledge description given in the proof of Proposition \ref{knowledgedescriptionprop} and take the 
corresponding function description we will obtain the reachability function description.
\end{remark}
We now give an alternate proof of Thoerem \ref{keyinvariancecondition}
\vskip.1in
\noindent
{\bf Theorem \ref{keyinvariancecondition}.}
{\it
If $g$ is a function from $\mathcal{C}$ to $\mathbb{R}$ and $E$ is a set of edges between vertices in $V(G)$ then 
$g$ is $E$-invariant if and only if $g \cdot v'_1 = g \cdot v'_2$ whenever $v'_1,v'_2$ are vertices of $G'_c(n)$ 
such there is an edge between $v'_1$ and $v'_2$ in $G'_c(n)$ whose edge label is in $E$.
}
\begin{proof}
The only if direction follows immediately from Proposition \ref{knowledgesetfunctionprop}. For the if direction, we first give a more stringent condtion for $E$-invariance. Using Theorem \ref{reductiontosimple}, we will then show that this condition follows from invariance on certain knowledge switching networks.
\begin{lemma}\label{weakerinvariancecondition}
If $g$ is a function from $\mathcal{C}$ to $\mathbb{R}$ and $E$ is a set of edges between vertices in $V(G)$ then 
$g$ is $E$-invariant if and only if $g \cdot J_1 = g \cdot J_2$ whenever $J_1,J_2$ are states of knowledge such that 
$J_1 \cup \{e\} \equiv J_2 \cup \{e\}$ for some $e \in E$ and all knowledge sets in $J_1$ and $J_2$ are either equal to $K_{t'}$ or have the form $K_V$ where $V \subseteq \Vmst$.
\end{lemma}
\begin{proof}
The only if direction follows immediately from Proposition \ref{stateofknowledgefunctionprop}. 
To prove the if direction, assume that $g \cdot J_1 = g \cdot J_2$ whenever $J_1,J_2$ are states of knowledge for $V(G)$ 
such that all knowledge sets in $J_1$ and $J_2$ are either equal to $K_{t'}$ or have the form 
$K_V$ where $V \subseteq \Vmst$ and there is an $e \in E$ for which it is possible to go from 
$J_1$ to $J_2$ in the knowledge game for directed connectivity using only the knowledge that $e$ is in $G$. 
Given a cut $C$ which can be crossed by an edge $e \in E$, take $J_1 = \{K_{L(C)}\}$ and $J_2 = \cup_{v \in R(C)}{\{K_{L(C) \cup \{v\}}\}}$. We have that $J_2(C) = 0$, $J_1(C) = 1$, all knowledge sets 
in $J_1$ and $J_2$ are either equivealent to $K_{t'}$ or have the form $K_V$ where $V \subseteq \Vmst$, 
and $J_1 \cup \{e\} \equiv J_2 \cup \{e\}$. By our assumption, $g \cdot J_1 = g \cdot J_2$.
 
Now consider any cut $C_2 \in \mathcal{C}$. If $L(C) \cap R(C_2)$ is nonempty then $J_1(C_2) = J_2(C_2) = 0$. 
If $R(C_2) \subsetneq R(C)$ then $J_1(C_2) = J_2(C_2) = 1$. Thus, 
if $C_2 \neq C$ then $J_1(C_2) = J_2(C_2)$. $J_2(C_2) - J_1(C_2) \neq 0$ if and only if $C_2 = C$. Putting everything together, 
$0 = g \cdot J_2 - g \cdot J_1 = 2^{-n}(J_2(C)-J_1(C))g(C)$ so $g(C) = 0$. Thus, $g(C) = 0$ for any $C$ which can be crossed by an edge $e \in E$, as needed.
\end{proof}
We now show that this condition follows from invariance on certain knowledge switching networks.
\begin{lemma}\label{justworks}
Let $J_1 = \{K_{11}, \cdots, K_{1{m_1}}\}$ and let $J_2 = \{K_{21}, \cdots, K_{2{m_2}}\}$. If $J_1 \cup \{e\} \equiv J_2 \cup \{e\}$ for some possible edge $e$ then we may write $J_2 - J_1$ as a sum of terms of the form $K_2 - K_1$ where $K_1 \cup \{e\} \equiv K_2 \cup \{e\}$ and both $K_1$ and $K_2$ are either of the form 
$\cup_{j \in S}{K_{1j}}$ where $S \subseteq [1,m_1], S \neq \emptyset$ or the form
$\cup_{k \in T}{K_{2k}}$ where $T \subseteq [1,m_2], T \neq \emptyset$.
\end{lemma}
\begin{proof}
\noindent We first give a proposition which allows us to express $J_2 - J_1$ in terms of these knowledge sets.
\begin{proposition}\label{andstoors}
If $J = \{K_1,K_2,\cdots,K_m\}$ where $m \neq 0$, then for any $C \in \mathcal{C}$,
$$J(C) = \sum_{S \subseteq [1,m], S \neq \emptyset}{(-1)^{|S|+1}((\cup_{i \in S}{K_i})(C))}$$
\end{proposition}
\begin{proof}
This is just the inclusion-exclusion principle. Note that $J(C) = 0$ if $K_i(C) = 0$ for every $i$ and $1$ otherwise. 
If $K_i(C) = 1$ for some $i$, then we can add or remove $i$ from $S$ without affecting 
$(\cup_{i \in S}{K_i})(C)$. But then all terms in the sum on the right 
cancel except $K_i(C)$, which is $1$.\\
If $K_i(C) = 0$ for all $i$, then for all non-empty subsets $S$ of $[1,m]$,  $(\cup_{i \in S}{K_i})(C) = 0$. 
Choosing an arbitrary $i$, we can add or remove $i$ from $S$ without affecting 
$(\cup_{i \in S}{K_i})(C)$, so we again have that everything cancels except $K_i(C)$, which is $0$.
\end{proof}
Lemma \ref{justworks} now follows directly from Theorem \ref{reductiontosimple}. We can easily create a 
sound monotone switching $G'$ which has a path $P'$ from $s'$ to $t'$ such that there are vertices $v'_i,v'_{i+1}$ 
on $P'$ with $J_{v'_i} = J_1$ and $J_{v'_{i+1}} = J_2$ and there is an edge $e'$ from $v'_i$ to $v'_{i+1}$ with 
label $e$. By Proposition \ref{andstoors} we have that 
$$J_2 - J_1 = \sum_{T \subseteq [1,m_2], T \neq \emptyset}{(-1)^{|T|+1}((\cup_{j \in T}{K_{2j}})(C))} - 
\sum_{S \subseteq [1,m_1], S \neq \emptyset}{(-1)^{|S|+1}((\cup_{i \in S}{K_{1i}})(C))}$$
By Theorem \ref{reductiontosimple}, if we let $E_{e'}$ be the set of directed edges corresponding to 
$e'$ in $H'_{red}(G',P')$, 
$$\sum_{e'_k \in E_{e'}}{e'_k} = \sum_{T \subseteq [1,m_2], T \neq \emptyset}{(-1)^{|T|+1}((\cup_{j \in T}{K_{2j}})(C))} - \sum_{S \subseteq [1,m_1], S \neq \emptyset}{(-1)^{|S|+1}((\cup_{i \in S}{K_{1i}})(C))}$$
where if $e'_k$ goes from $w'_1$ to $w'_2$ in $H'_{red}(G',P')$ then $e'_k = w'_2 - w'_1$.

Thus, $J_2 - J_1 = \sum_{e'_k \in E_{e'}}{e'_k}$ and the result follows.
\end{proof}
\begin{figure}[ht]
\centerline{\includegraphics[height=6cm]{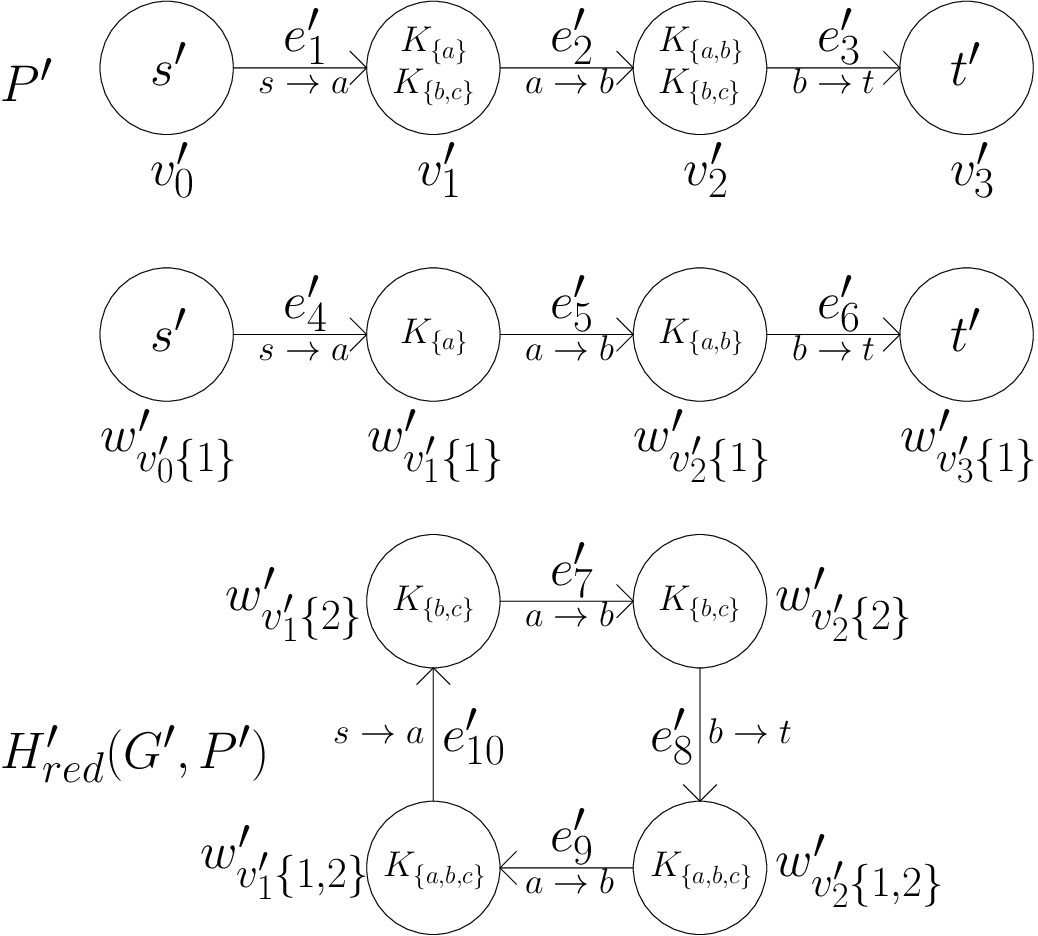}}
\caption[An illustartions of the ideas used to prove Lemma \ref{justworks}]{In this figure, we illustrate the ideas used in the proof of Lemma \ref{justworks}.
It can be verified that 
$e'_1 = e'_4 + e'_{10}$, $e'_2 = e'_5 + e'_7 + e'_9$, and 
$e'_3 = e'_6 + e'_8$.}
\label{advancedmonotonetocertainknowledgepicture}
\end{figure}
We are now ready to complete the proof of Theorem \ref{keyinvariancecondition}. Assume that $g(C) \neq 0$ for some cut $C$ which can be crossed 
by an edge $e \in E$. By Proposition \ref{weakerinvariancecondition}, there exist states of knowledge $J_1,J_2$ for $V(G)$ such that 
$J_1 \cup \{e\} \equiv J_2 \cup \{e\}$ and all knowledge sets in $J_1$ and $J_2$ are 
either equal to $K_{t'}$ or have the form $K_V$ where $V \subseteq \Vmst$, and $g \cdot J_1 \neq g \cdot J_2$. 
But then by Lemma \ref{justworks}, we may write $J_2 - J_1$ as a sum of terms of the form $K_2 - K_1$ where $K_1 \cup \{e\} \equiv K_2 \cup \{e\}$ and both $K_1$ and $K_2$ are either 
$K_{t'}$ or of the form $K_V$ where $V \subseteq \Vmst$. Since $g \cdot (J_2 - J_1) \neq 0$, there must 
be at least one such pair $K_1,K_2$ such that $g \cdot (K_2 - K_1) \neq 0$. But then taking $v'_1$ and $v'_2$ to be the 
corresponding vertices in $G'_c(n)$, there is an edge with label $e$ between $v'_1$ and $v'_2$ and 
$g \cdot v'_1 \neq g \cdot v'_2$, as needed.
\end{proof}
\section{Proof of Lemma \ref{specialboundlemma}}\label{specialboundlemmafullproof}
In this appendix, we prove the full version of Lemma \ref{specialboundlemma}. To simplify the proof, we use the partial ordering on knowledge sets given in subsection \ref{statesofknowledgepartialorder}.
 \vskip.1in
\noindent
{\bf Lemma \ref{specialboundlemma}.}
{\it
Let $G'$ be a certain knowledge switching network. For any certain knowledge description of $G'$ and 
any path $P = s \to v_1 \to \cdots \to v_{l-1} \to t$, if $G$ is the input graph with vertex set $V(G)$ and 
$E(G) = E(P)$, if $W'$ is a walk in $G'$ whose edge labels are all in $G$ from a vertex $v'_{start}$ with 
$K_{v'_{start}} \equiv K_{s'}$ to a vertex $v'_{end}$ with $K_{v'_{end}} \equiv K_{t'}$ then $W'$ passes 
through a vertex $v'$ such that $K_{v'} \not\equiv K_{t'}$, 
$V(K_{v'}) \subseteq \{v_1, \cdots, v_{l-1}\}$, and $|V(K_{v'})| \geq \lceil{\lg(l)}\rceil$. 
}
\begin{proof}
In this proof, we will split the path $P$ in two and use induction on each half. This will require 
projecting onto each half of $P$ in two different ways.
\begin{definition} \noindent
\begin{enumerate}
\item Call the vertices $L = \{v_1,\cdots,v_{\lceil{\frac{l-1}{2}}\rceil}\}$ the left half of $P$.
\item Call the vertices $R = \{v_{\lceil{\frac{l-1}{2}}\rceil + 1},\cdots,v_{l-1}\}$ the right half of $P$.
\end{enumerate}
\end{definition}
\begin{definition} \ 
\begin{enumerate}
\item We say an edge $e = u \to v$ is a left edge if $u,v \in L \cup \{s\}$
\item We say an edge $e = u \to v$ is a right edge if $u,v \in R \cup \{t\}$
\item We say an edge $e = u \to v$ is a left-jumping edge if $u = s$ and $v \in R$. Note that $t \notin R$.
\item We say an edge $e = u \to v$ is a right-jumping edge if $u = \in L$ and $v = t$. Note that $s \notin L$.
\end{enumerate}
\end{definition}
Our first projections focus on the progress we have made towards showing that there is a path from $s$ to $R \cup \{t\}$ and $L \cup \{s\}$ to $t$, respectively.
\begin{definition}
Given a vertex $v \in V(G)$,
\begin{enumerate}
\item Define $p_l(v) = v$ if $v \notin R$ and $p_l(v) = t$ if $v \in R$.
\item Define $p_r(v) = v$ if $v \notin L$ and $p_l(v) = s$ if $v \in L$.
\end{enumerate}
\end{definition}
\begin{definition}
Given an edge $e = u \to v$ where $u,v \in V(G)$,
\begin{enumerate}
\item Define $p_l(e) = p_l(u) \to p_l(v)$.
\item Define $p_r(e) = p_r(u) \to p_r(v)$.
\end{enumerate}
\end{definition}
\begin{definition}
Given a knowledge set $K$,
\begin{enumerate}
\item Define $p_l(K) = \{p_l(e): e \in K, p_l(e) \neq t \to t\}$.
\item Define $p_r(K) = \{p_r(e): e \in K, p_r(e) \neq s \to s\}$.
\end{enumerate}
\end{definition}
\begin{definition}
Given a certain knowledge switching network $G'$ together with a knowledge description of $G'$, 
define $p_l(G')$ to be the certain knowledge switching network formed from $G'$ with the following steps:
\begin{enumerate}
\item Replace all edge labels $e$ with $p_l(e)$
\item Replace all knowledge sets $K_{v'}$ in the certain knowledge description with $p_l(K_{v'})$.
\item Contract all edges in the switching network which now have label $t \to t$. When contracting an 
edge $e'$ with endpoints $v'$ and $w'$, we may choose either $K_{v'}$ or $K_{w'}$ to be the knowledge 
set for the resulting vertex.
\end{enumerate}
Similarly, given a certain knowledge switching network $G'$ together with a knowledge description of $G'$, 
define $p_r(G')$ to be the certain knowledge switching network formed from $G'$ with the following steps:
\begin{enumerate}
\item Replace all edge labels $e$ with $p_r(e)$
\item Replace all knowledge sets $K_{v'}$ in the certain knowledge description with $p_r(K_{v'})$.
\item Contract all edges in the switching network which now have label $s \to s$. When contracting an 
edge $e'$ with endpoints $v'$ and $w'$, we may choose either $K_{v'}$ or $K_{w'}$ to be the knowledge 
set for the resulting vertex.
\end{enumerate}
\end{definition}
\begin{proposition}\label{projectingisvalidprop}
Given a certain knowledge switching network $G'$ for directed connectivity on $V(G)$,
\begin{enumerate}
\item $p_l(G')$ is a certain knowledge switching network for directed connectivity on $V(G) \setminus R$. 
Furthermore, for any vertex $w' \in V(p_l(G'))$, for all of the vertices $v' \in V(G')$ which were 
contracted into $w'$, $K_{w'} \equiv p_l(K_{v'})$.
\item $p_r(G')$ is a certain knowledge switching network for directed connectivity on $V(G) \setminus L$. 
Furthermore, for any vertex $w' \in V(p_r(G'))$, for all of the vertices $v' \in V(G')$ which were 
contracted into $w'$, $K_{w'} \equiv p_r(K_{v'})$.
\end{enumerate}
\end{proposition}
\begin{proof}
We prove the first claim, the proof for the second claim is similar. To prove the first claim, it is sufficient to show the following.
\begin{enumerate}
\item $p_l(K_{s'}) = K_{s'}$
\item $p_l(K_{t'}) = K_{t'}$
\item For any knowledge sets $K_{u'},K_{v'}$ and any possible edge $e$ which is not a right edge, if $K_{u'} \cup \{e\} \equiv K_{v'} \cup \{e\}$ then $p_l(K_{u'}) \cup \{p_l(e)\} \equiv p_l(K_{v'}) \cup \{p_l(e)\}$.
\item For any knowledge sets $K_{u'},K_{v'}$, if $K_{u'} \equiv K_{v'}$ or $K_{u'} \cup \{e\} \equiv K_{v'} \cup \{e\}$ for some right edge $e$ then $p_l(K_{u'}) \equiv p_l(K_{v'})$
\end{enumerate}
The first two statements are trivial. For the third and fourth statements, we consider the effect of $p_l$ on each type of move in the modified certain knowledge game.
\begin{enumerate}
\item If we originally added or removed an edge $e$ from $K$ after directly seeing $e$, if $e$ was not a right edge then we now add or remove $p_l(e)$ from $p_l(K)$ after directly seeing $p_l(e)$. If $e$ was a right edge then we now do nothing.
\item If we originally added or removed an edge $v_3 \to v_5$ from $K$ after noting that $v_3 \to v_4, v_4 \to v_5 \in K$, then if $p_l(v_3),p_l(v_4),p_l(v_5)$ are all distinct we now add or remove $p_l(v_3 \to v_5)$ from $p_l(K)$. If $p_l(v_3),p_l(v_4),p_l(v_5)$ are not all distinct two of them must be equal to $t$. In all of these cases we now do nothing. If $p_l(v_3) = p_l(v_4) = t$ then $p_l(v_3 \to v_5) = p_l(v_4 \to v_5)$. This means that $p_l(K \cup \{v_3 \to v_5\}) = p_l(K) = p_l(K \setminus \{v_3 \to v_5\})$. Similar logic applies if $p_l(v_4) = p_l(v_5) = t$. Finally, if $p_l(v_3) = p_l(v_5) = t$ then $p_l(v_3 \to v_5) = t \to t$ so we again have that $p_l(K \cup \{v_3 \to v_5\}) = p_l(K) = p_l(K \setminus \{v_3 \to v_5\})$.
\item If we originally added or removed an edge $e \neq s \to t$ after noting that $s \to t \in K$, if $e$ was not a right edge we now 
add or remove an $p_l(e) \neq s \to t$ after noting that $s \to t \in p_l(K)$. If $e$ was a right edge then we now do nothing.
\end{enumerate}
Using Proposition \ref{sequencetoequivalence}, statements 3 and 4 follow directly from these observations.
\end{proof}
We now define a slightly different projection to each half. These projections will help us look at the progress towards removing obsolete information after obtaining a left-jumping or right-jumping edge.
\begin{definition}
Given a knowledge set $K$,
\begin{enumerate}
\item Define $p^*_l(K) = \{p_l(e): e \in K, p_l(e) \neq t \to t, p_l(e) \neq s \to t\}$.
\item Define $p^*_r(K) = \{p_r(e): e \in K, p_r(e) \neq s \to s, p_r(e) \neq s \to t\}$.
\end{enumerate}
\end{definition}
\begin{definition}
Given a certain knowledge switching network $G'$ together with a knowledge description of $G'$, 
define $p^*_l(G')$ to be the certain knowledge switching network formed from $G'$ with the following steps:
\begin{enumerate}
\item Delete $t'$ and all other vertices $v'$ such that $K_{v'} \equiv K_{t'}$ from $G'$
\item Delete all edges $e'$ such that $e'$ has an endpoint $v'$ and label $e$ and $K_{v'} \cup \{e\} \equiv K_{t'}$
\item Replace all edge labels $e$ with $p_l(e)$
\item Replace all knowledge sets $K_{v'}$ in the certain knowledge description with $p^*_l(K_{v'})$.
\item Contract all edges in the switching network which now have label $t \to t$. When contracting an 
edge $e'$ with endpoints $v'$ and $w'$, we may choose either $K_{v'}$ or $K_{w'}$ to be the knowledge 
set for the resulting vertex.
\item Add the vertex $t'$ to $G'$, assign it the knowledge set $K_{t'} = \{s \to t\}$, and add 
all labeled edges with endpoint $t'$ to $G'$ which are allowed by condition 3 of Definition 
\ref{certainknowledgedef}.
\end{enumerate}
Similarly, given a certain knowledge switching network $G'$ together with a knowledge description of $G'$, 
define $p^*_r(G')$ to be the certain knowledge switching network formed from $G'$ with the following steps:
\begin{enumerate}
\item Delete $t'$ and all other vertices $v'$ such that $K_{v'} \equiv K_{t'}$ from $G'$
\item Delete all edges $e'$ such that $e'$ has an endpoint $v'$ and label $e$ and $K_{v'} \cup \{e\} \equiv K_{t'}$
\item Replace all edge labels $e$ with $p_r(e)$
\item Replace all knowledge sets $K_{v'}$ in the certain knowledge description with $p^*_r(K_{v'})$.
\item Contract all edges in the switching network which now have label $s \to s$. When contracting an 
edge $e'$ with endpoints $v'$ and $w'$, we may choose either $K_{v'}$ or $K_{w'}$ to be the knowledge 
set for the resulting vertex.
\item Add the vertex $t'$ to $G'$, assign it the knowledge set $K_{t'} = \{s \to t\}$, and add 
all labeled edges with endpoint $t'$ to $G'$ which are allowed by condition 3 of Definition 
\ref{certainknowledgedef}.\end{enumerate}
\end{definition}
\begin{proposition}\label{lookingisvalidprop}
Given a certain knowledge switching network $G'$ for directed connectivity on $V(G)$,
\begin{enumerate}
\item $p^*_l(G')$ is a certain knowledge switching network for directed connectivity on $V(G) \setminus R$. 
Furthermore, for any vertex $w' \in V(p^*_l(G'))$, for all of the vertices $v' \in V(G')$ which were 
contracted into $w'$, $K_{w'} \equiv p^*_l(K_{v'})$.
\item $p^*_r(G')$ is a certain knowledge switching network for directed connectivity on $V(G) \setminus L$. 
Furthermore, for any vertex $w' \in V(p^*_r(G'))$, for all of the vertices $v' \in V(G')$ which were 
contracted into $w'$, $K_{w'} \equiv p^*_r(K_{v'})$.
\end{enumerate}
\end{proposition}
\begin{proof}
We prove the first claim, the proof for the second claim is similar. To prove the first claim, it is sufficient to show the following.
\begin{enumerate}
\item $p^{*}_l(K_{s'}) = K_{s'}$
\item For any knowledge sets $K_{u'},K_{v'}$ and any possible edge $e$ which is not a right edge, if $K_{u'} \cup \{e\} \equiv K_{v'} \cup \{e\} \not\equiv K_{t'}$ then $p^{*}_l(K_{u'}) \cup \{p_l(e)\} \equiv p^{*}_l(K_{v'}) \cup \{p_l(e)\}$.
\item For any knowledge sets $K_{u'},K_{v'}$, if $K_{u'} \equiv K_{v'} \not\equiv K_{t'}$ or $K_{u'} \cup \{e\} \equiv K_{v'} \cup \{e\} \not\equiv K_{t'}$ for some right edge $e$ then $p^{*}_l(K_{u'}) \equiv p^{*}_l(K_{v'})$
\end{enumerate}
The first statement is trivial. For the second and third statements, we consider the effect of $p^{*}_l$ on each type of move in the modified certain knowledge game.
\begin{enumerate}
\item If we originally added or removed an edge $e$ from $K$ after directly seeing $e$, if $e$ was not a right edge then we now add or remove $p_l(e)$ from $p_l(K)$ after directly seeing $p_l(e)$. If $e$ was a right edge then we now do nothing.
\item If we originally added or removed an edge $v_3 \to v_5$ from $K$ after noting that $v_3 \to v_4, v_4 \to v_5 \in K$, then if $p_l(v_3),p_l(v_4),p_l(v_5)$ are all distinct we now add or remove $p_l(v_3 \to v_5)$ from $p_l(K)$. Note that 
we cannot have $v_3 = s$ and $v_5 = t$ because we are assuming that we never have a knowledge set $K$ such that $K \equiv K_{t'}$. If $p_l(v_3),p_l(v_4),p_l(v_5)$ are not all distinct two of them must be equal to $t$. Following the same logic as before, in all of these cases we now do nothing. 
\item We do not have to consider moves where $s \to t \in K$ because we are assuming that we never have a knowledge set $K$ such that $K \equiv K_{t'}$.
\end{enumerate}
Using Proposition \ref{sequencetoequivalence}, statements 2 and 3 follow directly from these observations.
\end{proof}
Now that we have defined these projections, we give two more useful definitions and then prove Lemma \ref{specialboundlemma}.
\begin{definition} \noindent
\begin{enumerate}
\item We say a vertex $v'$ on a walk $W'$ satisfies the lemma for the left half if $K_{v'} \not\equiv K_{t'}$, 
$V(K_{v'}) \subseteq \{v_1, \cdots, v_{l-1}\}$, and $|V(K_{v'}) \cap L| \geq \lceil{\lg(l)}\rceil - 1$.
\item We say a vertex $v'$ on a walk $W'$ satisfies the lemma for the right half if $K_{v'} \not\equiv K_{t'}$, 
$V(K_{v'}) \subseteq \{v_1, \cdots, v_{l-1}\}$, and $|V(K_{v'}) \cap R| \geq \lceil{\lg(l)}\rceil - 1$.
\end{enumerate}
\end{definition}
\begin{definition} \noindent
\begin{enumerate}
\item We say a knowledge set $K$ is left-free if $K \not\equiv K_{t'}$ and $V(K) \cap L = \emptyset$.
\item We say a knowledge set $K$ is right-free if $K \not\equiv K_{t'}$ and $V(K) \cap R = \emptyset$.
\end{enumerate}
\end{definition}
We now prove Lemma \ref{specialboundlemma} by induction. The base case $l = 2$ is trivial. If $l > 2$ then 
given a walk $W'$ from $v'_{start}$ to $v'_{end}$ whose edge labels are all in $E(P)$, 
first modify $W'$ and $G'$ slightly as follows. Let $u'$ be the first vertex on $W'$ such that 
if $e$ is the label of the edge after $u'$ then $K_{u'} \cup \{e\} \equiv K_{t'}$. If $u'$ is not 
the vertex immediately before $v'_{end}$ then add an edge from $u'$ to $v'_{end}$ in $G'$ with label 
$e$ and replace the portion of the path from $u'$ to $v'_{end}$ with this single edge. Note that if the lemma 
is still satisfied now then it was satisfied originally. This modification ensures that 
we do not have to worry about moves in the modified certain knowledge game where we have $s \to t \in K$.

We now show that $W'$ must have at least one vertex $v'$ which satisfies the lemma for the left half. To see this, 
apply the projection $p_l$ to $G'$ and $W'$. $p_l(K_{v'_{start}}) \equiv K_{s'}$ and $p_l(K_{v'_{end}}) \equiv K_{t'}$, so by the inductive hypothesis there must be some vertex $w'$ on $p_l(W')$ such that $V(K_{w'}) \subseteq L$ and $|V(K_{w'})| \geq \lceil{\lg{l}-1}\rceil$. Choose a $v'$ which was contracted into $w'$ by $p_l$. $V(K_{v'}) \subseteq \{v_1, \cdots, v_{l-1}\}$ and $|V(K_{v'}) \cap L| = |V(K_{w'})| \geq \lceil{\lg{l}-1}\rceil$, so $v'$ satisfies the lemma for the left half, as needed. Following similar logic, $W'$ must also contain a vertex satisfying the lemma for the right half.

Now take $b'$ to be the last vertex on $W'$ which either satisfies the lemma for the left half or satisfies the lemma for the right half. Without loss of generality, we may assume that $b'$ satisfies the lemma for right half. We may also assume that $K_{b'}$ is left-free, as otherwise $b'$ satisfies Lemma \ref{specialboundlemma}. There are now two cases to consider. Either $K_{b'}$ contains a left-jumping edge, or it does not.

If $K_{b'}$ does not contain a left-jumping edge, then apply $p_l$ to the portion of $W'$ between $b'$ and $t'$. $p_l(K_{b'}) = \{\}$ and $p_l(K_{t'}) = K_{t'}$ so following similar logic as before there must be a vertex $a'$ on the portion of $W'$ between $b'$ and $t'$ which satisfies the lemma for the left half. However, this contradicts the definition of $b'$. 

If $K_{b'}$ does contain a left-jumping edge then choose a sequence of moves in the modified certain knowledge game for going along $W'$. Let $K$ be the first knowledge set we obtain such that $\bar{K}$ conatins a left-jumping edge and for every $K_2$ after $K$ but before $K_{b'}$, $\bar{K_2}$ contains a left-jumping edge. $K$ occurs when we are transitioning between some vertices $v'$ and $w'$ in $W'$ along an edge $e'$ with label $e$. 

Note that $p^{*}_l(K) \equiv K_{t'}$. This implies that $p^{*}_l(K_{v'}) \cup p_l(e) \equiv p^{*}_l(K_{w'}) \cup p_l(e) \equiv K_{t'}$. Now consider the portion of $p^{*}_l(W')$ from $p^{*}_l(v')$ to $p^{*}_l(b')$ and replace $p^{*}_l(v')$ with $t'$. Since $K_{b'}$ is left-free, $p^{*}_l(K_{b'}) = K_{s'}$. Using the inductive hypothesis, there must be a vertex $w'$ between $t'$ and $b'$ on $p^{*}_l(W')$ such that $V(K_{w'}) \subseteq L$ and $|V(K_{w'})| \geq \lceil{\lg{l}-1}\rceil$. Choose a vertex $a'$ which was contracted into $w'$. $a'$ satisfies the lemma for the left half. $K_{a'}$ occurs beetween $K$ and $K_{b'}$ as we move along $W$, so $\bar{K}_{a'}$ also contains a left-jumping edge which implies that $V(K_a)$ contains a vertex in $R$. Thus, $a'$ satisfies the conditions of Lemma \ref{specialboundlemma} and this completes the proof.
\end{proof}
\section{The power of non-monotone switching networks for directed connectivity}\label{powerofnonmonotone}
\noindent Unfortunately, proving lower size bounds on all switching networks solving the directed connectivity problem is much harder than 
proving lower size bounds on monotone switching networks solving the directed connectivity problem. 
Non-monotone switching networks can use the information that edges are not there in the input graph, which can be very powerful. 
In this section, we show that there are small sound non-monotone switching networks for directed connectivity 
on $n$ vertices which accept all of the inputs in $\mathcal{P}_{n}$, so the bound of Theorem \ref{bigresult} does not hold for non-monotone 
switching networks.
\begin{definition}
Given a set $I$ of input graphs on a set $V(G)$ of vertices with distinguished vertices $s,t$ where each graph 
in $I$ contains a path from $s$ to $t$, let $s(I)$ be the size of the smallest sound switching network 
for directed connectivity on $V(G)$ which accepts all of the input graphs in $I$.
\end{definition}
\begin{theorem}\label{pathsareeasy}
For all $n$, $s(\mathcal{P}_{n}) \leq n^3 + 2$.
\end{theorem}
\begin{proof}
\noindent The intuitive idea is as follows. If the input graph consists of just a path from $s$ to $t$, 
it is easy to find this path; we just have to follow it. If we are at some vertex $v_1$ and 
see that there is an edge from $v_1$ to $v_2$ and no other edges going out from $v_1$, 
then we can move to $v_2$ and we can forget about $v_1$ because the only place to go from $v_1$ is $v_2$. 
We only need to remember around $3\lg{n}$ bits of information. We need to remember what $v_1$ and $v_2$ 
are and we need to remember how many other possible edges going out from $v_1$ we have confirmed are 
not in $G$. We now give a rigorous proof:
\begin{definition}\label{pathfinderdef}
Define $G'_{pathfinder}(V(G))$ to be the non-monotone switching network for directed connectivity on $V(G)$ 
constructed as follows:
\begin{enumerate}
\item Start with $G'_c(n,2)$ (see Definition \ref{universalcertainknowledge})
\item For each pair of distinct ordered vertices $v_1,v_2 \in \Vmst$, add a path of length $n-1$ 
between $v'_{\{v_1\}}$ and $v'_{\{v_1,v_2\}}$ in parallel to the edge labeled $v_1 \to v_2$ between 
$v'_{\{v_1\}}$ and $v'_{\{v_1,v_2\}}$. Give the edges in this path the labels 
$\{\neg{(v_2 \to u)}: u \in V(G) \setminus \{s,v_1,v_2\}\}$
\end{enumerate}
\end{definition}
\begin{proposition}\label{pathfindersize}
For all $n$, $|V(G'_{pathfinder}(V(G)))| \leq n^3 + 2$
\end{proposition}
\begin{proof}
There are $n$ vertices of the form $v'_{\{v\}}$ where $v \in \Vmst$. For each of these vertices 
$v'_{\{v\}}$, there are $n-1$ added paths which have $v'_{\{v\}}$ as an endpoint and each of these 
paths adds $n-2$ vertices. Thus, 
$$|V(G'_{pathfinder}(V(G)))| - |V(G'_c(n,2))| = n(n-1)(n-2)$$ 
$|V(G'_c(n,2)) \setminus \{s',t'\}| = \binom{n}{2} + n + 2$ so 
$$|V(G'_{pathfinder}(V(G))) \setminus \{s',t'\}| = n(n-1)(n-2) + \binom{n}{2} + n + 2 \leq n^3 + 2$$ 
as needed.
\end{proof}
\begin{proposition}\label{pathfinderprop}
$G'_{pathfinder}(V(G))$ accepts all input graphs in $\mathcal{P}_{n}$.
\end{proposition}
\begin{proof}
If $G$ is an input graph with vertex set $V(G)$ and edges 
$E(G) = \{v_i \to v_{i+1}: i \in [0,l-1]\}$ where $v_0 = s$ and $v_l = t$ then we have a path from $s'$ to $t'$ whose 
edges are all consistent with $G$ as follows.
\begin{enumerate}
\item If we are at $s'$ then go to $v'_{\{v_1\}}$ along the edge labeled $s \to v_1$.
\item If we are at $v'_{\{v_i\}}$ for any $i \in [1,l-2]$ then go to $v'_{\{v_i,v_{i+1}\}}$ along the edge labeled $v_{i} \to v_{i+1}$. 
\item If we are at $v'_{\{v_i,v_{i+1}\}}$ for any $i \in [1,l-2]$, then for all $u \in V(G) \setminus \{s,v_i,v_{i+1}\}$, 
$v_{i+1} \to u \notin E(G)$. Thus we can go to $v'_{\{v_{i+1}\}}$ along the path between $v'_{\{v_{i+1}\}}$ and $v'_{\{v_i,v_{i+1}\}}$.
\item If we are at $v'_{\{v_{l-1}\}}$ then go to $t'$ along the edge labeled $v_{l-1} \to t$
\end{enumerate}
\end{proof}
\begin{lemma}\label{usingabsences}
$G'_{pathfinder}(V(G))$ is sound.
\end{lemma}
\begin{proof}
\begin{definition}
Given an input graph $G$, create an input graph $G_a$ as follows. Let 
$$E_a = \{v \to w: v,w \in V(G) \backslash \{s,t\}, v \neq w, \forall u \in V(G) \backslash \{s,v,w\}, w \to u \notin E(G)\}$$ 
Take $V(G_a) = V(G), E(G_a) = E(G) \cup E_a$.
\end{definition}
\begin{proposition}\label{augmentedinputusefulness}
If $G'_{pathfinder}(V(G))$ accepts an input graph $G$ then $G'_c(n,2)$ accepts the corresponding input graph $G_a$.
\end{proposition}
\noindent Using Proposition \ref{augmentedinputusefulness}, to prove Lemma \ref{usingabsences} it is sufficient to show that 
for any input graph $G$ there is a path from $s$ to $t$ in $G_a$ only if there is a path from $s$ to $t$ in $G$. 
To show this, assume that there is no path from $s$ to $t$ for some input graph $G$. Then let $V$ be the set of all vertices $v$ such 
that there is a path from $v$ to $t$ in $G$. Since there is no path from $s$ to $t$ in $G$, $s \notin V$. Let $C$ be the cut with 
$R(C) = V$. Note that there cannot be any edge in $G$ that crosses $C$.

Assume there is an edge in $G_a$ which crosses $C$. Then it must be an edge $v \to w$ in $E_a \setminus E(G)$ and we must have $v \in L(C)$, $w \in R(C)$. This implies that there is a path from $w$ to $t$. However, by the definition of $E_a$, $w \neq t$ and $\forall u \in V(G) \backslash \{s,v,w\}, w \to u \notin E(G)$. Thus, any path from $w$ to $t$ must go through $v$, so there must be a path from $v$ to $t$ and we should have that $v \in R(C)$. Contradiction.

There is no edge in $G_a$ which crosses $C$, so there is no path from $s$ to $t$ in $G_a$, as needed.
\end{proof}
Theorem \ref{pathsareeasy} now follows immediately from Proposition\ref{pathfindersize}, 
Proposition \ref{pathfinderprop}, and Lemma \ref{usingabsences}.
\end{proof}
\begin{example}
The switching network in Figure \ref{nonmonotone} is $G'_{pathfinder}(\{s,t,a,b\})$ with some edges removed. The top vertex has knowledge set $K_{\{a\}}$, the bottom vertex has knowledge set $K_{\{b\}}$ and the center vertex has knowledge set $K_{\{a,b\}}$
\end{example}
\end{appendix}
\received{February 2007}{March 2009}{June 2009}
\end{document}